\documentclass{aa}  

\usepackage{natbib}
\usepackage{graphicx}
\usepackage{txfonts}
\usepackage{hyperref}
\usepackage{mathrsfs}
\usepackage{color}

\hypersetup{
   colorlinks=true,
   citecolor=blue,
   linkcolor=blue,
   urlcolor=black
   }

\newcommand{\ed}{{\rm e}}
\newcommand{\dd}{{\rm d}}

\definecolor{emerald}{rgb}{0.3,0.85,0.2}
\definecolor{smcolor}{rgb}{0.7,0.3,0.0}

\newcommand{\jlc}[1]{#1}
\newcommand{\padc}[1]{#1}
\newcommand{\rec}[1]{#1}
\newcommand{\recc}[1]{#1}

\begin{document} 
  \title{Semidiurnal thermal tides in asynchronously rotating hot Jupiters}
  \author{P. Auclair-Desrotour\inst{1}
     \and
           J. Leconte\inst{1}  
          }

  \institute{Laboratoire d'Astrophysique de Bordeaux, Univ. Bordeaux, CNRS, B18N, allée Geoffroy Saint-Hilaire, 33615 Pessac, France\\
              \email{pierre.auclair-desrotour@u-bordeaux.fr, jeremy.leconte@u-bordeaux.fr}
             }

  \date{Received ...; accepted ...}

  \abstract
   {Thermal tides can torque the atmosphere of hot Jupiters into asynchronous rotation, while these planets are usually assumed to be locked into spin-orbit synchronization with their host star.}
   {In this work, our goal is to characterize the tidal response of a rotating hot Jupiter to the tidal semidiurnal thermal forcing of its host star, by identifying the structure of tidal waves responsible for variation of mass distribution, their dependence on the tidal frequency and their ability to generate strong zonal flows.}
   {We develop an \textit{ab initio} global modeling that generalizes the early approach of \cite{AS2010} to rotating and non-adiabatic planets. We derive analytically the torque exerted on the body and the associated timescales of evolution, as well as the equilibrium tidal response of the atmosphere in the zero-frequency limit. Finally, we integrate numerically the equations of thermal tides for three cases including dissipation and rotation step by step. }
   {The resonances associated with tidally generated gravito-inertial waves amplify significantly the resulting tidal torque in the range 1-30 days. This torque can drive globally the atmosphere into asynchronous rotation, as its sign depends on the tidal frequency. The resonant behaviour of the tidal response is enhanced by rotation, which couples the forcing to several Hough modes in the general case, while the radiative cooling tends to regularize it and diminish its amplitude.    }
   {}

  \keywords{hydrodynamics -- planet-star interactions -- waves -- planets and satellites: atmospheres -- planets and satellites: gaseous planets}

\maketitle


\section{Introduction} 

Modeling the general circulation of hot Jupiters is a key element in \jlc{constraining observationally} their properties (temperature structure, day-night heat transport, circulation regime). Particularly, it allows to establish a link between these properties and the Doppler-shift in the transmission spectra of the planets that can be measured in orbital phase curves of secondary eclipses \citep[e.g.][]{Rauscher2014}. Because of their proximity to their host star, hot Jupiters orbiting circularly are generally assumed to be locked into spin-orbit synchronous rotation, meaning that their rotation rate is exactly equal to their orbital frequency \citep[see][and references therein]{Showman2015}. The argument invoked for this assumption is the mechanism of gravitational tides, which torques the planet toward this state of equilibrium. Given the strength of the tidal torque, the timescale associated with this evolution \citep[a few million years, see][]{SG2002,OL2004} is short compared to that associated with the evolution of the planetary system. Thus, tidal forces should lock the planet into synchronous rotation before they circularize its orbit \citep[e.g.][]{Rasio1996}. However, other arguments have been given in recent works in favor of an asynchronous rotation, leading some authors to consider this configuration \citep[e.g.][]{Showman2009,Showman2015,Rauscher2014,TDD2014}.

These arguments invoke the transport of angular momentum between the planet's orbit and its atmosphere or interior \cite{SG2002}, the forcing of a fast superrotating equatorial jet in the atmosphere by the strong day-night heating contrast \citep[][]{Showman2015}, and the ability of thermal tides to generate asynchronous zonal flows \citep[][]{GO2009,AS2010}. This later mechanism was first introduced by \cite{GS1969} through an \textit{ad hoc} approach to explain the locking of Venus at the observed retrograde rotation rate. It results from the incoming stellar flux, which submits the atmosphere to a day-night periodic forcing. This forcing, like the tidal gravitational potential, generates density fluctuations leading to a global variation of mass distribution. The tidal torque induced by thermal tides can be in opposition with that induced by the gravitational forcing. In this case, the rotation of Venus-like planets evolves toward the asynchronous state of equilibrium where solid and atmospheric tidal torques compensate each other exactly \citep[see][]{ID1978,DI1980,CL01,CL2003,ADLM2017b}. These arguments were reinforced lately by results obtained with \textit{ab initio} models showing that the internal structure of the fluid layer and timescale associated with dissipative processes directly affect its tidal response \citep[][]{Leconte2015,ADLM2017a}. 

However, as discussed by \cite{GO2009}, there is no solid surface in hot Jupiters to support the load of a mass surplus in the atmosphere of the planet as in telluric Venus-like planets. As a consequence, \jlc{it would seem that} no net tidal bulge \jlc{could} appear and the tidal torque exerted on the atmosphere should be negligible. This seems to be \jlc{odds} with the conclusions of \cite{AS2010} about semidiurnal thermal tides, who argue for a strong amplification of the tidal torque in the range of forcing periods 1-30~days due to the propagation of resonant internal waves \jlc{\citep[see also][]{Lubow1997}}. In this early work, the \jlc{dynamical effect of} rotation of the planet was ignored, although it is directly related to the forcing frequency. Similarly,  \jlc{dissipative processes such as radiative cooling were not taken into account although they dissipate the energy of tidally generated gravity modes \citep[][]{Terquem1998}}. Thus, we propose here to revisit the study by \cite{AS2010} by using a similar \textit{ab initio} global modeling. In this linear approach, we include both the coupling induced by rotation between excited tidal modes and the tidal forcing, and the dissipation resulting from radiative/diffusive cooling, which is modeled by a Newtonian cooling. We aim at (1) clarifying the internal structure of the tidal response by identifying dominating modes, their characteristics and their behaviour, (2) quantifying the order of magnitude of timescales associated with the forcing of zonal-mean flows by the semidiurnal thermal tide, and (3) characterizing how the rotation and radiative cooling affect the thermal tidal response and the resulting tidal torque.

Hence, in Section~\ref{sec:tidal_waves_dyn}, we detail the physical setup of the model, establish the equations describing the structure of forced tidal waves, and discuss the used boundary conditions and gravitational and thermal tidal forcings. In Section~\ref{sec:tidal_torque}, the expressions of the tidal torques and quadrupoles are given, and the associated evolution timescales introduced. Then, in Section~\ref{sec:properties_waves}, we compute the tidal response of the planet due to the thermal component and the associated evolution timescales in three cases: (a) in a static and adiabatic planet \citep[case treated by][]{AS2010}, (b) in a static planet with radiative cooling, and (c) in a rotating planet with radiative cooling. We show that a resonant behaviour can arise from the reflection of gravity waves on the boundaries of the stably-stratified radiative zone. The radiative cooling tend to attenuate the amplitude of the response, while rotation increases it in the zero-frequency limit. In Section~\ref{sec:spectra_torque}, we compute the frequency spectra of the total tidal torque in the three above cases. In the static case, we reproduce the results previously obtained by \cite{AS2010} and show that rotation amplifies the resonant behaviour of the tidal response. \jlc{Synchronization timescales associated with the gravitational and thermal tidal components are computed in Section~\ref{sec:spin_evolution}.} We finally discuss the consequences of thermal tides on zonal-flows and the used approximations in Section~\ref{sec:discussion}, and give our conclusions in Section~\ref{sec:conclusions}.

\section{Tidal waves dynamics}
\label{sec:tidal_waves_dyn}

In this section, we establish the equations describing the tidal response of a rotating Jupiter submitted to both gravitational and thermal forcings. Thermal tides have been examined before in different ways \citep[][]{GO2009,AS2010,Leconte2015}. \jlc{Here, we use the formalism developed by \cite{ADLM2017a} with the physical setup of \cite{AS2010}.}


\subsection{Physical setup and background distributions}
\label{subsec:setup}
\begin{figure}
   \centering
   \includegraphics[width=0.45\textwidth]{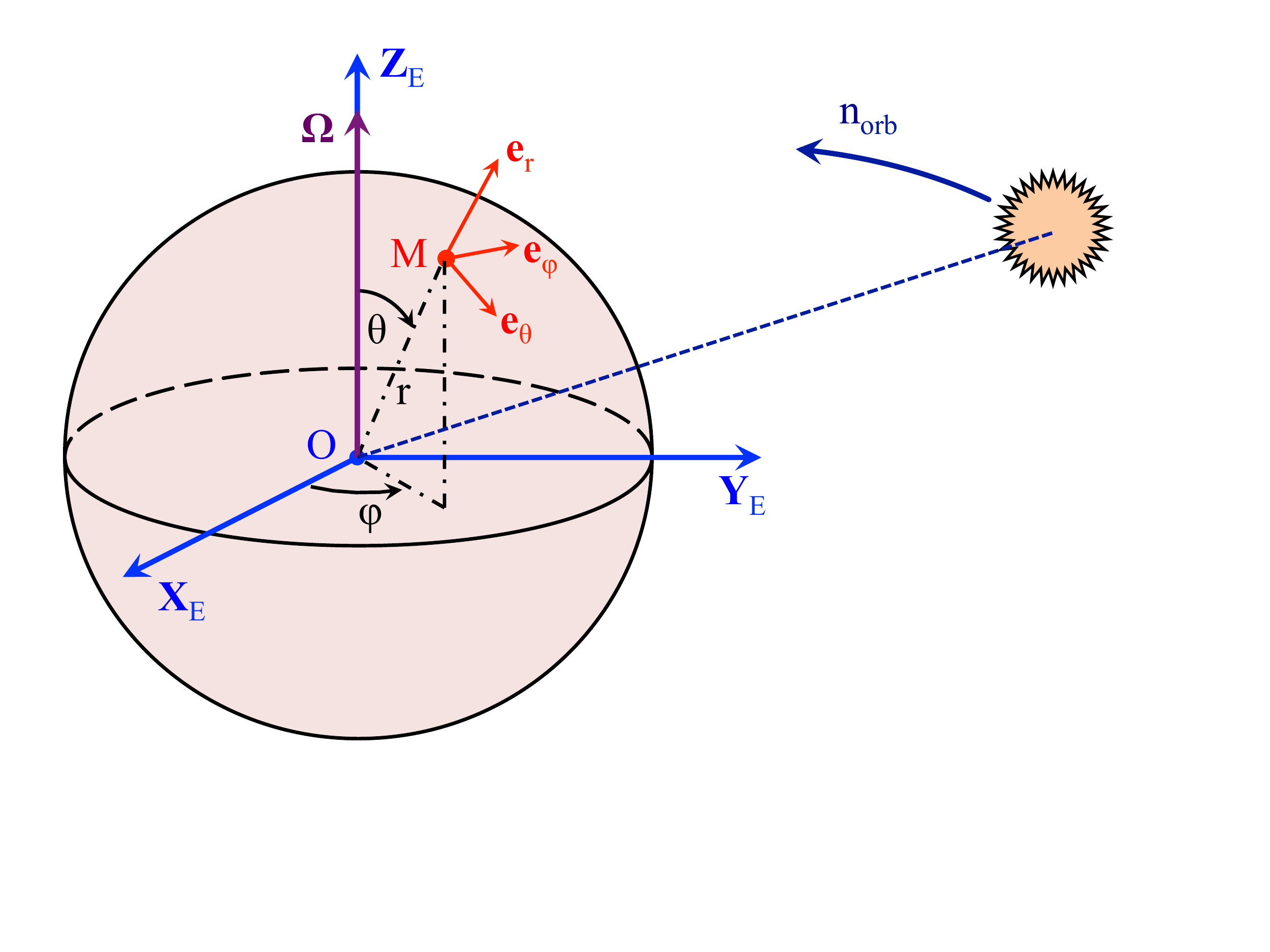}
   \caption{Reference frames and systems of coordinates. The notations $ \boldsymbol{\Omega} $ and $ n_{\rm orb} $ designate the rotation vector and the orbital angular velocity respectively.}
       \label{fig:system}%
\end{figure}

We consider a Jovian planet of mass $ M_{\rm p} $ and radius $ R_{\rm p} $, rotating on itself at the spin angular velocity $ \Omega $, and orbiting its host star at the dynamical frequency $ n_{\rm orb} $. In order to avoid mathematical complications related to internal circulation, we assume in this study that the planet is in uniform rotation (this simplification is discussed in Section~\ref{sec:discussion}). The associated co-rotating reference frame centered on the center of inertia of the body is denoted $ \mathcal{R}_{\rm E} : \left\{O, \textbf{X}_{\rm E} , \textbf{Y}_{\rm E} , \textbf{Z}_{\rm E} \right\} $, where $ \textbf{Z}_{\rm E} = \boldsymbol{\Omega}/ \left| \boldsymbol{\Omega} \right| $, $ \boldsymbol{\Omega} $ being the rotation vector of the planet, and $ \textbf{X}_{\rm E} $ and $ \textbf{Y}_{\rm E} $ designate two directions of the equatorial plane. To locate a point $ M $ of the planet, we use the spherical vectorial basis $ \left( \textbf{e}_r , \textbf{e}_\theta ,  \textbf{e}_\varphi \right)  $ and coordinates $ \left( r , \theta , \varphi \right) $, which refer to the radius, the colatitude and the longitude respectively (see Fig.~\ref{fig:system}). Hence, the position vector is expressed as $ \textbf{r} = r \, \textbf{e}_r $. 

The internal structure of the planet is defined by spatial distributions of pressure $ p_0 $, density $ \rho_0 $ and gravity $ g $. To simplify it, we ignore day-night temperature gradients and the effect of the centrifugal acceleration. This approximation is appropriate \jlc{provided that $ \Omega \ll \Omega_{\rm c} $}, $ \Omega_{\rm c} = \sqrt{g/r} $ being the \emph{critical Keplerian angular velocity}. The equation of hydrostatic balance is written

\begin{equation}
\frac{d^2 p_0}{dr^2} + \left[ \frac{2}{r} - \left( \frac{d \ln \rho_0}{d p_0} \right) \frac{d p_0}{dr} \right] \frac{d p_0}{dr} + 4 \pi \mathscr{G} \rho_0^2 = 0,
\label{hydro_balance}
\end{equation}

\noindent the parameter $ \mathscr{G} $ standing for the gravitational constant. Considering that the planet is basically composed of a deep convective envelope and a superficial stably-stratified atmosphere, we adopt the equation of state proposed by \cite{AS2010} to fully characterize the internal structure. This equation writes 

\begin{equation}
\rho_0 \left( p_0 \right) = \ed^{- p_0 / p_{\rm b}} \left( \frac{p_0}{a^2} \right) + \left( 1 - \ed^{- p_0 / p_{\rm b}} \right) \left( \frac{p_0}{K_{\rm c}} \right)^{\frac{1}{\Gamma_1}},
\label{eq_etat}
\end{equation}

\noindent where we have introduced the adiabatic exponent $ \Gamma_1 = \left( \partial \ln p_0 / \partial \ln \rho_0 \right)_S $ (the index $ S $ referring to specific macroscopic entropy), the pressure at the base of the stably-stratified layer $ p_{\rm b} $, \rec{the characteristic pressure specifying the entropy of the core $ K_{\rm c} = \mathscr{G} R_{\rm p}^2 $}, and the isothermal sound speed of the envelope $ a =  \left( p_{\rm b} K_{\rm c} \right)^{1/4} $. The transition between the two layers thus occurs at $ p = p_{\rm b} $. Eq.~(\ref{hydro_balance}) is then integrated upward using a Runge-Kutta scheme of the fourth order with the regularity condition $ d p_0 / dr = 0 $ at $ r = 0 $. We deduce from the gravity profile the profile of mass contained within the sphere of radius $ r $, expressed as

\begin{equation}
M \left( r \right) = 4 \pi \int_0^r r'^2 \rho_0 \left( r' \right) \dd r'.
\end{equation}

\noindent Hence, the pressure at the center of the planet is iterated to make $ M $ converge toward $ M_{\rm p} $ at the upper limit, \rec{denoted $R_{\rm e}$. Note that $ R_{\rm e} $ determines the pressure level at the upper limit of the atmosphere. It differs from $ R_{\rm p} $, which is the photospheric radius of the planet, and can be arbitrarily chosen as soon as $ R_{\rm e} > R_{\rm p} $. In this work, we set $ R_{\rm e} = 1.01 \ R_{\rm p} $ in order to have $ p_0 \left( R_{\rm e} \right) / p_\star < 10^{-6} $ and avoid side effects at the upper boundary.} 

Other background distributions are deduced straightforwardly from $ p_0 $, $ \rho_0 $ and $ g $. The vertical profiles of pressure height and sound velocity are expressed as 

\begin{equation}
\begin{array}{rcl}
\displaystyle H = \frac{p_0}{g \rho_0} & \mbox{and} & \displaystyle c_{\rm s} = \sqrt{\frac{\Gamma_1 p_0}{\rho_0}}. 
\end{array}
\end{equation}

\noindent The stratification of the fluid with respect to convection is characterized by the Brunt-Väisälä frequency $ N $, defined by 

\begin{equation}
N^2  = - g \left[ \frac{d \ln \rho_0}{d r} - \frac{1}{\Gamma_1} \frac{d \ln p_0}{dr} \right].
\label{N2def}
\end{equation}

\noindent Finally, we assume that the fluid is a perfect gas uniform in composition, of molar mass $ \mathcal{M}_{\rm g} $ and specific gas constant $ \mathcal{R}_{\rm s} = \mathcal{R}_{\rm GP} / \mathcal{M}_{\rm g} $, the notation $ \mathcal{R}_{\rm GP} $ referring to the perfect gas constant. Denoting $ \kappa = \left(\Gamma_1 - 1 \right) / \Gamma_1 $, we get the background profiles of temperature and \jlc{heat} capacity per unit mass,

\begin{equation}
\begin{array}{rcl}
\displaystyle T_0 = \frac{g H}{\mathcal{R}_{\rm s}}  & \mbox{and} & \displaystyle C_p = \frac{\mathcal{R}_{\rm s}}{\kappa}.
\end{array}
\end{equation}

In this work, following \cite{LM1967} \citep[see also][]{Dickinson1968}, we take into account the effect of internal dissipation on tidal waves by using a \emph{Newtonian cooling}, i.e. by considering that the local thermal losses $ J_{\rm d} $ due to radiative and diffusive processes are proportional to the temperature variation $ \Delta T $. As demonstrated by \cite{Iro2005} with numerical simulations in the case HD209458b, this approach is justified if $ \Delta T / T_0 $ does not exceed 5\%, which corresponds well to the framework of our linear modeling. The local dissipated power per unit mass can thus be written 

\begin{equation}
J_{\rm d} = C_p \sigma_0 \Delta T,
\label{Newtonian_cooling}
\end{equation}

\noindent where $ \sigma_0 $ designates the effective radiative frequency associated with diffusion and radiation. This parameter is the reciprocal of the \rec{thermal timescale $ \tau_0 $. The tidal response does not depend much on the vertical profile of $ \tau_0 $. However, it depends on its order of magnitude in the region where the stellar heating is absorbed, typically around the $ p_\star $ pressure level. Therefore, to set $ \tau_0 $, we choose to use the vertical profile computed numerically by \cite{Iro2005} in the case of HD 209458b using an advanced model of thermal transfers (see Fig.~4 in their article). This profile shows two tendencies: \recc{$\tau_0 \propto p_0^{1/2}$ for $ p_0 \lesssim p_\star $ and $ \tau_0 \propto p_0^{2}$ for $ p_0 \gtrsim p_\star $}. We thus approximate it by the empirical scaling law}

\begin{equation}
\tau_0 = \frac{\tau_{\star}}{2} \left[  \left( \frac{p_0}{p_{\star}} \right)^{\frac{1}{2}}  +  \left( \frac{p_0}{p_{\star}} \right)^2 \right],
\end{equation}

\noindent the parameter $ \tau_{\star} $ standing for the radiative time at the base of the heated layer, at the characteristic pressure $ p_\star $ \citep[$ \tau_{\star} \sim 1-10 $ days and $ p_\star \approx 1 $ bar in the case of HD 209458b, see][]{SG2002,Cho2003,Iro2005}. This modeling mimics the two regimes of thermal time, with a transition occurring at $ p_0 \sim p_\star $. For $ p_0 < p_\star $, $  \tau_0 $ increases slowly with pressure. Below the $ p_\star $-level, \jlc{pressure broadening increases the opacity rapidly}. As a consequence, the effect of radiative cooling becomes negligible. Introducing $ \sigma_{\star} = 2 \pi / \tau_{\star} $, we define the radiative frequency as

\begin{equation}
\sigma_0 = 2 \sigma_{\star} \left[  \left( \frac{p_0}{p_{\star}} \right)^{\frac{1}{2}}  +  \left( \frac{p_0}{p_{\star}} \right)^2 \right]^{-1}.
\label{sigma0_scaling}
\end{equation}

\noindent In the following, $ \tau_{\star} $ will be used as a control parameter to specify the efficiency of the radiative cooling\jlc{, the $ \tau_\star = + \infty $ limit corresponding to the adiabatic case.}

\begin{figure*}
   \centering
   \includegraphics[width=0.45\textwidth,trim = 1.5cm 1.8cm 1.5cm 1.5cm,clip]{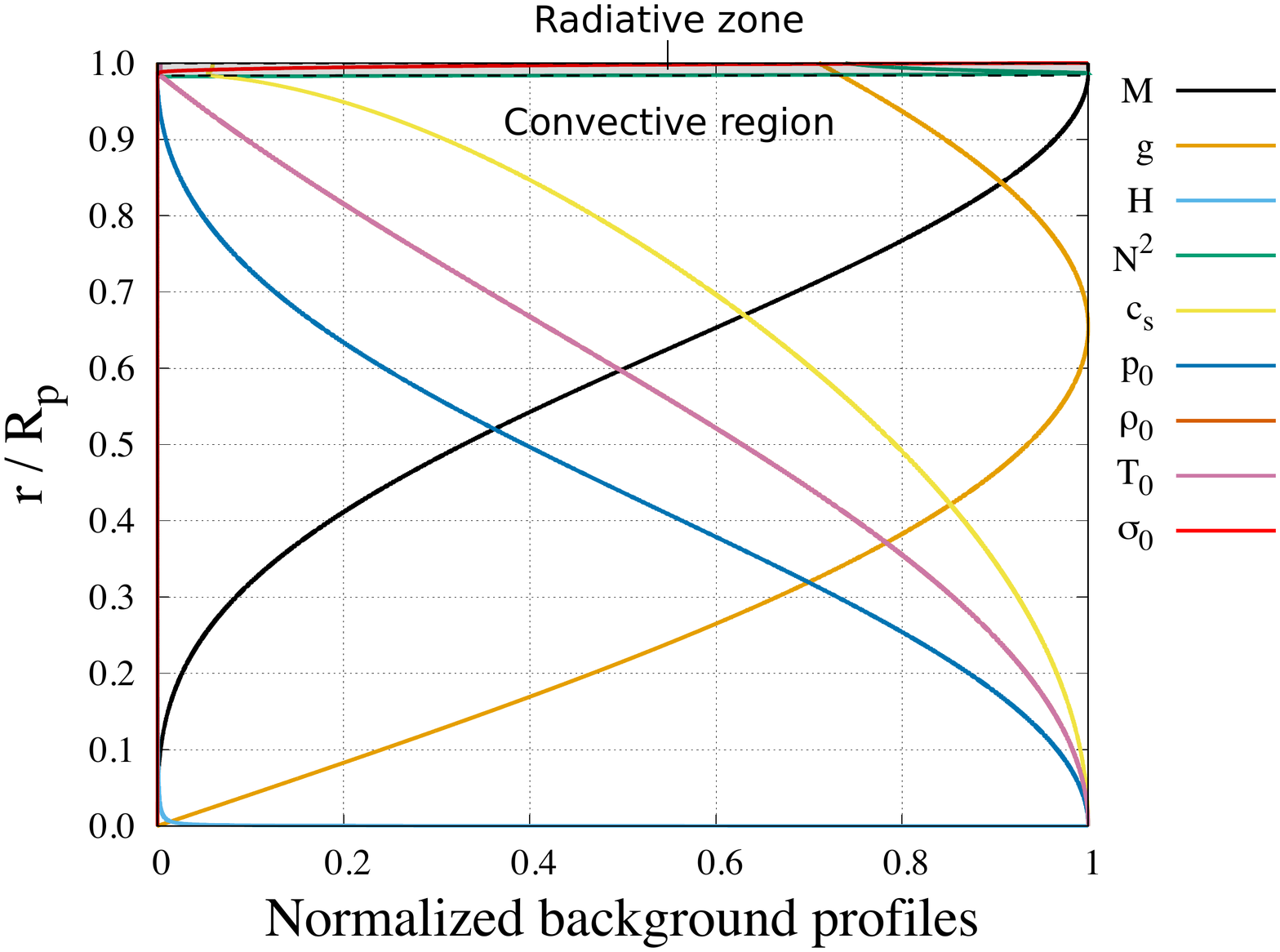} \hspace{1cm}
   \includegraphics[width=0.45\textwidth,trim =  1.5cm 1.8cm 1.5cm 1.5cm,clip]{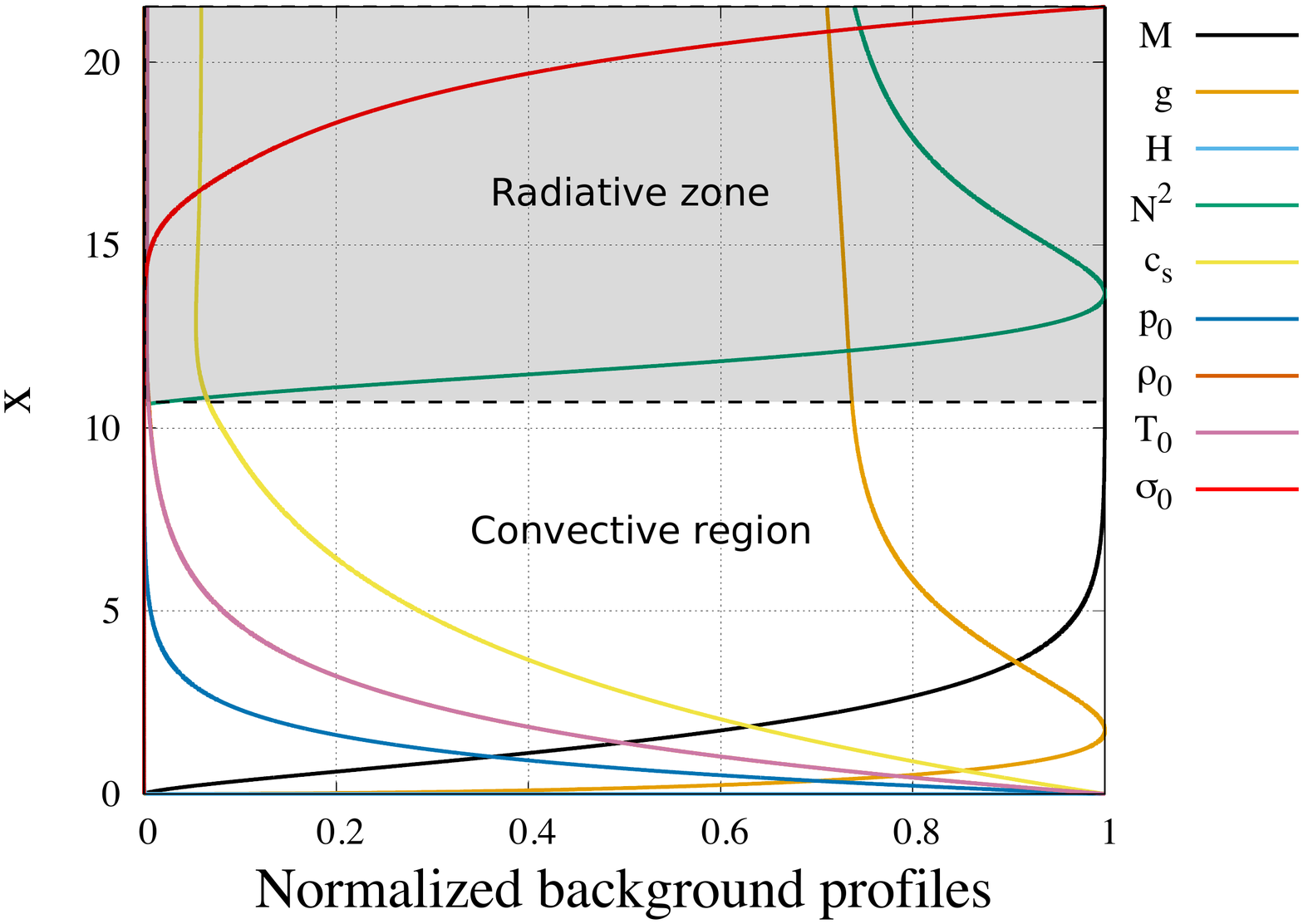}
   \caption{Background profiles normalized by their maxima as functions of the normalized radius $ r / R_{\rm p} $ (left panel) and of the pressure altitude $ x $ (right panel). Values of parameters are those given in Section~\ref{subsec:setup}, that is $ M_{\rm p} = 0.7 M_{\rm J} $, $ R_{\rm p} = 1.27 R_{\rm J} $, $ K_{\rm c} = \mathscr{G} R_{\rm J}^2 $, $ \Gamma_1 = 1.4 $, $ p_{\rm b} = 100 $ bar, $ p_\star = 1 $ bar and $ \tau_\star = 1 $ day. \jlc{The grey (white) area corresponds to the radiative (convective) region.} }
       \label{fig:bgd}%
\end{figure*}

As the thermal forcing generating thermal tides is absorbed in the upper layers of the atmosphere \citep[in the case of HD~209458b, 99.99\% of the incoming stellar flux is absorbed before reaching the 7 bar level, see][]{Iro2005}, the tidal perturbation mainly affects the stably-stratified region. Therefore, it is convenient to choose as radial coordinate the reduced altitude defined by

\begin{equation}
x = \int_0^r \frac{\dd r'}{H \left( r' \right)}
\end{equation}

\noindent rather than the radius $ r $ \citep[][]{CL70}. Indeed, this change of coordinates expands the domain of the stably stratified atmosphere \citep[see][for gravitational atmospheric tides in Jovian planets]{IL1993a,IL1993b,IL1994}, which allows us to increase the vertical resolution in this region. \jlc{Another} way to expand the heated radiative zone would be to choose the optical depth $ z_{\rm op} $ measured from $ r = + \infty $ as radial coordinate, as done by \cite{GO2009}. With this coordinate, the upper limit of the atmosphere corresponds to $ z_{\rm op} = 0 $ and the center of the planet to $ z_{\rm op} = + \infty $.

In order to compare our results to those obtained previously by \cite{AS2010}, we use the same values of physical parameters. Hence, denoting $ M_{\rm J} $ and $ R_{\rm J} $ the Jupiter's mass and radius, we set the mass and radius of the planet to $ M_{\rm p} = 0.7 \, M_{\rm J} $ and $ R_{\rm p} = 1.27 \, R_{\rm J} $, the core compressibility to $ K_{\rm c} = \mathscr{G} R_{\rm J}^2 $, the adiabatic exponent to $ \Gamma_1 = 1.4 $, the pressure at the base of the radiative atmosphere to $ p_{\rm b} = 100 $ bar and the pressure at the base of the heated layer to $ p_\star = 1 $ bar. The corresponding background profiles are plotted on Fig.~\ref{fig:bgd} as functions of $ r $ (left panel) and $ x $ (right panel) with $ \tau_\star = 1 $ day. By comparing the two plots, we observe that the proportion of the vertical domain occupied by the stably-stratified atmosphere ($ N^2 \neq 0 $) switches from a few percents with $ r $ to almost a half of the total domain with $ x $, as mentioned above. The base of the stably-stratified layer thus corresponds approximately to $ x = 11 $. 

\subsection{Structure and regimes of tidal waves}
\label{ssec:structure_regimes_waves}

\jlc{To establish the structure of tidal waves, we summarize the main lines of the formalism detailed in \cite{ADLM2017a}.} The planet is submitted to the tidal gravitational and thermal forcings of its host star. The gravitational forcing is due to the tidal gravitational potential $ U $, such that the tidal force is defined by $ \textbf{F} = \nabla U $ \citep[we follow the convention of][]{Zahn1966a}, and the tidal heating to the heat power per unit mass $J$. These forcings generate tidal winds of velocity $ \textbf{V} = \left( V_r , V_\theta , V_\varphi \right) $ as well as fluctuations of pressure ($ \delta p $), density ($ \delta \rho $) and temperature ($ \delta T $), which are assumed to be small compared to background quantities in this linear approach. Thus, conserving only terms of the first order in $ \textbf{V} $, $ \delta p $, $ \delta \rho $ and $ \delta T $ and neglecting the fluctuations of the self-gravitational potential \citep[this is the so-called \emph{Cowling approximation}, see][]{Cowling1941}, we write the momentum equation \citep[][]{ADLM2017a}

\begin{equation}
\partial_t \textbf{V} + 2 \boldsymbol{\Omega} \times \textbf{V} = - \frac{1}{\rho_0} \nabla \delta p - \frac{g}{\rho_0} \delta \rho \, \textbf{e}_r + \nabla U,
\label{momentum_equation}
\end{equation}

\noindent where $ t $ designates the time and $ \partial_X = \partial / \partial_X  $ the partial derivative along the $ X $ coordinate. The dynamics are completed by the equation of mass conservation,

\begin{equation}
\partial_t \delta \rho + \nabla \cdot \left( \rho_0 \textbf{V} \right) = 0,
\end{equation}

\noindent the equation of energy,

\begin{equation}
\frac{1}{\rho_0} \left( \frac{1}{c_{\rm s}^2} \partial_t \delta p - \partial_t \delta \rho \right) + \frac{N^2}{g} V_r = \frac{1}{T_0} \left( \frac{J}{C_p} - \sigma_0 \delta T \right),
\label{energy_equation}
\end{equation}

\noindent and the equation of perfect gas,

\begin{equation}
\frac{\delta p}{p_0} = \frac{\delta T}{T_0} + \frac{\delta \rho}{\rho_0}. 
\label{perfect_gas}
\end{equation}

\noindent It shall be noted that this set of primitive equations is very similar to that used by \cite{AS2010}. We have only added the effect of rotation by taking into account the Coriolis acceleration ($2 \boldsymbol{\Omega} \times \textbf{V}$) in the momentum equation, Eq.~(\ref{momentum_equation}), and the effect of radiative/diffusive processes through the Newtonian cooling term ($ \sigma_0 \delta T $) in the equation of energy, Eq.~(\ref{energy_equation}). 

We now look for solutions of Eqs.~(\ref{momentum_equation}-\ref{perfect_gas}) both periodic in time and longitude. Thus, a perturbed quantity $ f $ is expanded into Fourier series, 

\begin{equation}
f \left( t , \textbf{r} \right) = \sum_{m,\sigma} f^{m,\sigma} \left( x , \theta \right) \ed^{i \left( \sigma t + m \varphi \right)}, 
\label{Fourier}
\end{equation}

\noindent where we have introduced \jlc{the imaginary number $ i $,} the tidal frequency $ \sigma $ and the longitudinal wavenumber $ m $ of a given mode (typically \rec{$ m =  2 $} and \rec{$ \sigma =  2 \left( \Omega - n_{\rm orb} \right) $} for the stellar semidiurnal tide studied in the next section). In the following, the superscripts $ \left( m, \sigma \right) $ will be omitted where no confusion \jlc{arises}. 

The latitudinal projection of the rotation vector in the Coriolis acceleration (Eq.~(\ref{momentum_equation})) induces a coupling between the horizontal and the vertical projections of the equation of dynamics. Deriving analytically the structure of the tidal response requires to eliminate this coupling. Therefore, we assume the \emph{traditional approximation} \citep[e.g.][]{Unno1989}, which consists in ignoring $ 2 \Omega \sin \theta $ terms. This assumption is usually considered as appropriate in the regime of super-inertial waves, defined by $ 2 \Omega \ll \left| \sigma \right| $, where rotation hardly affects the fluid tidal response. In strongly stratified regions ($  \left| \sigma \right|  \ll N $), the previous condition becomes $ 2 \Omega \lesssim \left| \sigma \right| $, \citep[][]{Mathis2009,Prat2017}, which allows us to treat sub-inertial waves in this case (see also Section~\ref{sec:discussion}). However, as discussed by \cite{OL2004}, the traditional approximation gives an inaccurate representation of the fluid tidal response within the inertial regime ($ 2 \Omega > \left| \sigma \right| $) in convective regions, because it leads to an overestimated tidal dissipation. In this case, these authors argue that it is better to assume the \emph{static approximation}, which consists in simply ignoring rotation, as \cite{AS2010} do. 

\rec{In the next sections, we will focus on thermal tides and ignore the component of the tidal response generated by the tidal potential. The traditional approximation can be assumed in this case because thermal tides only affect the thin stably-stratified atmosphere of the planet, which approximately stands for $\sim 2\%$ of the planet radius (see Fig.~\ref{fig:bgd}, left panel). Tidal waves induced by the absorption of the stellar heating are expected to propagate within the stably stratified zone and not go through its threshold, as they are mainly restored by the Archimedean force. We will verify this postulate a posteriori in Section~\ref{sec:properties_waves}, by plotting internal tidal density variations as a function of the latitude and pressure levels. The traditional approximation would lead to strongly inaccurate results if we considered the tidal component generated by the tidal gravitational potential because this forcing affect the whole planet, and particularly the thick convective region, where the assumptions mentioned above are violated. However, we choose to keep the terms associated with the gravitational tidal forcing in the following analytic development for the sake of generality, given that the present model can be applied to the case of the thin stably-stratified atmospheric layers of a terrestrial planet, where the traditional approximation can be assumed.}


Substituting Eq.~(\ref{Fourier}) into Eqs.~(\ref{momentum_equation}-\ref{perfect_gas}) and introducing the reduced pressure fluctuation $ y = \delta p / \rho_0 $, we obtain the set of equations defining the latitudinal and radial distributions of the perturbed quantities,

\begin{align}
\label{eq1}
i \sigma V_\theta^{m,\sigma} - 2 \Omega \cos \theta V_\varphi^{m,\sigma} & =  - \frac{1}{r} \partial_\theta \left( y^{m,\sigma} - U^{m,\sigma} \right), \\
i \sigma V_\varphi^{m,\sigma} + 2 \Omega \cos \theta V_\theta^{m,\sigma} & =  - \frac{i m}{r \sin \theta} \left( y^{m,\sigma} - U^{m,\sigma} \right), \\
\label{eq3}
i \sigma V_r^{m,\sigma} & = - \frac{1}{\rho_0} \partial_r \left( \rho_0 y^{m,\sigma} \right) - \frac{g}{\rho_0} \delta \rho^{m,\sigma} + \partial_r U^{m,\sigma}, 
\end{align}

\begin{align}
i \sigma \delta \rho + \frac{1}{r^2} \partial_r \left( r^2 \rho_0 V_r \right) &= - \rho_0 \nabla_\perp \cdot \textbf{V}^{m,\sigma}, \\
\label{eq5}
\frac{i \sigma + \Gamma_1 \sigma_0}{c_{\rm s}^2} y^{m,\sigma} + \frac{N^2}{g} V_r^{m,\sigma} & =  \frac{ i \sigma + \sigma_0 }{\rho_0} \delta \rho^{m,\sigma} + \frac{J^{m,\sigma}}{T_0 C_p},
\end{align}

\noindent where $ \nabla_{\perp} $ designates the horizontal part of the divergence operator. 

\begin{figure*}
   \centering
   \includegraphics[width=0.27\textwidth]{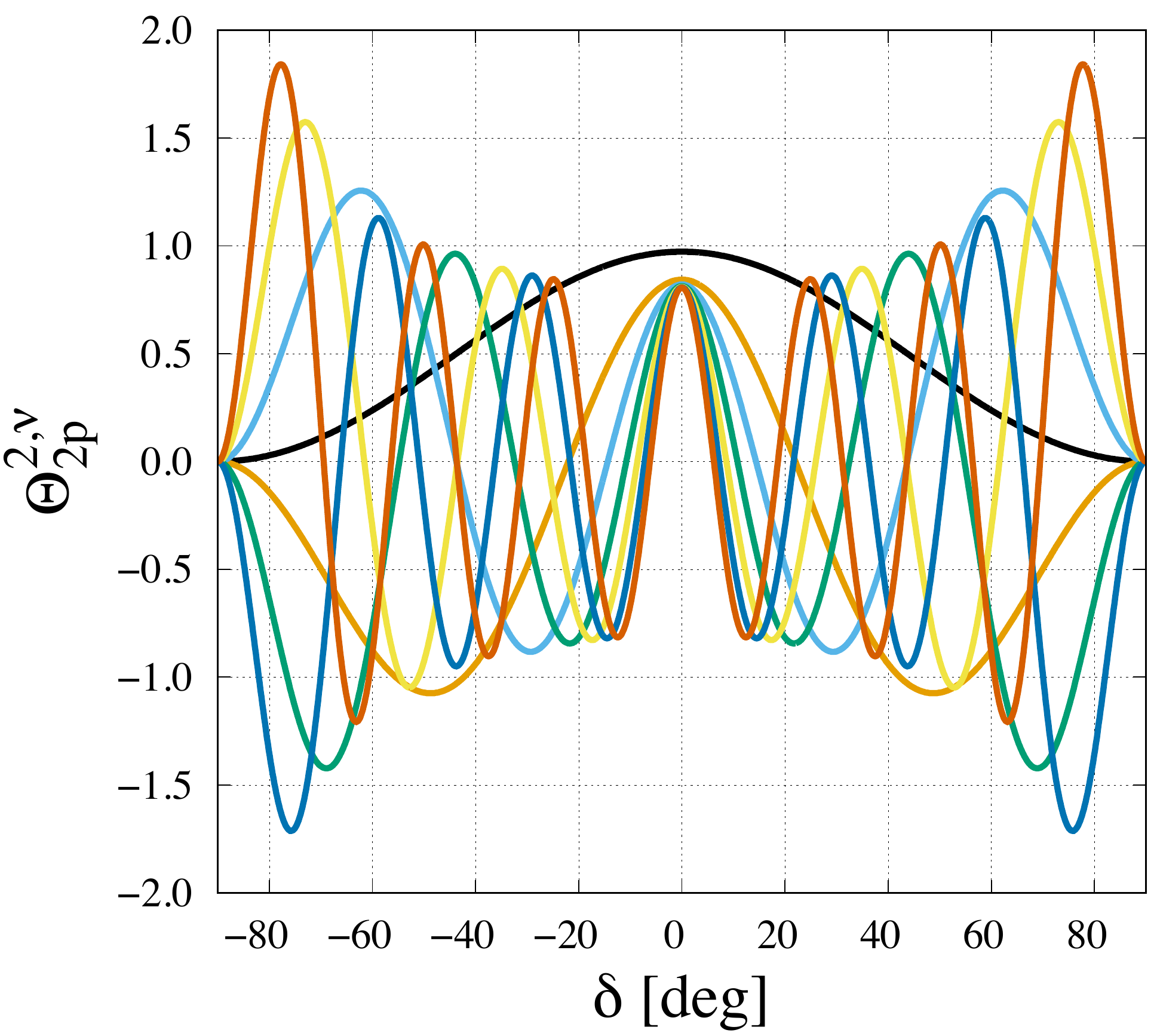} \hspace{0.5cm}
   \includegraphics[width=0.27\textwidth]{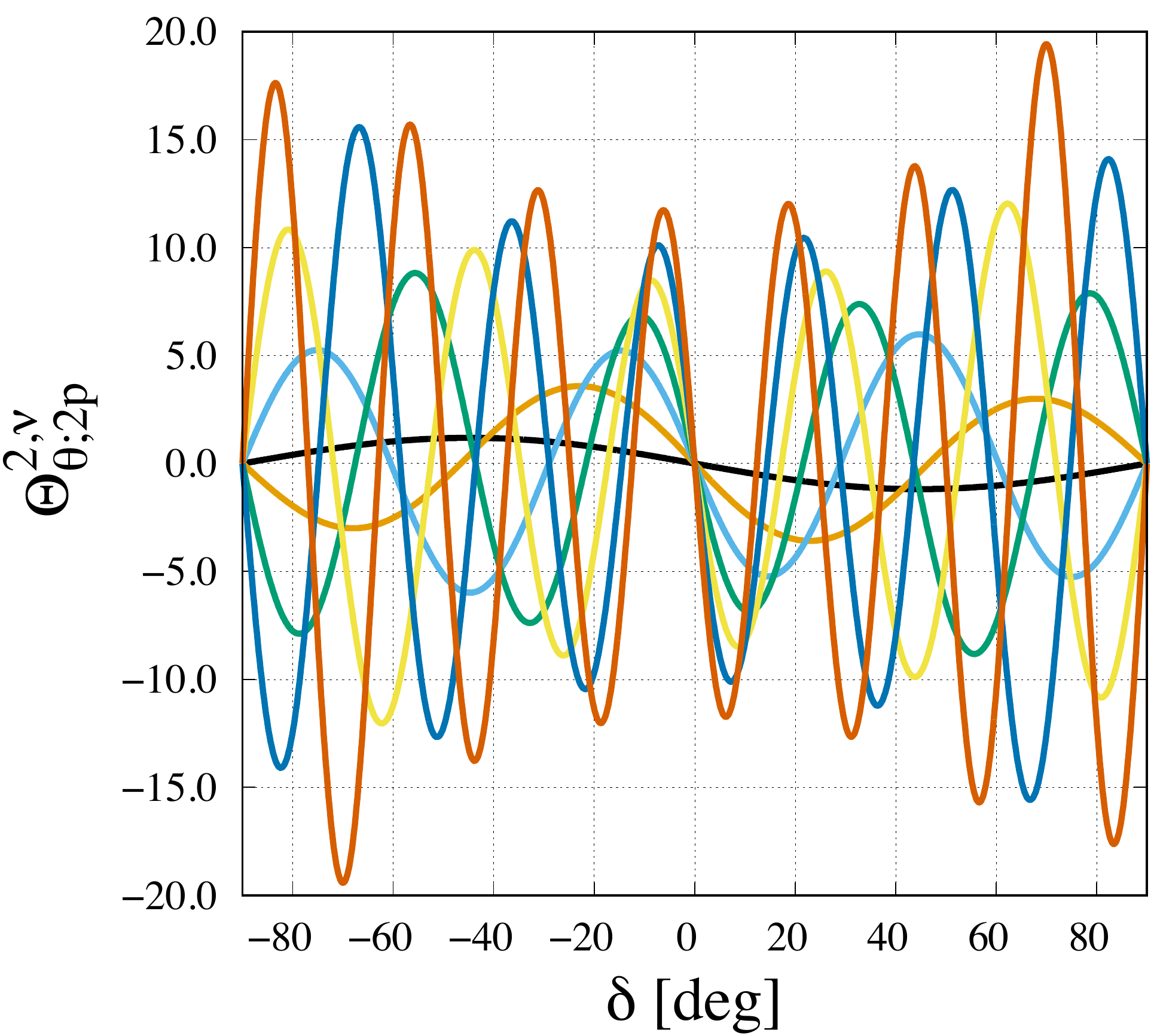} \hspace{0.5cm}
   \includegraphics[width=0.27\textwidth]{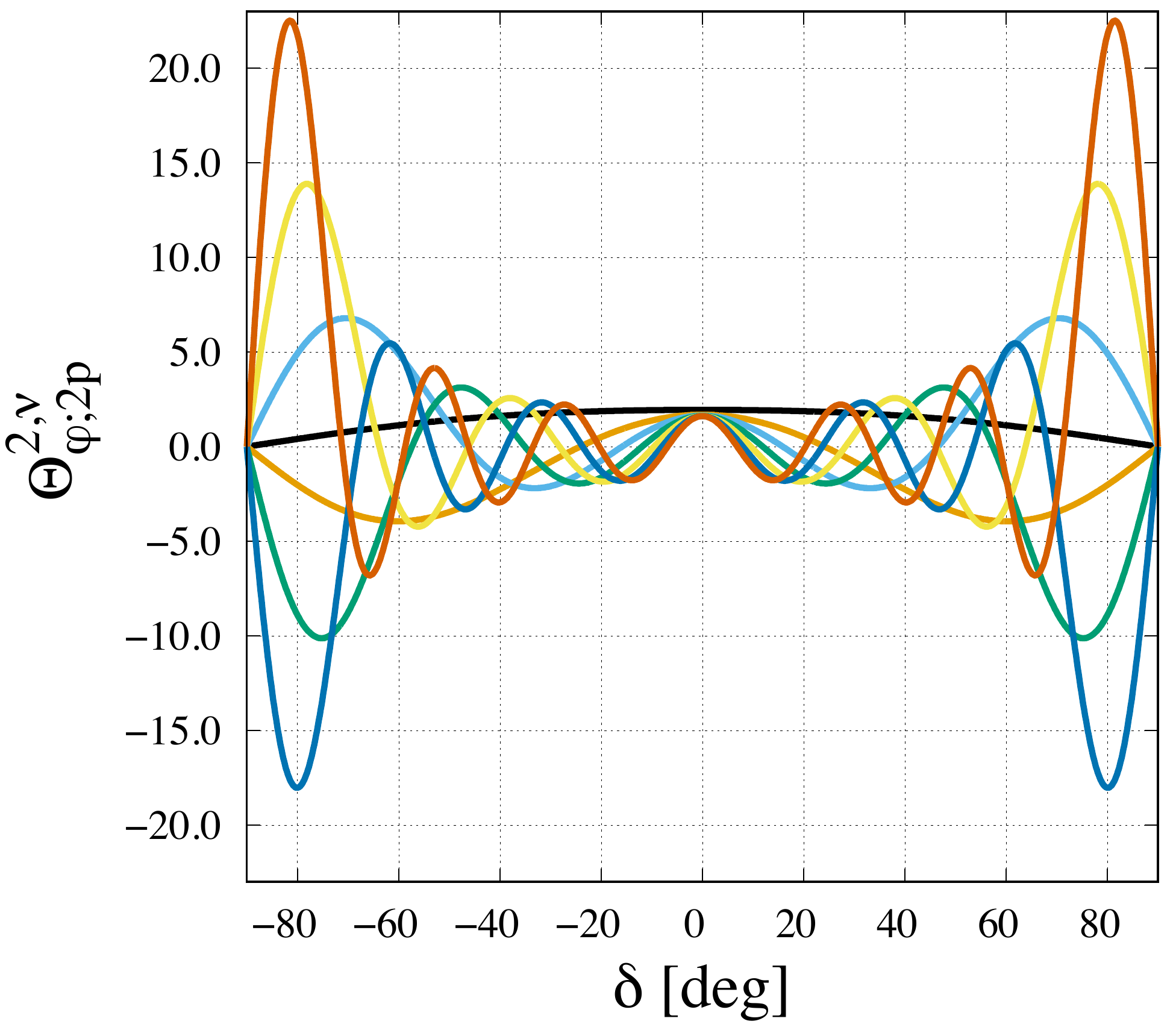} \hspace{0.5cm}
   \includegraphics[width=0.07\textwidth]{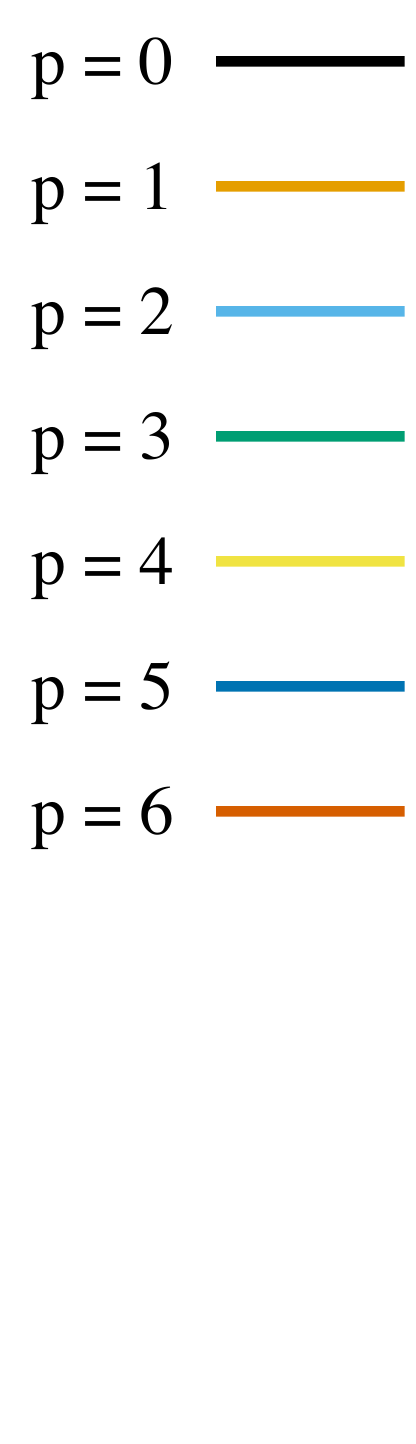} \\
   \includegraphics[width=0.27\textwidth]{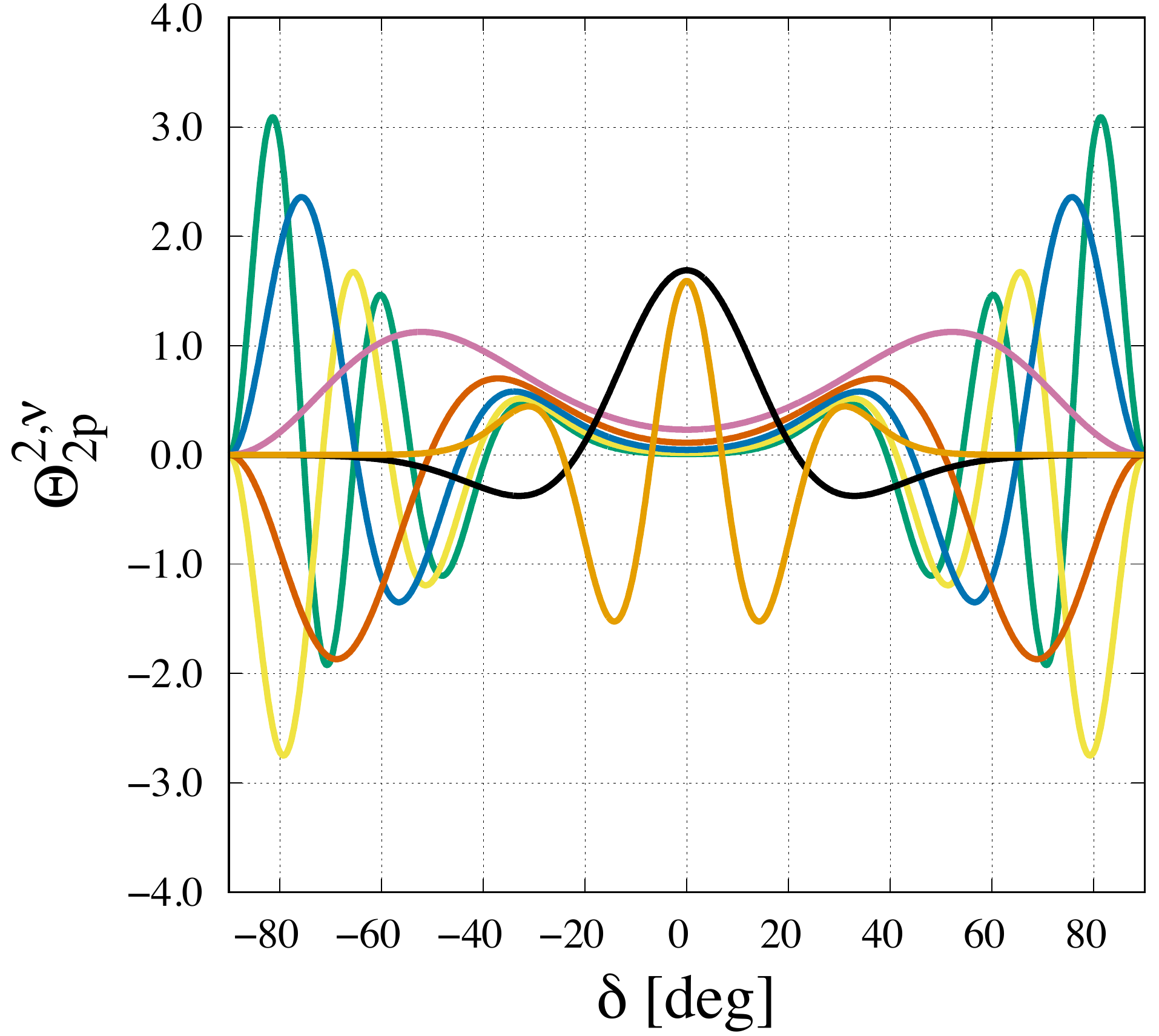} \hspace{0.5cm}
   \includegraphics[width=0.27\textwidth]{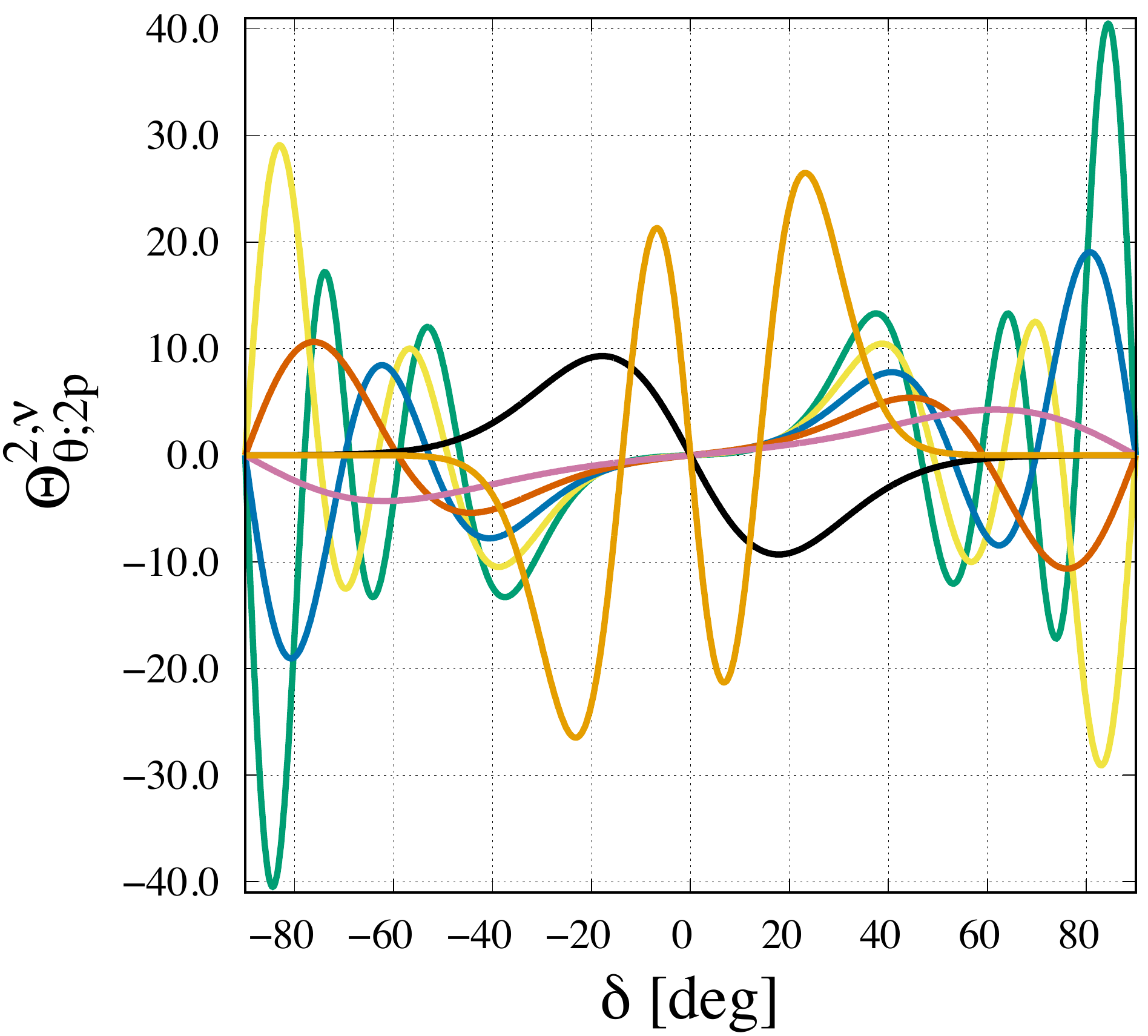} \hspace{0.5cm}
   \includegraphics[width=0.27\textwidth]{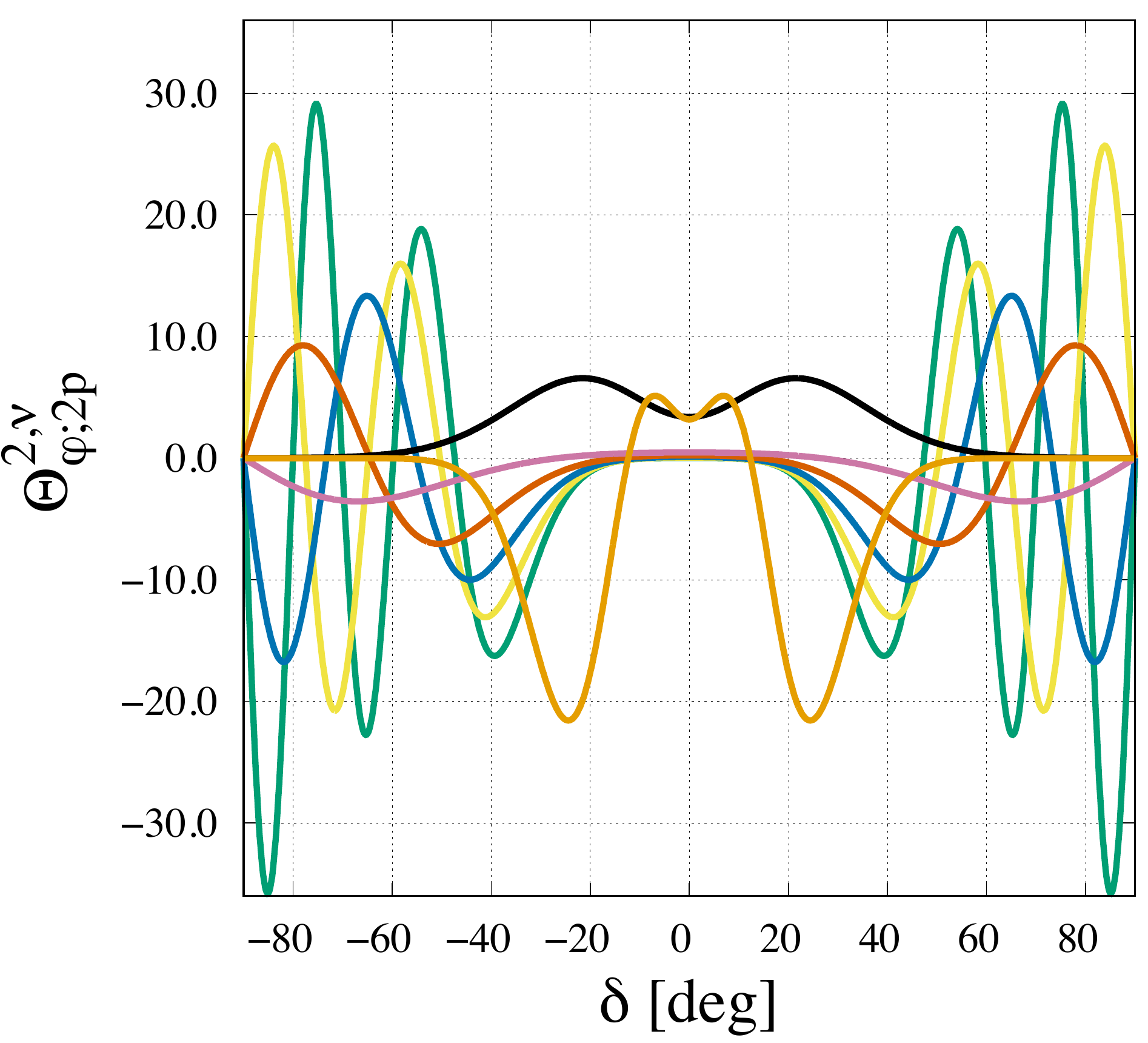} \hspace{0.45cm}
   \includegraphics[width=0.075\textwidth]{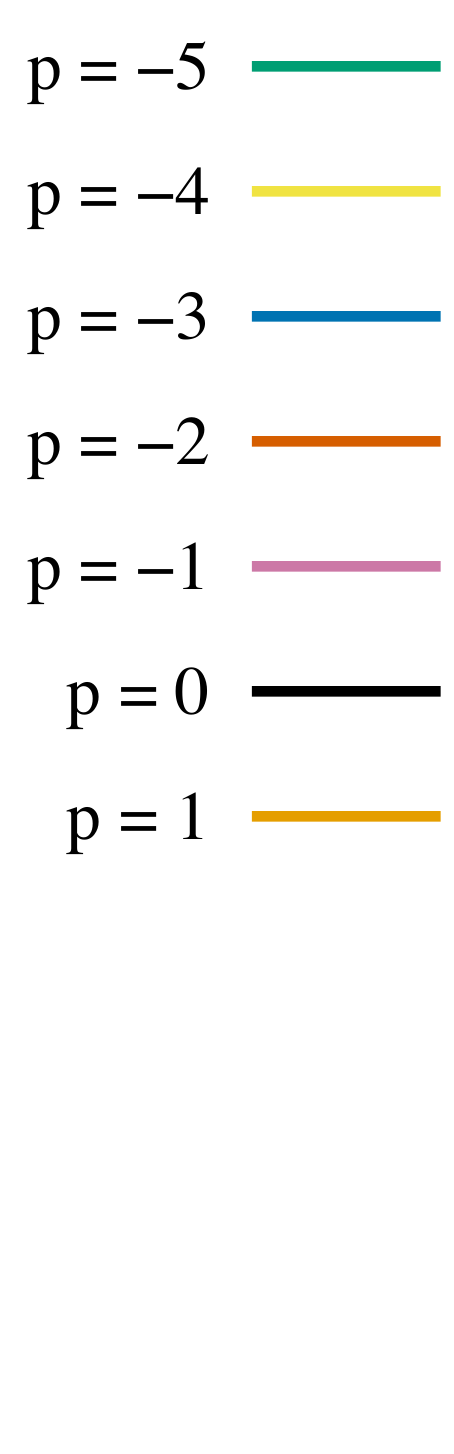}
   \caption{Normalized symmetrical ($ n $ even) Hough functions (left) and associated horizontal functions (middle and right) as functions of latitude (degrees) for $ m = 2 $ and two values of $ \nu = 2 \Omega / \sigma $ representative of inertial regimes. {\it Top:} $ \nu = 0.2 $ (super-inertial regime); weak impact of rotation on the tidal response. {\it Bottom:} $ \nu = 2 $ (sub-inertial regime); strong impact of rotation. Gravity modes are designated by subscripts $ p \geq 0 $ while Rossby modes correspond to $ p < 0 $. }
       \label{fig:Hough}%
\end{figure*}

The traditional approximation allows us to proceed to the separation of the $ x $ and $ \theta $ coordinates in solutions, and to expand Fourier coefficients of Eq.~(\ref{Fourier}) into series of \emph{Hough functions} \jlc{\citep[][]{Hough1898}}. Thus, any Fourier coefficient $  f^{m,\sigma} $ of the tidal gravitational potential, thermal forcing, pressure, density, temperature fluctuations and vertical velocity is written

\begin{equation}
f^{m,\sigma} \left( x , \theta \right) = \sum_n f_n^{m,\sigma} \left( x \right) \Theta_n^{m,\nu} \left( \theta \right),
\label{series_Hough1}
\end{equation}

\noindent while Fourier coefficients of horizontal velocities are expressed as

\begin{equation}
V_X^{m,\sigma} \left( x , \theta \right) = \sum_n V_{X ; n}^{m,\nu} \left( x \right) \Theta_{X ; n}^{m,\nu} \left( \theta \right), 
\label{series_Hough2}
\end{equation}

\noindent with $ X = \theta , \varphi $. In Eqs.~(\ref{series_Hough1}) and (\ref{series_Hough2}), we have introduced the spin parameter $ \nu $ defined by

\begin{equation}
\nu \left( \sigma \right) = \frac{2 \Omega}{\sigma},
\label{nu}
\end{equation}

\noindent  the latitudinal wavenumber $ n $, such that $ n \in \mathbb{N} $ if $ \left| \nu \right| \leq 1 $ and $ n \in \mathbb{Z} $ otherwise \citep[with the notations of][]{LS1997}, the so-called Hough functions $ \Theta_n^{m,\nu} $, and the associated horizontal functions $ \Theta_{\theta ; n}^{m,\nu} $ and $ \Theta_{\varphi ; n}^{m,\nu} $. Let us introduce the operator

\begin{align}
\mathcal{L}^{m,\nu} = & \frac{1}{\sin \theta} \frac{d}{d \theta} \left( \frac{\sin \theta}{1 - \nu^2 \cos^2 \theta} \frac{d}{d \theta} \right) \\
 & - \frac{1}{1 - \nu^2 \cos^2 \theta} \left( m \nu \frac{1 + \nu^2 \cos^2 \theta}{1 - \nu^2 \cos^2 \theta} + \frac{m^2}{\sin^2 \theta} \right). 
\end{align}

\noindent Hough functions are the solutions of the eigenvalue-eigenfunction problem defined by the \emph{Laplace's tidal equation} \citep[][]{Laplace1798,CL70,LS1997,Wang2016},

\begin{equation}
\mathcal{L}^{m,\nu} \Theta^{m,\nu} = - \Lambda^{m,\nu} \Theta^{m,\nu},
\label{Laplace_eq}
\end{equation}

\noindent integrated with regularity boundary conditions. They are associated with the eigenvalues $ \Lambda_n^{m,\nu} $ and constitute an orthogonal basis through the scalar product

\begin{equation}
\langle \Theta_n^{m,\nu} , \Theta_j^{m,\nu}  \rangle = \int_0^\pi \Theta_n^{m,\nu} \left( \theta \right) \Theta_j^{m,\nu} \left( \theta \right) \sin  \theta \, \dd \theta . 
\end{equation}

\noindent For convenience, we use the normalized Hough functions, such that $ \langle \Theta_n^{m,\nu} , \Theta_j^{m,\nu} \rangle = \delta_{n,j} $ for any $ n $ and $ j $, the notation $ \delta_{n,j} $ standing for the Kronecker symbol. Moreover we use the notation $ A_{n,l}^{m,\nu} = \langle P_n^m , \Theta_n^{m,\nu}  \rangle $ for the projection coefficients of Hough functions on the normalized associated Legendre polynomials, such that

\begin{equation}
\Theta_n^{m,\nu} = \sum_{l \geq  m} A_{n,l}^{m,\nu} P_l^m \left( \cos \theta \right). 
\label{Thetan_Pln}
\end{equation}

\noindent The associated horizontal functions intervening in Eq.~(\ref{series_Hough2}) are straightforwardly deduced from the $ \Theta_n^{m,\nu} $. They are defined by

\begin{align}
& \Theta_{\theta ; n}^{m,\nu} = \frac{1}{1 - \nu^2 \cos^2 \theta} \left( \frac{d}{d \theta} + m \nu \cot \theta \right) \Theta_n^{m,\nu}, \\
& \Theta_{\varphi ; n}^{m,\nu} = \frac{1}{1 - \nu^2 \cos^2 \theta} \left( \nu \cos \theta \frac{d}{d \theta} + \frac{m}{\sin \theta} \right) \Theta_n^{m,\nu}. 
\end{align}

Symmetric Hough functions ($ n $ even) and the associated horizontal functions are plotted on Fig.~\ref{fig:Hough} as functions of the colatitude for $ m = 2 $ (main component of the semidiurnal tide) for typical cases of the super-inertial ($ \left| \nu \right| \leq 1 $) and sub-inertial ($ \left| \nu \right| >1 $) regimes, namely $ \nu = 0.2 $ and $ \nu = 2 $. We can observe on these graphs the two families of Hough modes characterizing the horizontal structure of tidal waves \citep[e.g.][]{Unno1989,LS1997}: 
\begin{enumerate}
  \item \emph{Gravity modes} ($ n \geq 0 $), also referred to as \emph{g modes}. These modes are defined for $ \nu \in \mathbb{R} $ and associated with positive eigenvalues. When $ \nu \rightarrow 0 $, they converge toward the associated Legendre polynomials $ P_l^m $ \rec{(with $ l = m + n $)}, which are the solutions of the Laplace's tidal equation in the non-rotating case. Similarly, the associated eigenvalues converge toward those of Legendre associated polynomials, \rec{that is $ \Lambda_n^{m,0} = \left( m + n \right) \left( m + n + 1 \right) $}. 
  \item \emph{Rossby modes} ($ n < 0 $), or \emph{r modes}. These modes exist in the sub-inertial regime only ($ \left| \nu \right| > 1 $). In this regime, gravity modes are confined within an equatorial band that becomes narrower while $ \left| \nu \right| $ increases and Rossby modes spread from one pole to another (see Fig.~\ref{fig:Hough}, left panels).
\end{enumerate}

%

To obtain the vertical profiles of other perturbed quantities, we have to solve the system of Eqs.~(\ref{eq3}) to (\ref{eq5}). After some manipulations and the introduction of the displacement $  \boldsymbol{\xi} $, such that $  \textbf{V} = \partial_t  \,\boldsymbol{\xi}  $, this system is reduced to the system of ODEs 

\begin{align}
\label{ODE_eq1}
 \frac{d y_n}{dx} & = A_1 y_n  + B_1 r^2 \xi_{r ; n} + C_1, \\
 \label{ODE_eq2}
 \frac{d}{dx} \left( r^2 \xi_{r ; n} \right) & = A_2 y_n  + B_2 r^2 \xi_{r ; n} + C_2,,
\end{align}

\noindent with the coefficients 

\begin{align}
 & A_1 \left( x \right) = \frac{H N^2}{g} + \frac{i \kappa \sigma_0}{\sigma - i \sigma_0}, \\
 \label{B1vert}
 & B_1 \left( x \right) = - \frac{H}{r^2} \left( \frac{\sigma}{\sigma - i \sigma_0} N^2 - \sigma^2 \right) , \\
 & C_1 \left( x \right) = \frac{d U_n}{dx} - i \frac{\kappa}{\sigma - i \sigma_0} J_n , 
\end{align}

\begin{align}
\label{A2vert}
 & A_2 \left( x \right) = \frac{H \Lambda_n}{\sigma^2} \left( 1 - \varepsilon_{\rm s ; n} \right), \\
 & B_2 \left( x \right) =\frac{1}{\Gamma_1} - \frac{i \sigma_0}{\sigma - i \sigma_0} \frac{H N^2}{g}, \\
 & C_2 \left( x \right) = - \frac{H \Lambda_n}{\sigma^2} U_n - i \frac{\kappa r^2}{g \left( \sigma - i \sigma_0 \right)} J_n.
\end{align}

The parameter $ \varepsilon_{\rm s ; n} $ that appears in Eq.~(\ref{A2vert}) is an acoustic parameter comparing the tidal frequency to the \emph{Lamb frequency} of the $ n $-mode, 

\begin{equation}
\sigma_{\rm s ;n} = \sqrt{\Lambda_n} \frac{c_{\rm s}}{r},
\label{sigmasn}
\end{equation}

\noindent  which is the characteristic cutoff frequency of horizontally propagating acoustic modes (we also introduce the general acoustic cutoff frequency $ \sigma_{\rm s} = c_{\rm s}/r $). It writes

\begin{equation}
\varepsilon_{\rm s ; n} \left( x \right) = \frac{\sigma - i \Gamma_1 \sigma_0}{\sigma - i \sigma_0} \frac{\sigma^2}{\sigma_{\rm s ; n}^2}. 
\end{equation} 

\begin{figure}
   \centering
   \includegraphics[width=0.45\textwidth]{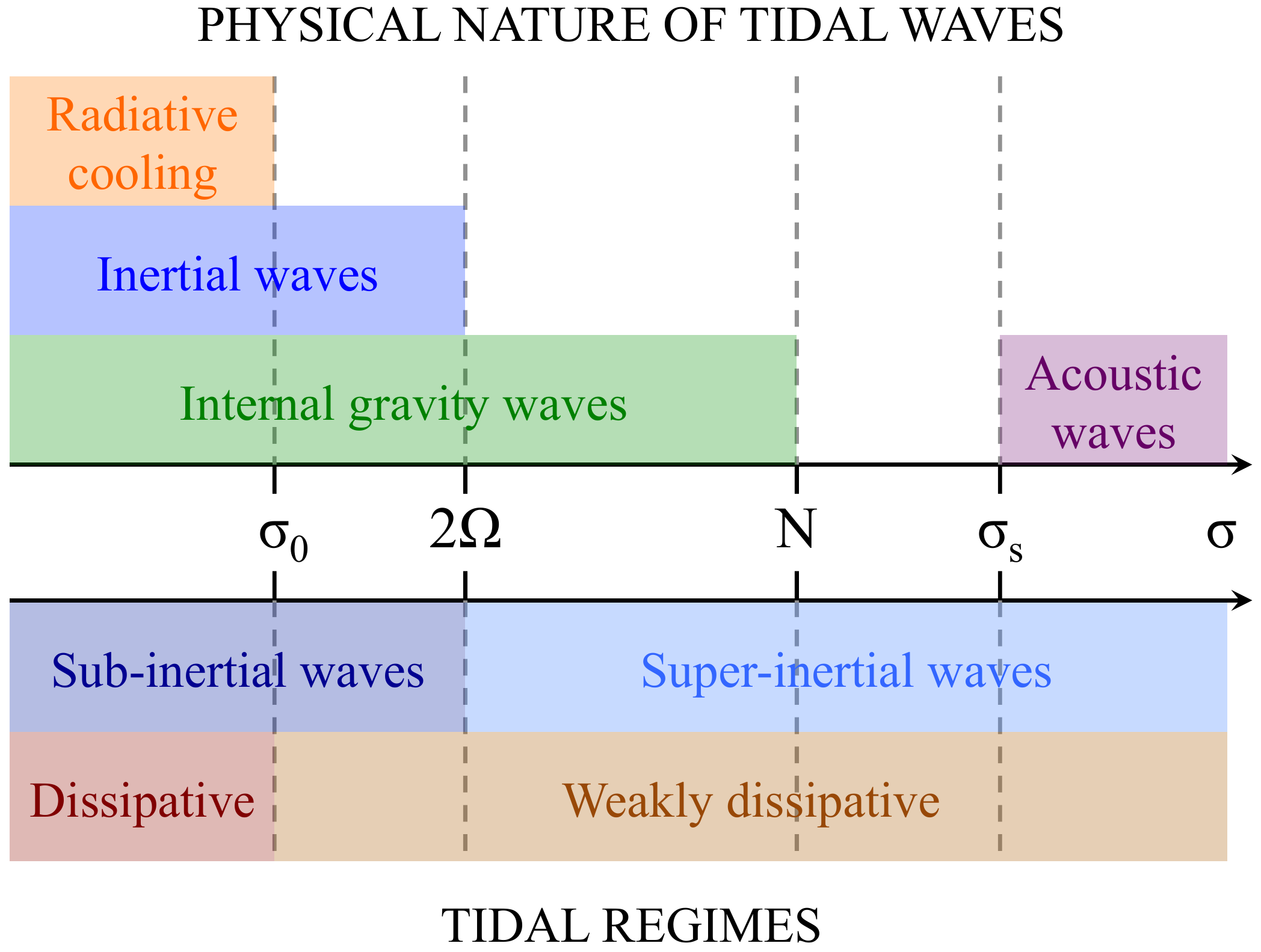}
   \caption{Frequency spectrum of tidal regimes and waves characterizing the fluid tidal response. The parameter $ \sigma $ designates the forcing frequency (Eq.~(\ref{Fourier})), $ \sigma_0 $ the thermal frequency (Eq.~(\ref{sigma0_scaling})), $ 2 \Omega $ the inertia frequency, $ N $ the Brunt-Väisälä frequency (Eq.~(\ref{N2def})), and $ \sigma_{\rm s} $ the characteristic acoustic cutoff frequency (Eq.~(\ref{sigmasn})).}
       \label{fig:spectre_regimes}%
\end{figure}

\noindent Therefore, $ \varepsilon_{\rm s ; n} $ weights the contribution of acoustic waves to the fluid tidal response, this contribution being negligible if $ \left|\varepsilon_{\rm s ; n} \right| \ll 1 $. Similarly, by considering Eqs.~(\ref{B1vert}) and (\ref{nu}), we can note that the ratio $ N^2 / \sigma^2 $ measures the contribution of internal gravity waves, i.e. waves restored by the Archimedean force, and the spin parameter ($\nu$) the contribution of inertial waves, restored by the Coriolis acceleration. This draws up a global picture of possible tidal regimes. As shown by Fig.~\ref{fig:spectre_regimes}, where the frequency spectrum of these regimes is given, the nature of the tidal response is thus totally determined by the hierarchy of characteristic frequencies of the physical system ($\sigma$, $ \sigma_{\rm s} $, $ N $, $ 2 \Omega $ and $ \sigma_0 $).

The last step consists in reducing the system of Eqs.~(\ref{ODE_eq1}-\ref{ODE_eq2}) to a single vertical structure equation, which is first expressed as

\begin{equation}
\frac{d^2 y}{dx^2} + A \frac{d y}{dx} + B y = C,
\label{vertical_structure}
\end{equation}

\noindent with the $ x $-dependent coefficients \citep[see][Eq.~(41)]{ADLM2017a}

\begin{align}
   A \left( x \right)  = & - \frac{\sigma - 2 i \sigma_0}{\sigma - i \sigma_0} \frac{H N^2}{g} - \frac{1}{\Gamma_1} - K_\circ,
\end{align}

\begin{align}
B \left( x \right) = & H^2 \hat{k}_{\perp ; n}^2 \left( \frac{\sigma}{\sigma - i \sigma_0} \frac{N^2}{\sigma^2} - 1 \right) \left( 1 - \varepsilon_{\rm s ; n} \right) \\
 & - \left(  \frac{d}{dx} +  \frac{i \sigma_0}{\sigma - i \sigma_0} \frac{H N^2}{g} -  \frac{1}{\Gamma_1} - K_\circ \right) \left( \frac{H N^2}{g} + \frac{i \kappa \sigma_0}{\sigma - i \sigma_0} \right),  \nonumber
\end{align}

\begin{align}
C \left( x \right) = & \left( \frac{d}{dx} + \frac{i \sigma_0}{\sigma - i \sigma_0} \frac{H N^2}{g} - \frac{1}{\Gamma_1}  - K_\circ \right) \left( \frac{d U_n}{dx} - i \frac{\kappa}{\sigma - i \sigma_0} J_n \right)  \nonumber \\ 
 & + H^2 \hat{k}_{\perp ; n}^2 \left(  \frac{\sigma }{\sigma - i \sigma_0} \frac{N^2}{\sigma^2} - 1 \right) \left[ U_n - i \frac{ \left(\Gamma_1 - 1 \right)  \varepsilon_{\rm s ; n}  }{\sigma - i \Gamma_1 \sigma_0}  J_n  \right] .
\end{align}

\noindent In these expressions, we have identified the horizontal wavenumber $ \hat{k}_{\perp ; n} $ of the $ \left( n,m,\sigma \right) $-mode, defined by

\begin{equation}
\hat{k}_{\perp ; n}^2 = \frac{\Lambda_n}{r^2}, 
\end{equation}

\noindent and a sphericity term, denoted $ K_\circ $, that is written 

\begin{equation}
K_\circ = \frac{r^2}{H} \left( \frac{\sigma}{\sigma - i \sigma_0} \frac{N^2}{\sigma^2} -  1 \right)^{-1} \frac{d}{dx} \left[ \frac{H}{r^2} \left( \frac{\sigma}{\sigma - i \sigma_0} \frac{N^2}{\sigma^2} - 1 \right) \right].
\end{equation}

We finally introduce the change of variable $ y_n = \Phi_n \Psi_n $, where $ \Phi_n $ is the function defined by

\begin{equation}
\Phi_n \left( x \right) = \exp \left[ - \frac{1}{2} \int_0^x A \left( x' \right) \dd x' \right] .
\end{equation}

\noindent This allows us to write the vertical structure equation as a Schrödinger-like equation, which describes the forced response of a harmonic oscillator. Eq.~(\ref{vertical_structure}) thus becomes 

\begin{equation}
\frac{d^2 \Psi_n}{dx^2} + \hat{k}_n^2 \Psi_n = \Phi_n^{-1} C,
\label{vertical_structure2}
\end{equation}

\noindent where $ \hat{k}_n $ designates the normalized vertical wavenumber of the $ \left( n , m , \sigma \right) $-mode (the vertical non-normalized wavenumber being $ \hat{k}_{r ; n} = \hat{k}_n / H $) and is expressed as \citep[see][Eq.~(47)]{ADLM2017a}

\begin{align}
\label{kzn2}
\hat{k}_n^2 = & H^2 \hat{k}_{\perp ; n}^2 \left( \frac{\sigma}{\sigma - i \sigma_0} \frac{N^2}{\sigma^2} - 1 \right) \left( 1 - \varepsilon_{\rm s ; n} \right) \\
 & - \left(  \frac{d}{dx} +  \frac{i \sigma_0}{\sigma - i \sigma_0} \frac{H N^2}{g} -  \frac{1}{\Gamma_1} - K_\circ \right) \left( \frac{H N^2}{g} + \frac{i \kappa \sigma_0}{\sigma - i \sigma_0} \right) \nonumber  \\
 & + \frac{1}{2} \frac{d}{dx} \left[  \frac{\sigma - 2 i \sigma_0}{\sigma - i \sigma_0} \frac{H N^2}{g} + K_\circ  \right]   \nonumber \\
 & - \frac{1}{4} \left( \frac{\sigma - 2 i \sigma_0}{\sigma - i \sigma_0} \frac{H N^2}{g} - \frac{1}{\Gamma_1} - K_\circ  \right)^2. \nonumber
 \label{kz2}
\end{align} 

\noindent We recognize in the first term of $ \hat{k}_n^2 $ the vertical wavenumber of gravito-inertial waves propagating within a homogeneous fluid. This term predominates in the stably-stratified radiative zone in the low-frequency asymptotic limit. It is responsible for the rapid increase of the vertical wavenumber, which scales as $ \hat{k}_n \propto \sigma^{-1} $ \jlc{(if $ \left| \sigma \right| \gg \sigma_0 $)} or $ \hat{k}_n \propto \sigma^{-1/2} $ \jlc{(if $ \left| \sigma \right| \ll \sigma_0 $)} when $ \left| \sigma \right| \ll N^2 $. Other terms result from radial variations of background distributions. They may be important around transition zones in the internal structure of the planet, such as the base of the atmosphere where the gradient of $ N^2 $ is strong (see Fig.~\ref{fig:bgd}). 

The tidally generated variation of mass distribution at the origin of the tidal torque is finally derived from $ \Psi_n $. It is written 

\begin{align}
\delta \rho_n = & - \frac{\rho_0}{gH} \frac{N^2}{N^2 - \sigma \left( \sigma - i \sigma_0 \right)} \left[ \Phi_n \left( \frac{d \Psi_n}{dx} + \mathcal{B}_n \Psi_n \right)   \right. \\
 & \left. - \frac{d U_n}{dx} + i \frac{\kappa \sigma}{N^2} J_n \right], \nonumber
 \label{deltaqn}
\end{align}

\noindent with 

\begin{align}
 \mathcal{B}_n \left( x \right) = & \frac{1}{2} \left[ \frac{1}{\Gamma_1} \frac{\sigma - i \, \Gamma_1 \sigma_0}{\sigma - i \sigma_0} \left( 2 \frac{\sigma \left( \sigma - i \sigma_0 \right)}{N^2} - 1 \right) \right. \\ 
  & \left. - \frac{\sigma}{\sigma - i \sigma_0} \frac{H N^2}{g} + K_\circ \right], \nonumber
\end{align}

\noindent The polarization relations of other perturbed quantities are given in Appendix~\ref{app:polarization_relations}. It shall be noted here that all of the results derived in this sections remain true for any vertical profiles of background quantities as far as the fluid is a perfect gas at hydrostatic equilibrium. 

\subsection{Thermal and gravitational forcings}

As mentioned above, the incoming stellar flux is absorbed in the upper layers of the atmosphere. Thus, following \cite{AS2010} and introducing the stellar zenith angle $ \phi_\star $, we define the distribution of total heat per unit mass as 

\begin{equation}
\left\{
\begin{array}{ll}
  \displaystyle J_\star \left( p_0 , \phi_\star \right) = \frac{g}{p_\star} F_\star \ed^{-p/p_\star} \cos \phi_\star & \displaystyle {\rm for} \  \phi_\star \in \left[ 0 , \frac{\pi}{2} \right] , \\
  \displaystyle J_\star \left( p_0 , \phi_\star \right) = 0 & \displaystyle {\rm for} \ \phi_\star \in \left]  \frac{\pi}{2} , \pi  \right].
\end{array}
\right.
\label{heat_distribution}
\end{equation}

\noindent In Eq.~(\ref{heat_distribution}), $ F_\star $ designates the incoming flux at the sub-stellar point, expressed as $ F_\star = \sigma_{\rm SB} T_\star^4 \left( R_\star / r_\star \right)^2  $, where $ \sigma_{\rm SB} $ is the Stefan-Boltzmann constant, $ r_\star $ the star-planet distance, and $ R_\star $ and $ T_\star $ the stellar radius and mean surface temperature respectively. \rec{If we assume that $ R_\star \ll r_\star $, the gravitational potential generated by the star in the accelerated reference frame of the planet is written}

\begin{equation}
U_\star \left( \textbf{r} , \textbf{r}_\star \right) = \frac{\mathscr{G} M_\star}{\left| \textbf{r} - \textbf{r}_\star \right|} - \frac{\mathscr{G} M_\star}{r_\star^2} r \cos \theta,
\end{equation}

\noindent the vector $ \textbf{r}_\star $ being the planet-star vector (such that $ \left| \textbf{r}_\star \right| = r_\star $) and $ M_\star $ the mass of the star. In the general case, the stellar heating and gravitational forcing are expanded into Fourier series and spherical harmonics, that is 

\begin{align}
&  U_\star = \sum_{l,m,\sigma} U_l^{m,\sigma} \left( x \right) P_l^m \left( \cos \theta \right) \ed^{i \left( \sigma t + m \varphi \right)}, \\
& J_\star = \sum_{l,m,\sigma} J_l^{m,\sigma} \left( x \right) P_l^m \left( \cos \theta \right) \ed^{i \left( \sigma t + m \varphi \right)},
\end{align}

\noindent the $ P_l^m $ standing for the normalized associated Legendre polynomials\footnote{The normalized associated Legendre polynomials are defined  by \citep[][]{AS1972}
\[  
P_l^m \left( X \right) = \sqrt{\frac{\left( 2 l +  1 \right) \left( l - m \right) !}{2 \left( l + m \right) !}} \left( - 1 \right)^m \left( 1 - X^2 \right)^{m/2} \frac{d^m}{dX^m} \left[  P_l \left( X \right) \right],
\]
where the $ P_l $ stand for the Legendre polynomials,
\[
P_l \left( X \right) = \frac{1}{2^l l !} \frac{d^l}{dX^l} \left[ \left( X^2 - 1 \right)^l \right].
\]
}. 

However, to simplify the forcing, we suppose that the planet orbits its host star circularly and that its equatorial plane is coplanar with the orbital plane. As a consequence, the expansion of the tidal gravitational potential in spherical harmonics is reduced to \rec{the semidiurnal quadrupolar component ($ l = m = 2 $)}, expressed as \citep[][]{AS2010,MLP2009}

\begin{equation}
U_2^{2, \sigma} = \sqrt{\frac{3}{5}} \frac{M_\star}{M_\star + M_{\rm p}} n_{\rm orb}^2 r^2,
\label{U22}
\end{equation}

\noindent with $ \sigma = 2 \left( \Omega - n_{\rm orb} \right) $. Similarly, \rec{the quadrupolar component} of the thermal forcing is given by

\begin{equation}
J_2^{2 , \sigma} = \frac{5}{16} \sqrt{\frac{3}{5}} \frac{g}{p_\star} F_\star \ed^{-p/p_\star} 
\label{J22}
\end{equation}

\begin{figure}
   \centering
   \includegraphics[width=0.49\textwidth,trim = 1.5cm 2.2cm 2.0cm 1.cm, clip]{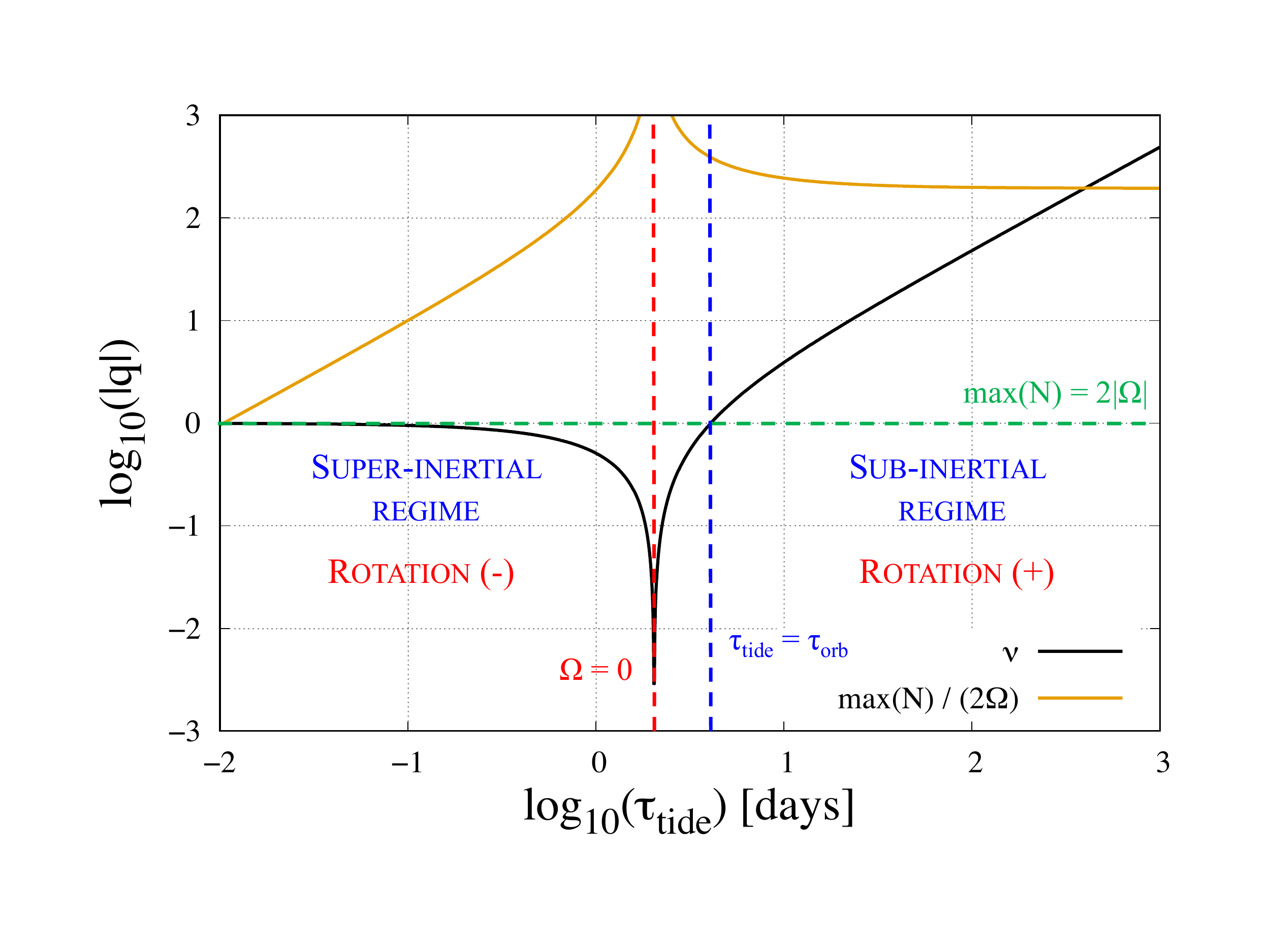}
   \caption{Logarithms of the spin parameter ($ \nu $) and stability ratio $ {\rm max} \left( N \right) / \left( 2 \Omega \right) $ as functions of the forcing period $ \tau_{\rm tide} = 2 \pi / \left| \sigma \right| $ (days) in logarithmic scale. The blue dashed line designates the transition between the inertial and sub-inertial regimes (see Fig.~\ref{fig:spectre_regimes}), which occurs for $ \tau_{\rm tide} = \tau_{\rm orb}  $, the parameter $ \tau_{\rm orb} = 2 \pi /n_{\rm orb} $ being the orbital period. The red dashed line corresponds to the $\Omega = 0 $ case. ``Rotation $ (-) $'' and ``Rotation $(+) $'' stand for retrograde and prograde rotation respectively. The green dashed line delimits the boundary of the asymtotic domain where the stable stratification dominates rotation in the radiative zone ($ N \gg 2 \left| \Omega \right| $), which is the condition of applicability of the traditional approximation in the sub-inertial regime.  }
   \label{fig:nu_Ttide}%
\end{figure}

\noindent Note that the semidiurnal spin parameter is simply expressed as 

\begin{equation}
\nu = 1 + \frac{2 n_{\rm orb}}{\sigma},
\label{nu_sigma}
\end{equation}

\noindent which highlights the parameter measuring the impact of rotation on the tidal response, $ \left|  n_{\rm orb} / \sigma \right| $. For a sub-synchronous rotation ($ \sigma < 0 $), the inertial regime corresponds to $ \left|  n_{\rm orb} / \sigma \right| \ll 1 $ (see Fig.~\ref{fig:nu_Ttide}). In this case, the planet is rapidly rotating. The transition from the inertial regime to a sub-inertial one occurs for $ \left|  n_{\rm orb} / \sigma \right| \sim 1 $. Beyond this critical value, $ \nu \propto 2  n_{\rm orb} / \sigma $ and diverges at the spin-orbit synchronous rotation ($ \sigma \rightarrow 0 $), where the effects of rotation are stronger than the forcing. In the zero-frequency limit, the traditional approximation is only appropriate for strongly stably-stratified layers. Thus, the effects of rotation can be taken account provided that the region where tidal waves propagate satisfies the condition $ 2 \Omega \ll N $ \citep[e.g.][]{Mathis2009,ADLM2017a}. 

\subsection{Boundary conditions}
\label{ssec:BC}

To integrate the vertical structure equation, Eq.~(\ref{vertical_structure2}), two boundary conditions must be choosen. \rec{We set one condition at the lower boundary of the atmosphere, and one at its upper boundary. \cite{AS2010} chooses to set the lower boundary at the center of the planet ($ x = 0 $). Hence, they use} a standard regularity condition requiring all variables to be finite \citep[e.g.][]{Unno1989}. In the static case, this condition writes

\begin{equation}
\sigma^2 \xi_{r ; n} = \frac{l}{r} \left( y_n - U_n \right).
\end{equation}

\noindent It can be adapted to the rotating case in the super-inertial regime ($ \left| \nu \right| \leq 1 $), where the previous expression becomes

\begin{equation}
\sigma^2 \xi_{r ; n} = \frac{\sqrt{1 + 4 \Lambda_n} - 1}{2 r} \left( y_n - U_n \right).
\end{equation}

\noindent However, it cannot be applied for $ \left| \nu \right| > 1 $, Rossby modes being associated with negative $ \Lambda_n $. Therefore, in this work, \rec{we choose to use} a rigid-wall condition enforcing the fact that fluid particles cannot go through \rec{the lower boundary}. It is simply expressed as $ \xi_r =  0 $, i.e.

\begin{equation}
\frac{d \Psi_n}{dx} + \mathcal{A}_n \Psi_n = - \Phi_n^{-1} \left( - \frac{d U_n}{dx} + i \frac{\kappa}{\sigma - i \sigma_0} J_n \right). 
\end{equation}

\noindent with (see Appendix~\ref{app:polarization_relations})

\begin{equation}
\mathcal{A}_n \left( x \right) =   \frac{1}{2} \left( \frac{1}{\Gamma_1} \frac{\sigma - i \, \Gamma_1 \sigma_0}{\sigma - i \sigma_0}  - \frac{\sigma }{\sigma - i \sigma_0} \frac{H N^2 }{g} + K_\circ \right).
\end{equation}

\rec{Note that the lower boundary shall not necessarily be set at the center of the planet since we focus on thermal tides, which only affect the stably stratified atmospheric layer of the planet. The convective region is not perturbed by the tidal thermal forcing. The lower boundary could thus be set at any pressure level greater than that corresponding to the base of the stably stratified region, that is $ p_{\rm b} \approx 10^2 $ bar. Nevertheless, setting the lower limit of the atmosphere at $ p_{\rm b} $ would cause reflections of waves and induce side effects, as discussed by \cite{OL2004} and illustrated by Appendix~\ref{app:lower_boundary}. This is the reason why this boundary shall be located at higher pressure levels. Here, considering that the tidal perturbation is confined to the atmosphere and does not affect the convective region, we set the lower boundary as far as possible from the basis of the stably stratified zone, that is at the center of the planet. This allows us to avoid artefacts related to reflections. We verify a posteriori by checking other conditions that, in this case, the lower boundary condition has no repercussion of the tidal response generated by the thermal forcing.}


\rec{The upper limit of the atmosphere is located in the tidally forced region. Thus, the associated boundary condition partly determines} the frequency-dependent behaviour of the fluid, as \rec{demonstrated} by \cite{AS2010}. In the adiabatic case, it is possible to apply the \emph{radiation condition}, that is to consider that waves carry energy upward without reflections \citep[e.g.][]{Wilkes1949,CL70}. This condition regularizes the tidal response by eliminating resonances due to gravity waves of short vertical wavelengths in the vicinity of spin-orbit synchronous rotation \citep[e.g.][]{AS2010,ADLM2017a}. The vertical energy flow associated with the $ \left( n , m , \sigma \right) $-mode is expressed as 

\begin{align}
\mathfrak{F}_{r ; n} = &  \frac{1}{2} \frac{\sigma \rho_0}{H} \Im \left\{  \left( \frac{\sigma}{\sigma - i \sigma_0} N^2 - \sigma^2 \right)^{-1} \left[ \left| \Phi_n \right|^2 \left( \Psi_n^* \frac{d \Psi_n}{dx} + \mathcal{A}_n \left| \Psi_n \right|^2 \right)    \right.   \right. \nonumber \\ 
 & \left. \left.   +  \Phi_n^* \Psi_n^* \left( - \frac{d U_n}{dx} + i \frac{\kappa}{\sigma - i \sigma_0} J_n \right)  \right]  \right\}.
\end{align} 

\noindent Assuming that the tidal sources are located below the altitude of the upper boundary (denoted $ R_{\rm e} $ or $x_{\rm e} $ depending on the used vertical coordinate), we ignore the last term. For $ \sigma_0 = 0 $, the only \jlc{potentially negative} term is thus $ \Psi_n^* \left( d \Psi_n / dx \right) $. We then suppose that the vertical wavelength of tidal waves is shorter than the length scale of background distributions, so that the solution of Eq.~(\ref{vertical_structure2}) writes $ \Psi_n \left( x \right) = \mathscr{A} \ed^{i \hat{k}_n x} + \mathscr{B} \ed^{- i \hat{k}_n x} $ at the upper boundary ($ x = x_{\rm e} $). The radiation condition hence consists in conserving only the first term of this solution associated with $ \mathfrak{F}_{r ; n} > 0 $, that is in setting

\begin{equation}
\begin{array}{rcl}
 \displaystyle  {\rm sign} \left( \Re \left\{ \hat{k}_n \right\}   \right) =  {\rm sign} \left( \frac{\sigma}{N^2 - \sigma^2} \right) & \mbox{and} & \displaystyle \mathscr{B} = 0,
\end{array}
\end{equation}

\noindent \jlc{the symbol $ \Re $ referring to the real part of a complex number.}

As it may be noted, this condition is only convenient in the adiabatic case and for propagative modes $ \hat{k}_n^2 \geq 0 $. If $ \sigma_0 \neq 0 $, both terms of the solution can contribute to a positive upward energy flux. Moreover, it is not appropriate to waves of wavelengths comparable to the pressure height scale. For these two reasons, we choose to apply at $ x = x_{\rm e} $ the standard \emph{free-surface condition}, $ \delta p = g \rho_0 \xi_r $ \citep[e.g.][]{Unno1989}, which can be enforced for any $ \hat{k}_n $. This condition is formulated as

\begin{equation}
\frac{d \Psi_n}{dx} + \left[ \mathcal{A}_n + \frac{H}{g} \left( \frac{\sigma}{\sigma - i \sigma_0} N^2 - \sigma^2 \right) \right] \Psi_n = \Phi_n^{-1} \left( \frac{ i \kappa J_n}{\sigma - i \sigma_0} - \frac{d U_n}{dx} \right).
\end{equation}

\noindent In the absence of dissipation, it leads to a highly frequency-resonant behaviour around $ \sigma = 0 $, given that gravity waves can be reflected backward at the upper boundary \citep[][]{AS2010,ADLM2017a}. \jlc{In reality, due to the strong stellar heating, the energy given by the tidal forcing is partly radiated toward space, which makes internal tidal waves evanescent and consequently strongly attenuate the amplitude of the perturbation. This effect is modeled by \cite{GO2009} with a \emph{Marshak condition}, which enforces a radiative energy loss at the upper limit of the atmosphere. In our study, we model dissipative processes with the Newtonian cooling defined by Eqs.~(\ref{Newtonian_cooling}) and (\ref{sigma0_scaling}). As it will be observed in Section~\ref{sec:properties_waves}, the frequency dependence of the tidal response is regularized by the damping for $ \left| \sigma \right| \lesssim \sigma_0 $ (see Fig.~\ref{fig:spectre_regimes}).}

\section{Tidal torque and quadrupole}
\label{sec:tidal_torque}

As the goal of this work is to examine the ability of thermal tides to modify the planet's rotation and to generate zonal flows, we introduce in this section the expressions used to compute the tidal torque exerted on the fluid shell with respect to the spin axis of the planet. This torque is obtained by projecting the tidally induced variation of mass distribution on the tidal force ($ \textbf{F} = \boldsymbol{\nabla} U $). Thus, its $ \left( m , \sigma \right)  $-component is defined by \citep[][]{Zahn1966a,ADLM2017a}

\begin{equation}
\mathcal{T}^{m,\sigma} = \Re \left\{ \frac{1}{2} \int_\mathscr{V} \partial_\varphi U^{m,\sigma} \left( \delta \rho^{m,\sigma} \right)^* \dd \mathscr{V}  \right\},
\label{torque_def}
\end{equation}

\noindent where $ \mathscr{V} $ designates the volume of the fluid shell \jlc{and $ ^* $ the conjugate of a complex number}. By substituting in Eq.~(\ref{torque_def}) the expansions of fluctuations in Hough functions given by Eq.~(\ref{series_Hough1}), we end up with 

\begin{equation}
\mathcal{T}^{m,\sigma} =  \pi m \sum_{l \geq  m } W_l^{m,\sigma} \Im \left\{ Q_l^{m,\sigma}  \right\}.
\end{equation}

\noindent In this expression, we have introduced \jlc{the imaginary part $ \Im $,} the factors $  W_l^{m,\sigma} $ such that $ U_l^{m,\sigma} = W_l^{m,\sigma} r^l $ and the tidal multipole moments $ Q_l^{m,\sigma} $, which are expressed as 

\begin{equation}
Q_l^{m,\sigma} = \sum_n A_{n,l}^{m,\nu} \int_0^R r^{2+l} \delta \rho_n^{m,\sigma} \dd r .
\label{Qlmsigma}
\end{equation}

The coefficients $ A_{n,l}^{m,\nu} $ quantify the distortion of the tidal response caused by rotation. In the static case, $ A_{n,l}^{m,0} = \delta_{l, m +n} $, and we retrieve the expression given by \cite{AS2010}. These authors propose an expansion of this expression in terms of other perturbed quantities in order to avoid numerical errors\footnote{This point is discussed in Appendix~\ref{app:multipole_moment}.}. We adapt their formula to the rotating case in Appendices~\ref{app:low_frequencies} and \ref{app:multipole_moment}, and get Eq.~(\ref{Qnlm}). In the following, in order to validate the obtained results, we will compute the tidal quadrupole moment by using both Eq.~(\ref{Qlmsigma}) and Eq.~(\ref{Qnlm}). The evolution timescale of the planet's global rotation rate due to the $ \left( m , \sigma \right) $-mode is expressed as

\begin{equation}
\tau_\Omega = \left| \frac{\mathcal{I}_Z \left( \Omega - n_{\rm orb} \right)  }{\mathcal{T}^{m,\sigma}} \right|,
\end{equation}

\noindent where $ \mathcal{I}_Z $ is the moment of inertia with respect to the $ Z $-axis, that is written as 

\begin{equation}
\mathcal{I}_Z = \frac{8 \pi}{3} \int_0^R r^4 \rho_0 dr. 
\end{equation}

However, the global torque and timescale provide no information about the local strength of the thermal tidal forcing. To know which layers are forced and where strong jets can be generated, we use the longitudinal tidal force per unit volume averaged over the longitude,

\begin{equation}
\mathcal{F}_\varphi = \Re \left\{ \frac{1}{4 \pi r \sin \theta}  \int_0^{2 \pi}  \partial_\varphi U^{m,\sigma} \left( \delta \rho^{m,\sigma} \right)^* \dd \varphi \right\}.
\end{equation}

Hence, the characteristic timescale necessary to generate a jet of velocity $ V_{\rm jet} $ in the absence of viscous coupling writes

\begin{equation}
\tau_{\rm evol} = \left| \frac{\rho_0 V_{\rm jet}}{\mathcal{F}_\varphi}  \right|.
\label{tau_jet}
\end{equation}

In the case of the quadrupolar semidiurnal tide, which is the object of this study, the perturbation \rec{is forced by the component defined by $ l = m = 2 $ and $ \sigma = m \left( \Omega - n_{\rm orb} \right) $.} Therefore, the total torque induced by the semidiurnal tide reduces to (see (Eq.~\ref{U22}))

\begin{equation}
\mathcal{T}_{\rm SDT}^{\sigma} =  2 \pi \sqrt{\frac{3}{5}} \left( \frac{M_\star}{M_\star + M_{\rm p}} \right) n_{\rm orb}^2  \Im \left\{ Q_2^{2,\sigma} \right\},
\label{torque_SDT}
\end{equation}

\noindent with the quadrupole moment

\begin{equation}
Q_2^{2,\sigma} = \sum_n A_{n,2}^{m,\nu} \int_0^R  r^4 \delta \rho_n^{2,\sigma}  \dd r . 
\label{Q22}
\end{equation}

\section{Properties of thermally forced tidal waves}
\label{sec:properties_waves}

We now explore the properties of the tidal response and its dependence on dissipative mechanisms and rotation by applying the linear analysis to three different configurations. The aim of this section is to investigate the consequences of the thermal tidal torque on the general circulation of the atmosphere, and particularly its ability to generate strong zonal flows. Thus, we isolate the thermal tide by setting $ U = 0 $ in the equations describing the vertical structure of the fluid tidal response (Section~\ref{ssec:structure_regimes_waves}). The tidal gravitational potential given by Eq.~(\ref{U22}) is nevertheless used to compute the induced gravitational tidal torque. 

\begin{table}[h]
\centering
 \textsf{\caption{\label{parameters} Parameters used for the three cases examined in Section~\ref{sec:properties_waves}. The acronyms RZ and HZ stand for \emph{Radiative Zone} and \emph{Heated Zone} respectively.  }}
\begin{small}
    \begin{tabular}{ l l l l}
      \hline
      \hline
      \textsc{Parameter} & \textsc{Symbol} & \textsc{Value} & \textsc{Unit} \\ 
      \hline 
      Planet mass & $ M_{\rm p} $ & $ 0.7 $ & $ M_{\rm J}$  \\
      Planet radius & $ R_{\rm p} $ & $ 1.27 $ & $ R_{\rm J}$  \\
      Orbital period & $ \tau_{\rm orb} $ & $ 4.08 $ & d \\
      Distance to star & $ a_\star $ & $ 0.05 $ & AU \\
      Star mass & $ M_\star $ & $  1.0 $ & $ M_\sun $ \\
      Star radius & $ R_\star $ & $ 1.0 $ & $ R_\sun $ \\
      Star surface temperature & $ T_\star $ & $ 1.0 $ & $ T_\sun $ \\
      Adiabatic exponent & $ \Gamma_1 $ & $ 1.4 $ & -- \\
      Pressure at the base of the RZ & $ p_{\rm b} $ & $ 100 $ & bar \\
      Pressure at the base of the HZ &$ p_\star $ & $ 1.0 $ & bar \\ 
      \hline
    \end{tabular}
\end{small}
 \end{table}

\begin{figure*}
   \centering
   \includegraphics[height=0.4cm]{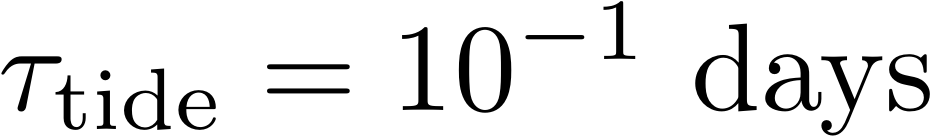} \hspace{0.8cm}
   \includegraphics[height=0.4cm]{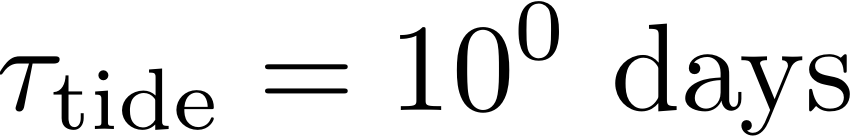} \hspace{0.8cm}
   \includegraphics[height=0.4cm]{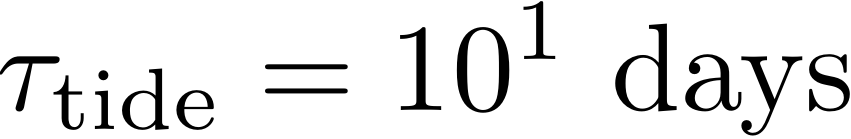} \hspace{0.8cm}
   \includegraphics[height=0.4cm]{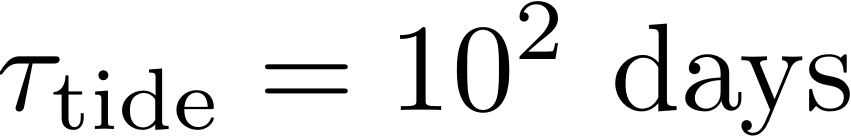} \hspace{0.8cm}
   \includegraphics[height=0.4cm]{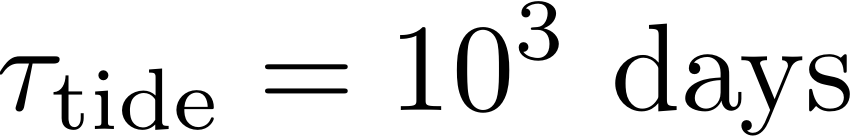} \hspace{0.8cm}\\
   \raisebox{1.0cm}{\includegraphics[width=0.02\textwidth]{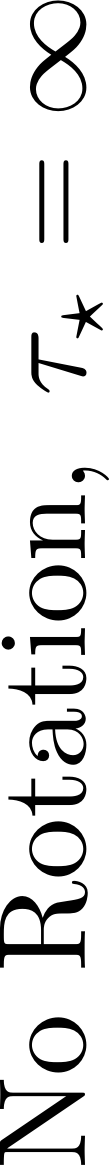}} \hspace{0.1cm}
   \raisebox{1.5\height}{\includegraphics[width=0.015\textwidth]{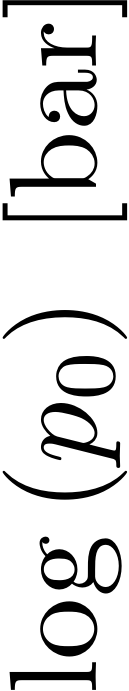}}
   \includegraphics[width=0.18\textwidth,trim = 2.5cm 2.2cm 3.cm 0.cm, clip]{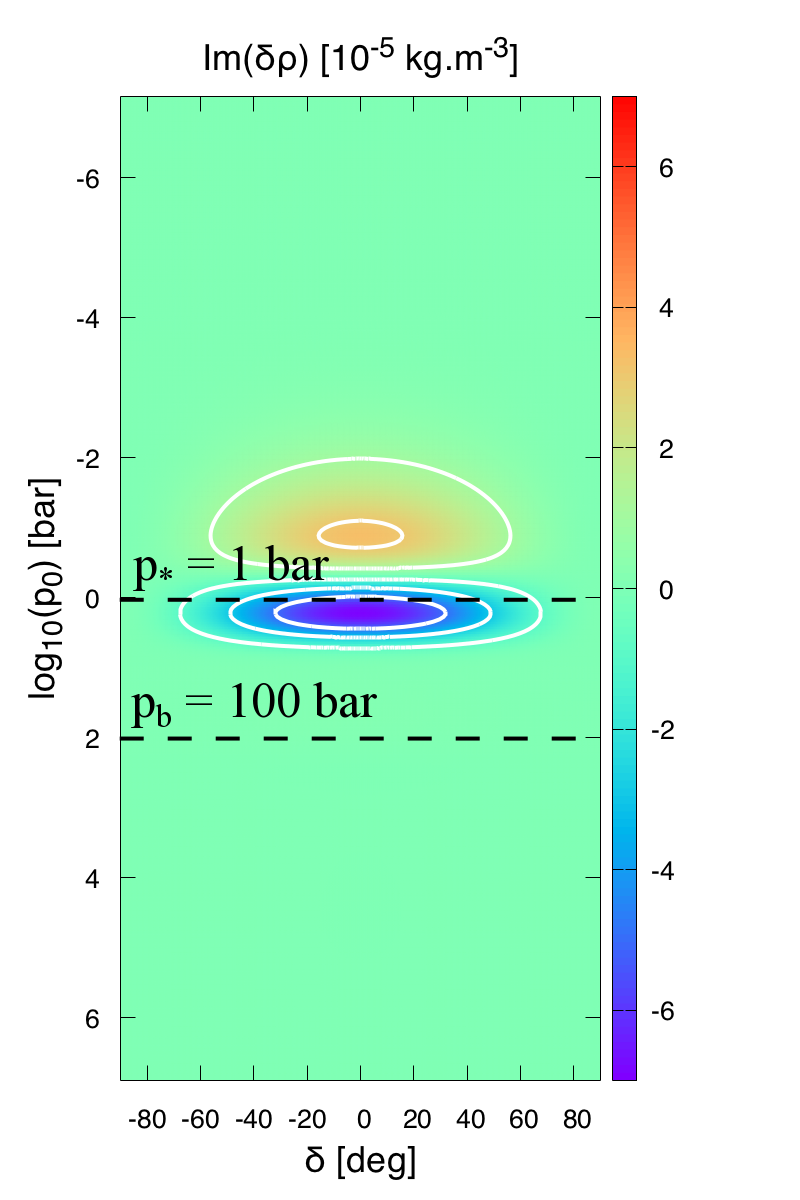}
   \includegraphics[width=0.18\textwidth,trim = 2.5cm 2.2cm 3.cm 0.cm, clip]{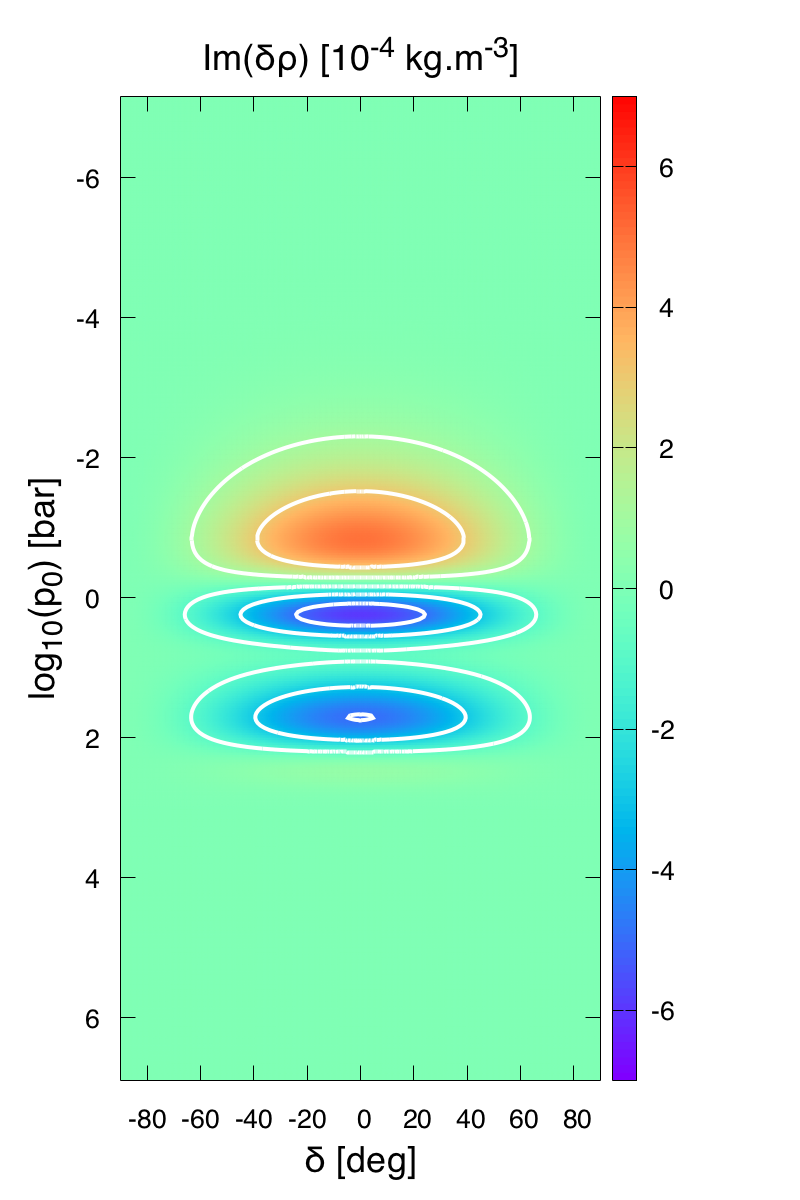}
   \includegraphics[width=0.18\textwidth,trim = 2.5cm 2.2cm 3.cm 0.cm, clip]{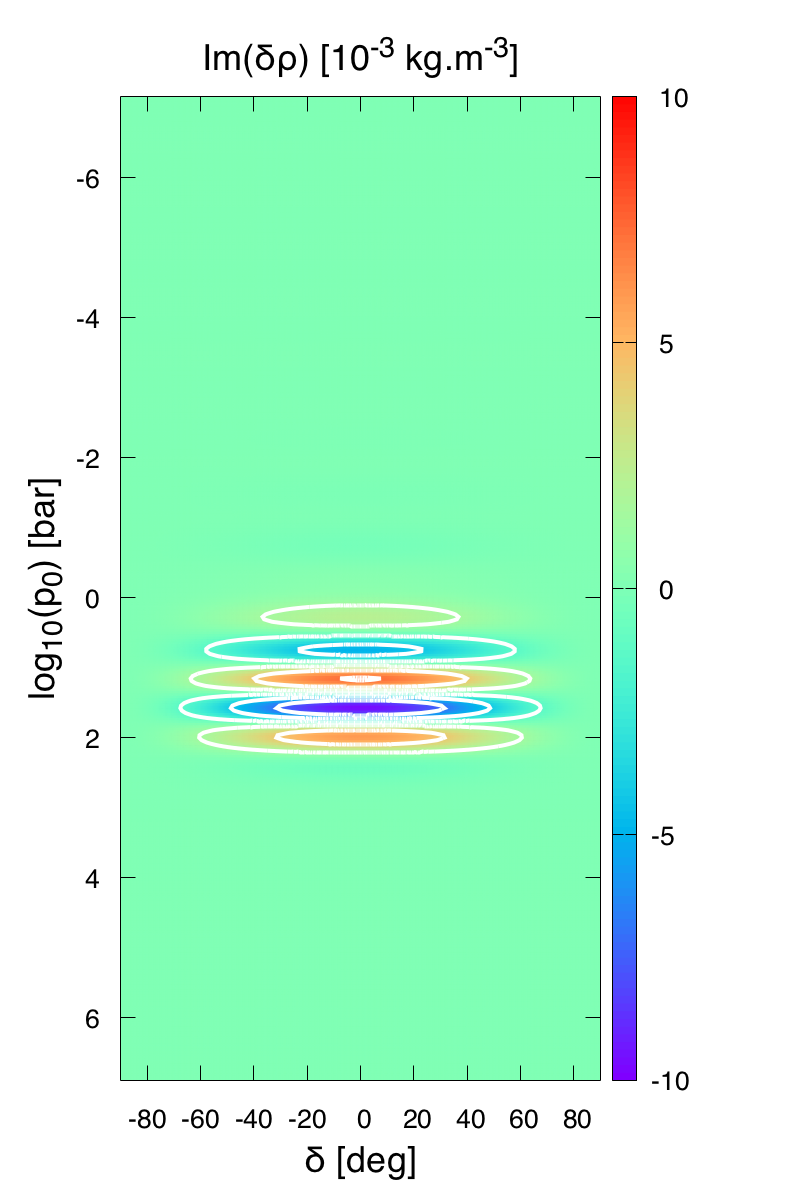} 
   \includegraphics[width=0.18\textwidth,trim = 2.5cm 2.2cm 3.cm 0.cm, clip]{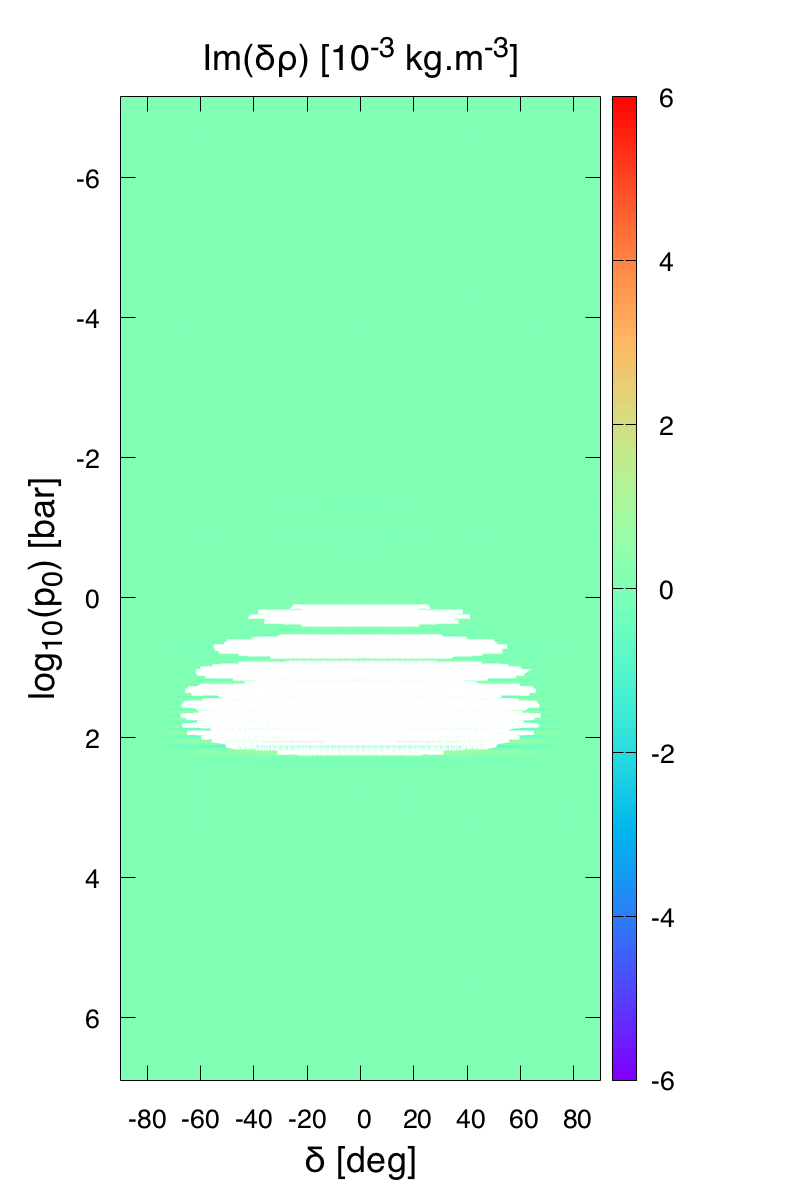} 
   \includegraphics[width=0.18\textwidth,trim = 2.5cm 2.2cm 3.cm 0.cm, clip]{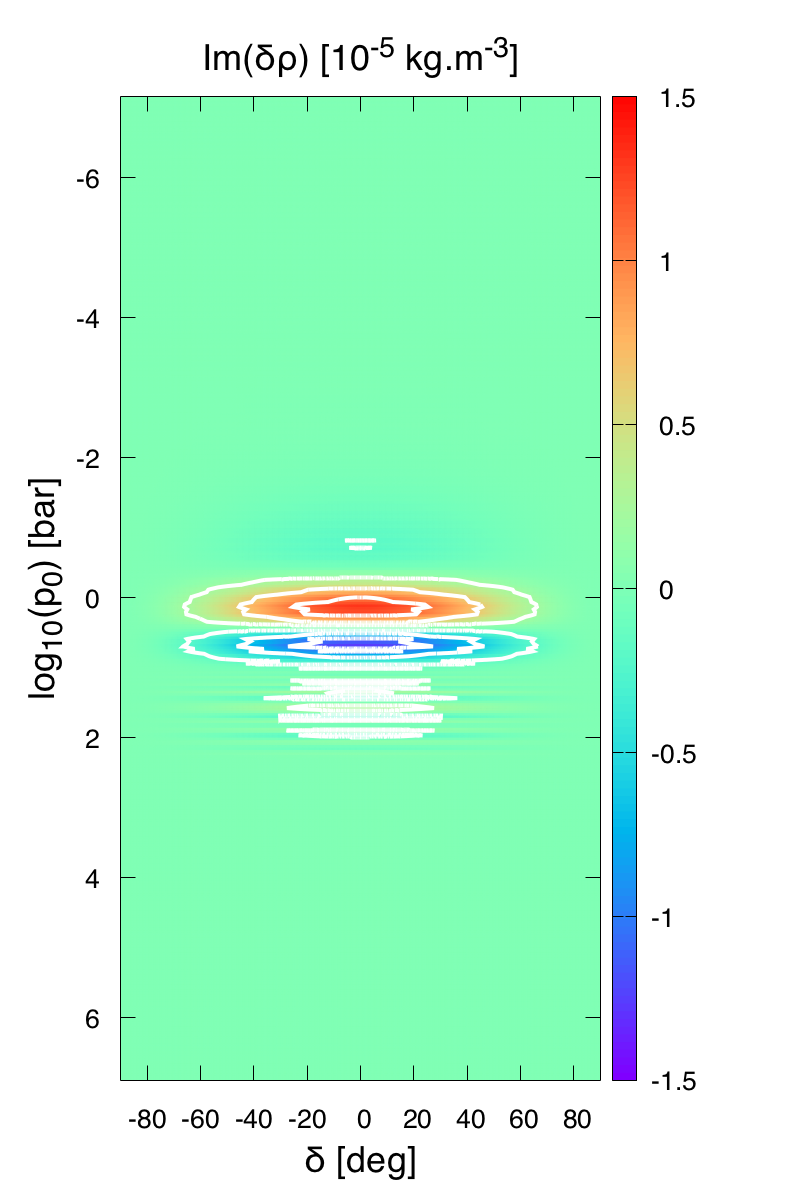}\\
   \raisebox{1.0cm}{\includegraphics[width=0.02\textwidth]{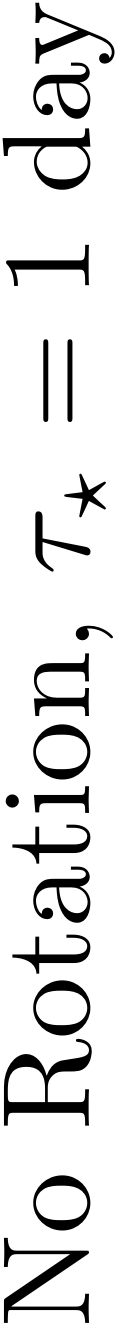}} \hspace{0.1cm}
   \raisebox{1.5\height}{\includegraphics[width=0.015\textwidth]{auclair-desrotour_fig6g.pdf}}
   \includegraphics[width=0.18\textwidth,trim = 2.5cm 2.2cm 3.cm 0.cm, clip]{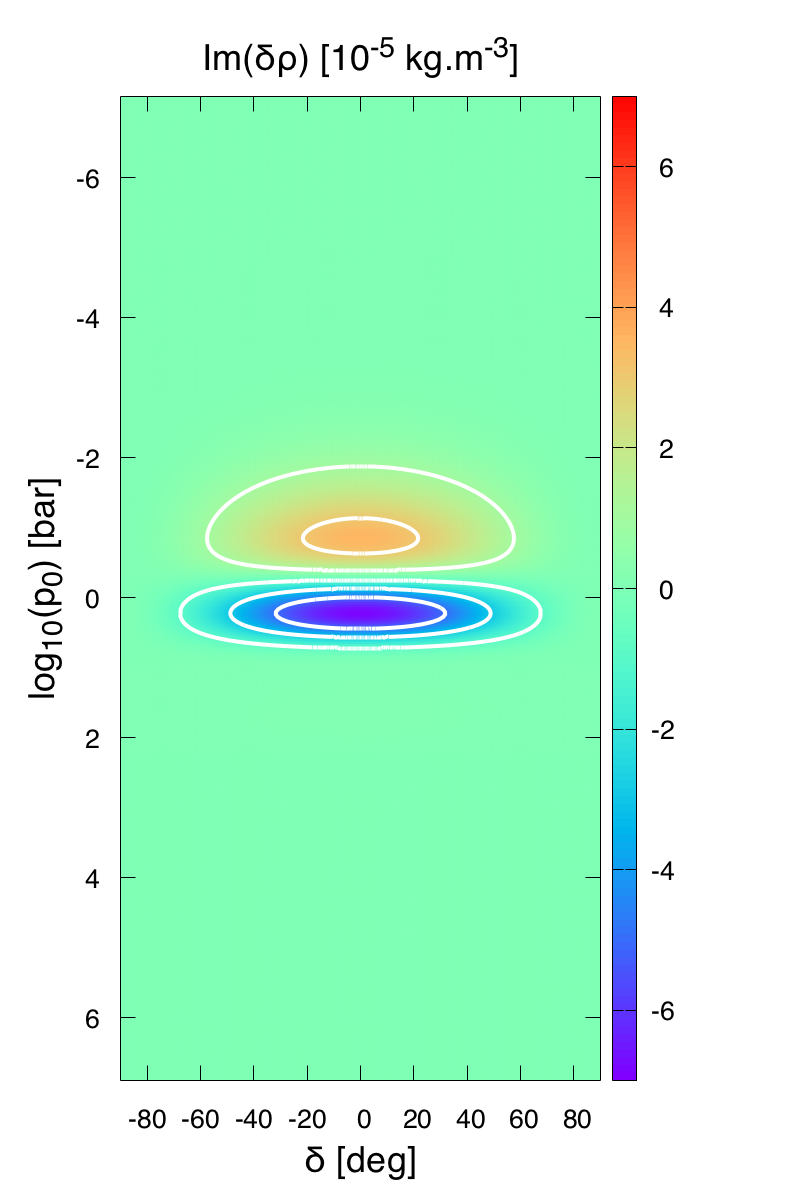}
   \includegraphics[width=0.18\textwidth,trim = 2.5cm 2.2cm 3.cm 0.cm, clip]{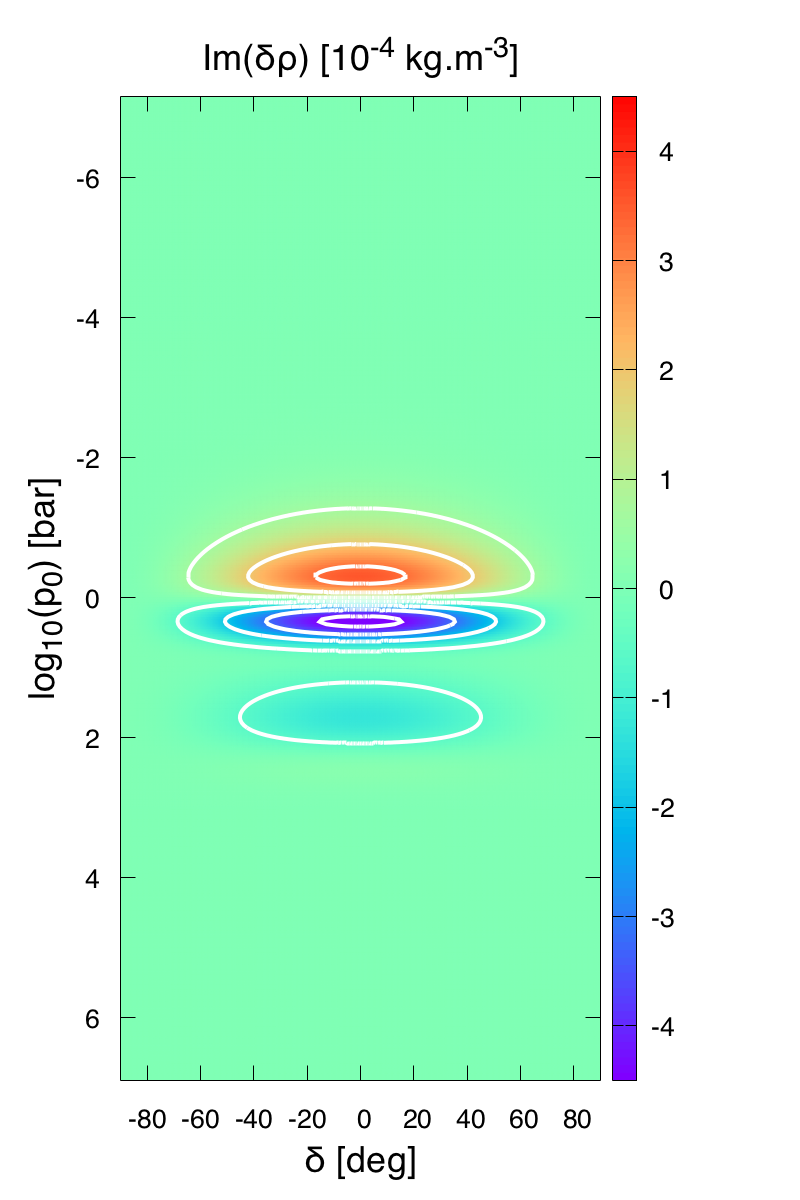}
   \includegraphics[width=0.18\textwidth,trim = 2.5cm 2.2cm 3.cm 0.cm, clip]{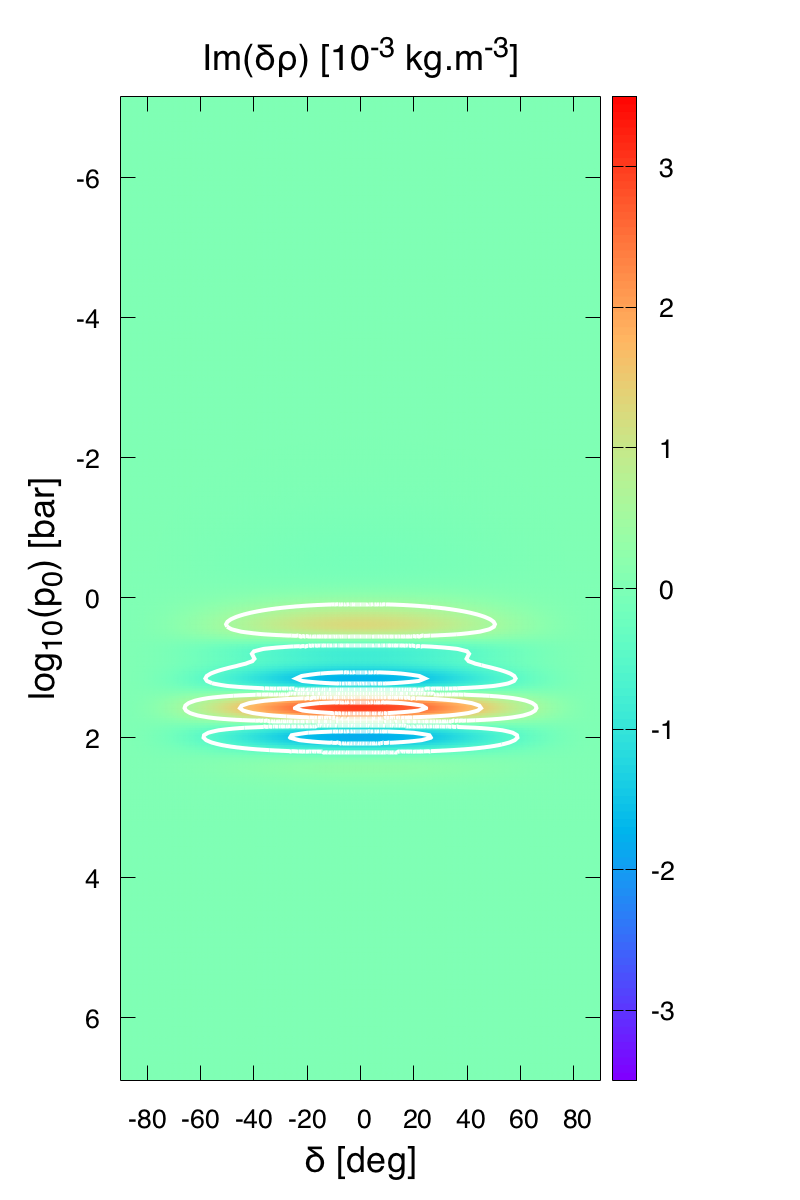} 
   \includegraphics[width=0.18\textwidth,trim = 2.5cm 2.2cm 3.cm 0.cm, clip]{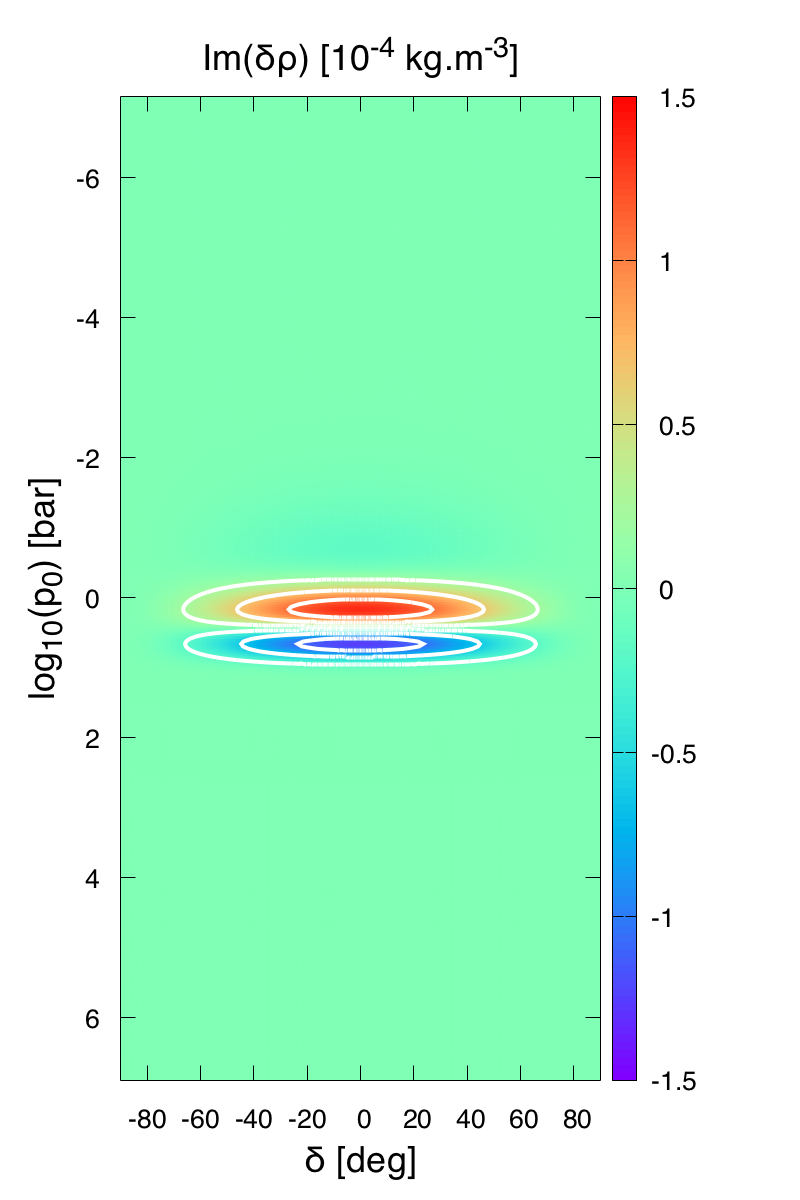} 
   \includegraphics[width=0.18\textwidth,trim = 2.5cm 2.2cm 3.cm 0.cm, clip]{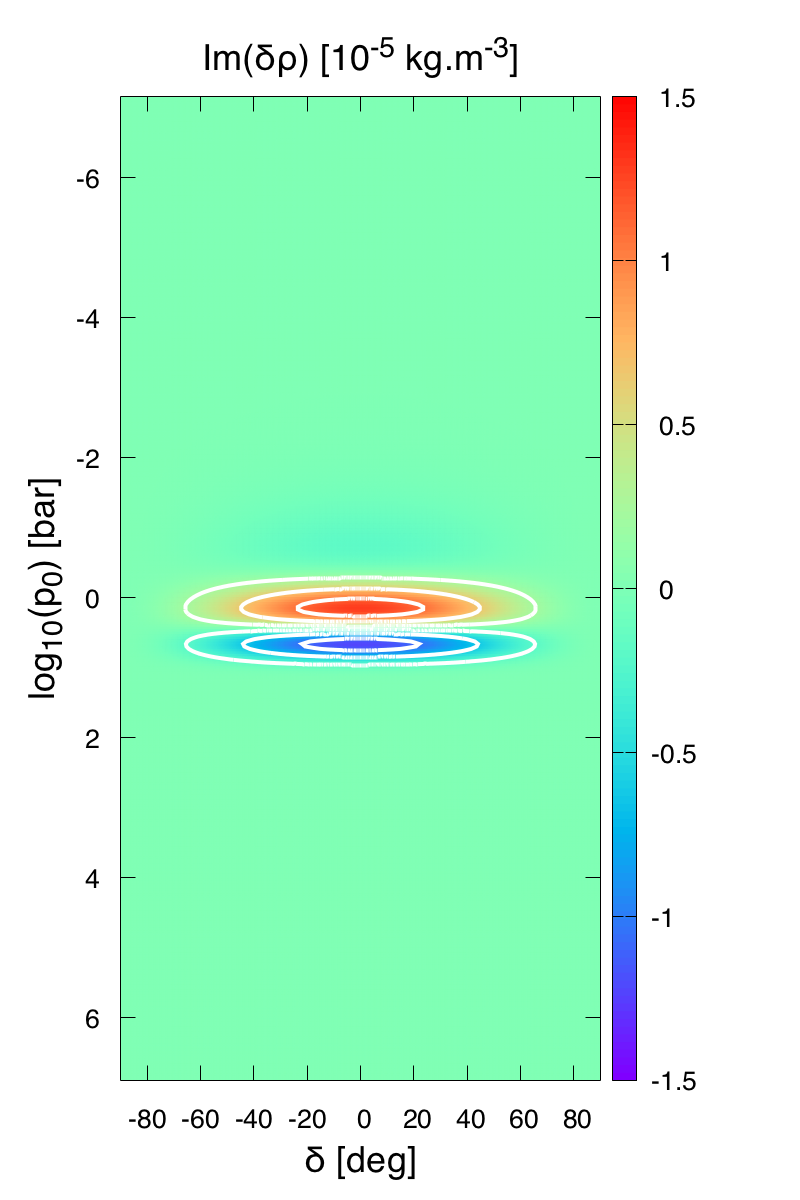} \\
   \raisebox{1.0cm}{\includegraphics[width=0.02\textwidth]{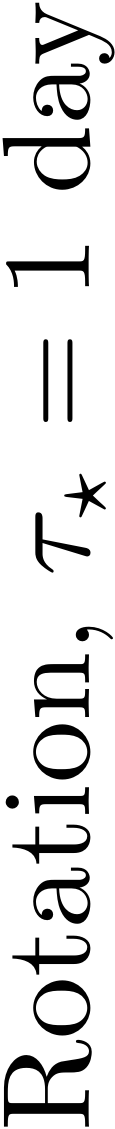}} \hspace{0.1cm}
   \raisebox{1.5\height}{\includegraphics[width=0.015\textwidth]{auclair-desrotour_fig6g.pdf}}
   \includegraphics[width=0.18\textwidth,trim = 2.5cm 2.2cm 3.cm 0.cm, clip]{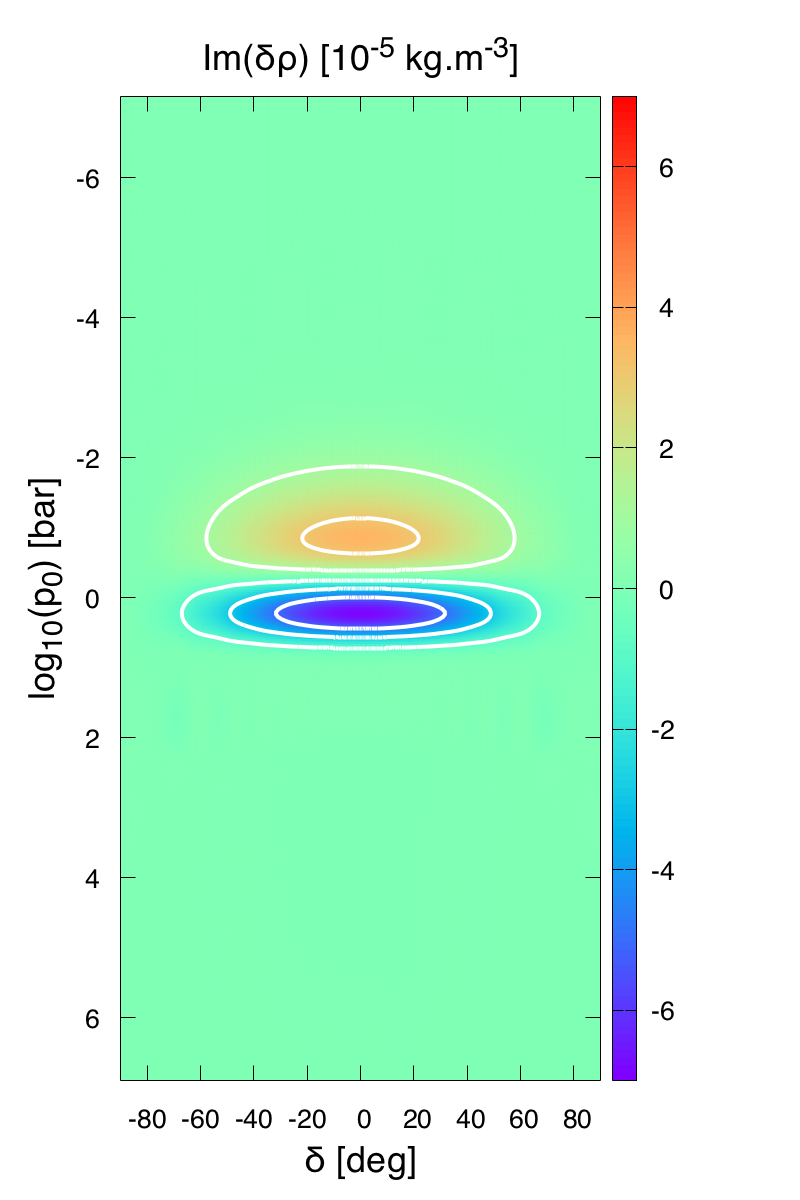}
   \includegraphics[width=0.18\textwidth,trim = 2.5cm 2.2cm 3.cm 0.cm, clip]{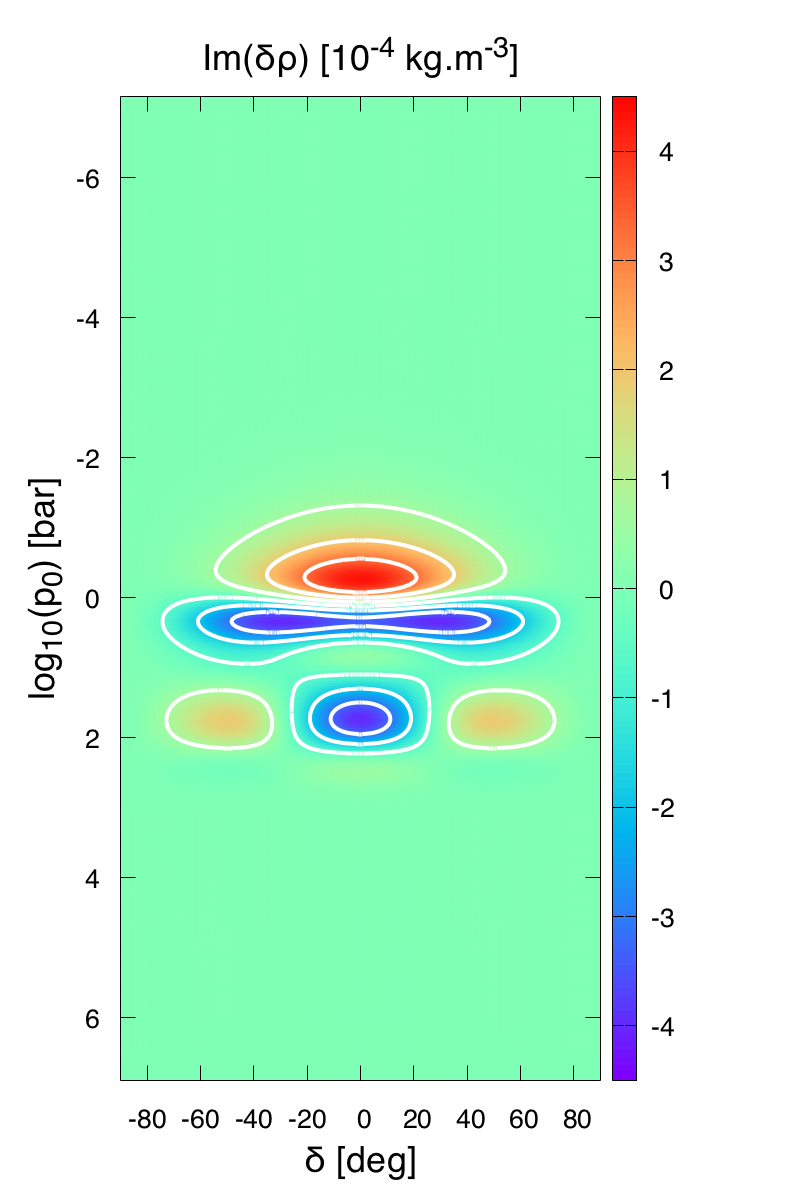}
   \includegraphics[width=0.18\textwidth,trim = 2.5cm 2.2cm 3.cm 0.cm, clip]{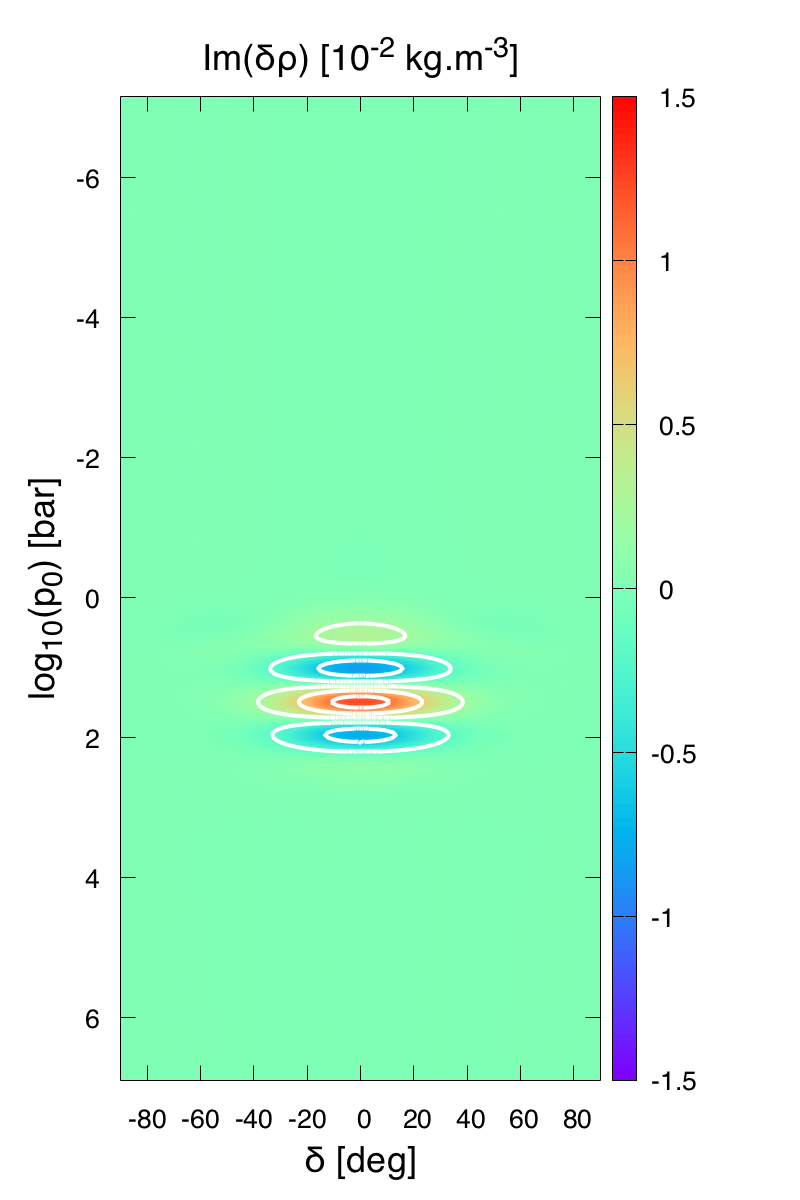} 
   \includegraphics[width=0.18\textwidth,trim = 2.5cm 2.2cm 3.cm 0.cm, clip]{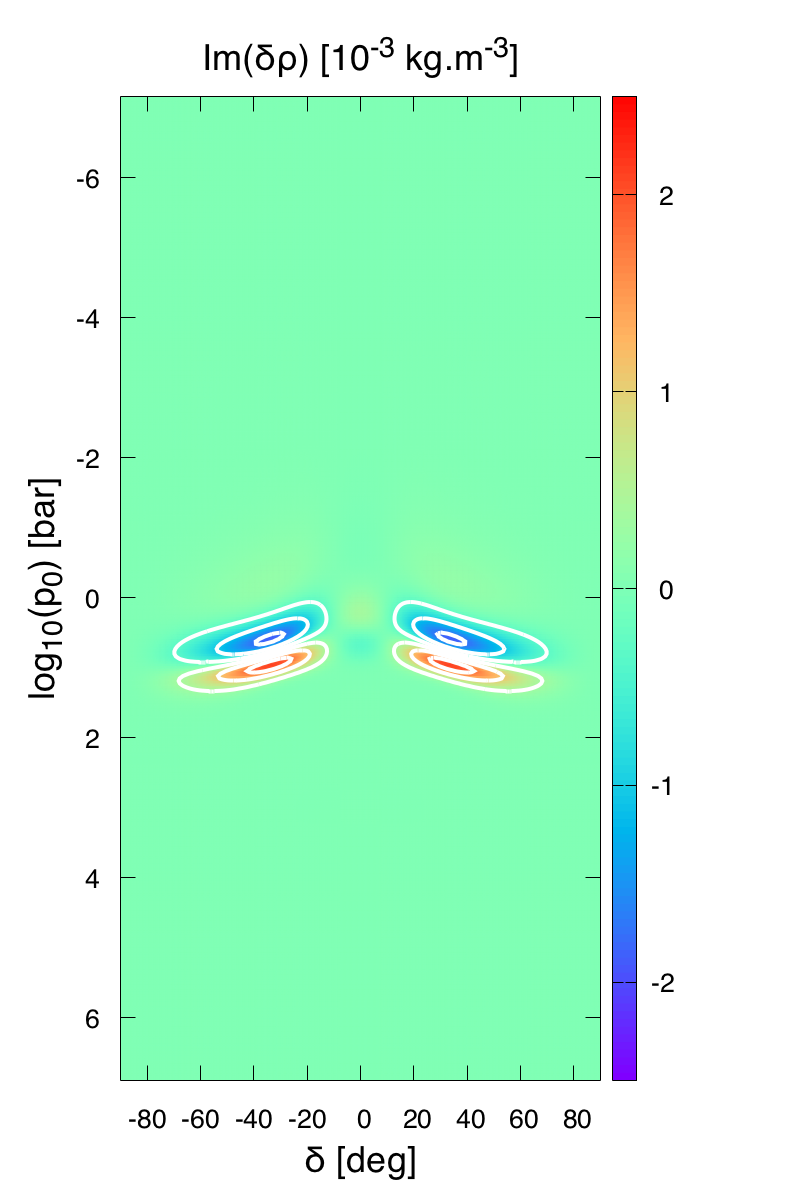} 
   \includegraphics[width=0.18\textwidth,trim = 2.5cm 2.2cm 3.cm 0.cm, clip]{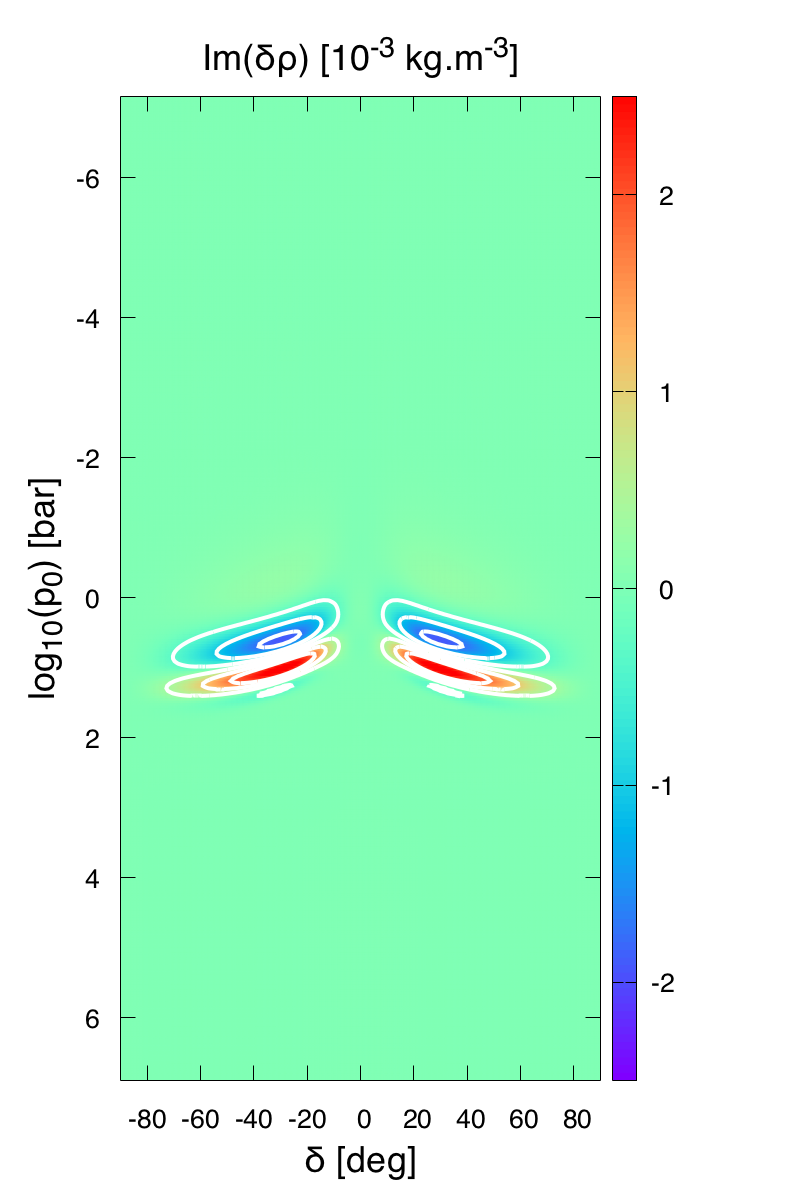} \\
   \hspace{1.9cm}
   \includegraphics[height=0.3cm]{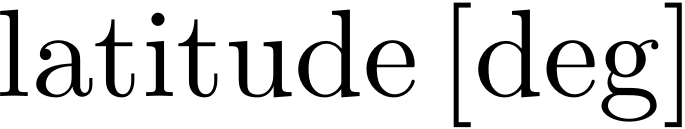} \hspace{1.6cm}
   \includegraphics[height=0.3cm]{auclair-desrotour_fig6y.pdf} \hspace{1.6cm}
   \includegraphics[height=0.3cm]{auclair-desrotour_fig6y.pdf} \hspace{1.6cm}
   \includegraphics[height=0.3cm]{auclair-desrotour_fig6y.pdf} \hspace{1.6cm}
   \includegraphics[height=0.3cm]{auclair-desrotour_fig6y.pdf} \hspace{1.3cm}
   \caption{Imaginary part of density fluctuations generated by the quadrupolar semidiurnal thermal tide a functions of the latitude (horizontal axis, degrees) and background pressure in logarithmic scale (vertical axis, bars). Density fluctuations are plotted using Eq.~(\ref{deltaqn}) for several decades of the forcing period ($ \tau_{\rm tide} = 2 \pi / \sigma $), \jlc{$ \log \left(  \tau_{\rm tide} \right) = -1, \ldots , 3 $} (from left to right), and the three studied cases. {\it Top:} case treated by \cite{AS2010}, i.e. \jlc{adiabatic without rotation} (no-Coriolis approximation). {\it Middle:} \jlc{non-adiabatic without rotation} ($ \tau_\star= 1 $ day). {\it Bottom:} \jlc{non-adiabatic} (\jlc{$ \tau_\star = 1 $~day}) with rotation (traditional approximation). The horizontal structure of the tidal response (Eq.~(\ref{Laplace_eq})) is computed for 250 Hough modes using the spectral method described by \cite{Wang2016}, with projections on 375 associated Legendre polynomials. The vertical structure equation (Eq.~(\ref{vertical_structure})) is integrated on a regular mesh composed of 1000 points (10000 points for the case $\tau_\star =  +\infty $) using the implicit fourth order finite differences scheme detailed in Appendix~\ref{app:num_scheme}.   }
       \label{fig:structure_ondes}%
\end{figure*}

\begin{figure*}
   \centering
   \includegraphics[height=0.4cm]{auclair-desrotour_fig6a.pdf} \hspace{0.8cm}
   \includegraphics[height=0.4cm]{auclair-desrotour_fig6b.pdf} \hspace{0.8cm}
   \includegraphics[height=0.4cm]{auclair-desrotour_fig6c.pdf} \hspace{0.8cm}
   \includegraphics[height=0.4cm]{auclair-desrotour_fig6d.pdf} \hspace{0.8cm}
   \includegraphics[height=0.4cm]{auclair-desrotour_fig6e.pdf} \hspace{0.8cm}\\
   \raisebox{1.0cm}{\includegraphics[width=0.02\textwidth]{auclair-desrotour_fig6f.pdf}} \hspace{0.1cm}
   \raisebox{1.5\height}{\includegraphics[width=0.015\textwidth]{auclair-desrotour_fig6g.pdf}}
   \includegraphics[width=0.18\textwidth,trim = 2.5cm 2.2cm 3.cm 0.cm, clip]{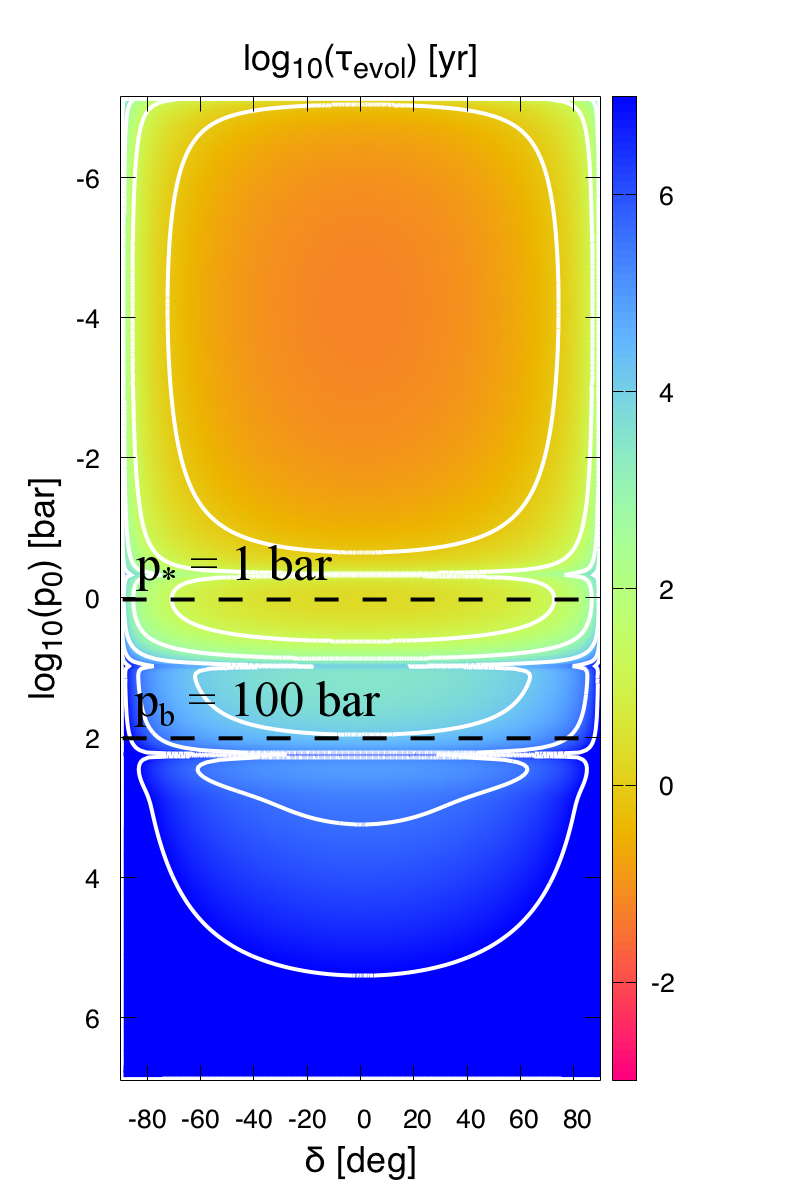}
   \includegraphics[width=0.18\textwidth,trim = 2.5cm 2.2cm 3.cm 0.cm, clip]{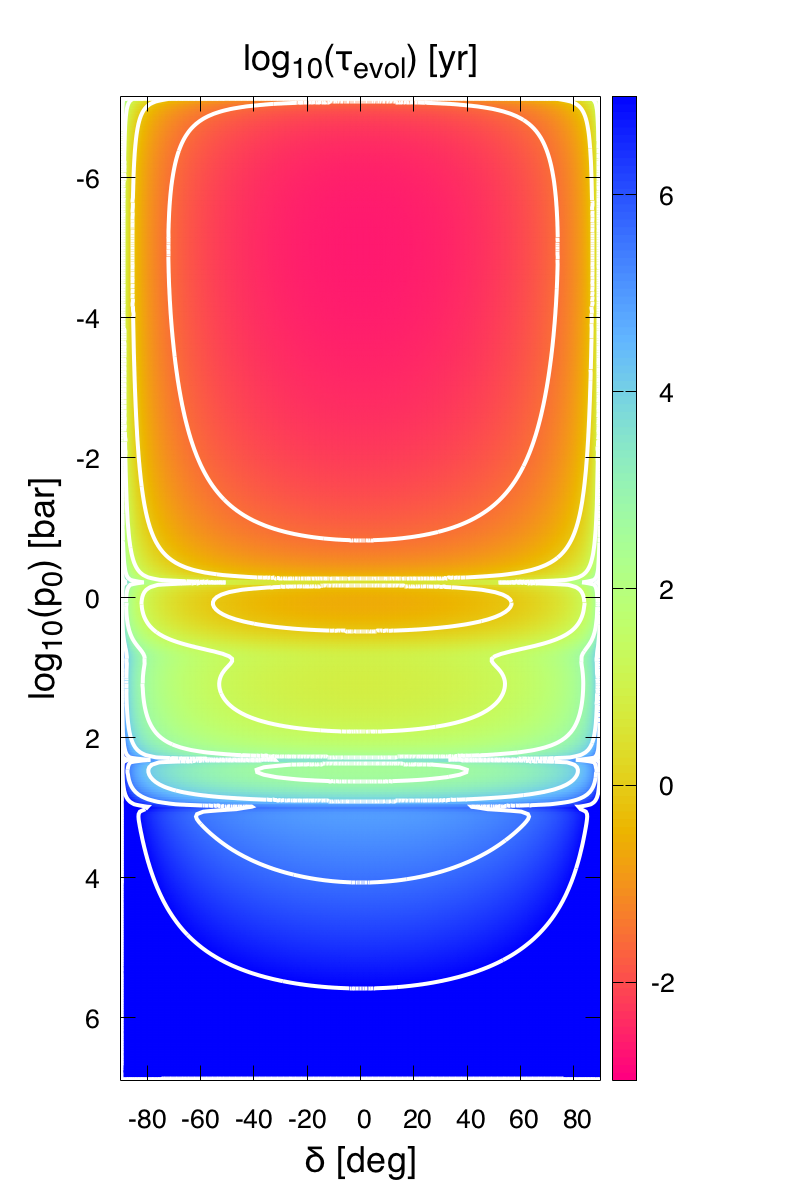}
   \includegraphics[width=0.18\textwidth,trim = 2.5cm 2.2cm 3.cm 0.cm, clip]{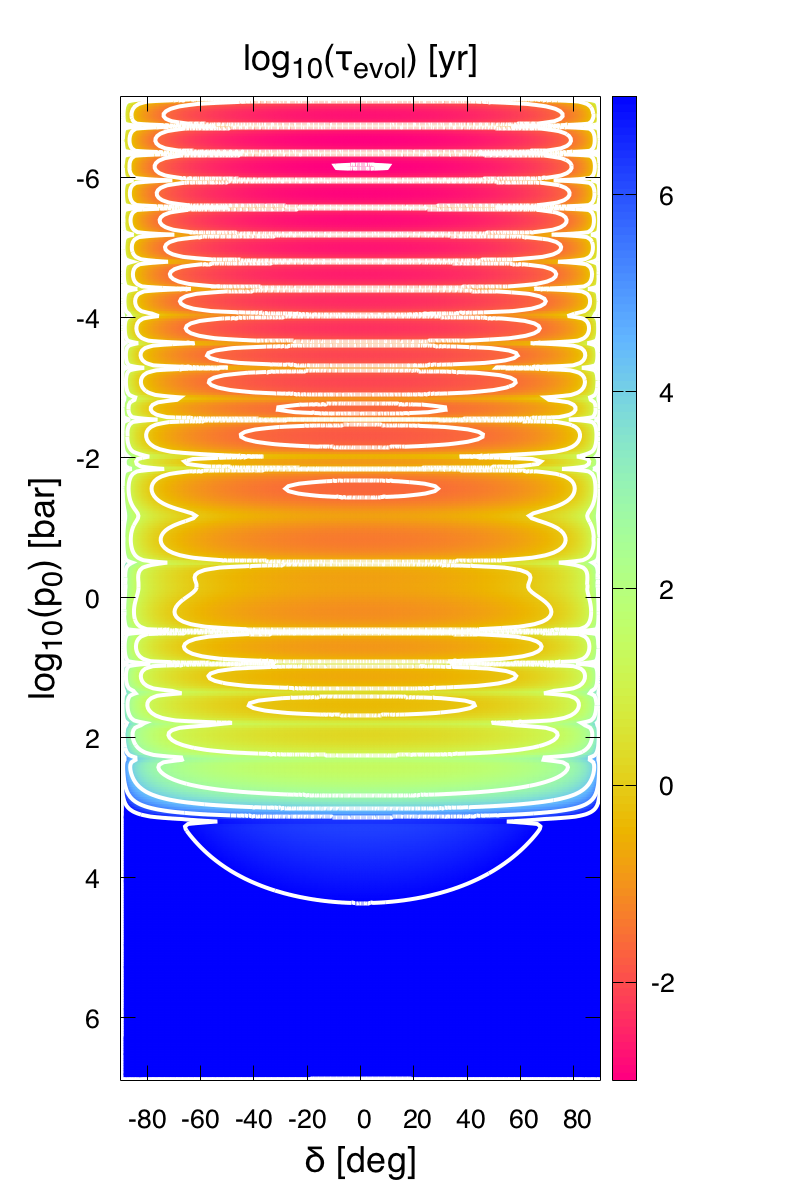} 
   \includegraphics[width=0.18\textwidth,trim = 2.5cm 2.2cm 3.cm 0.cm, clip]{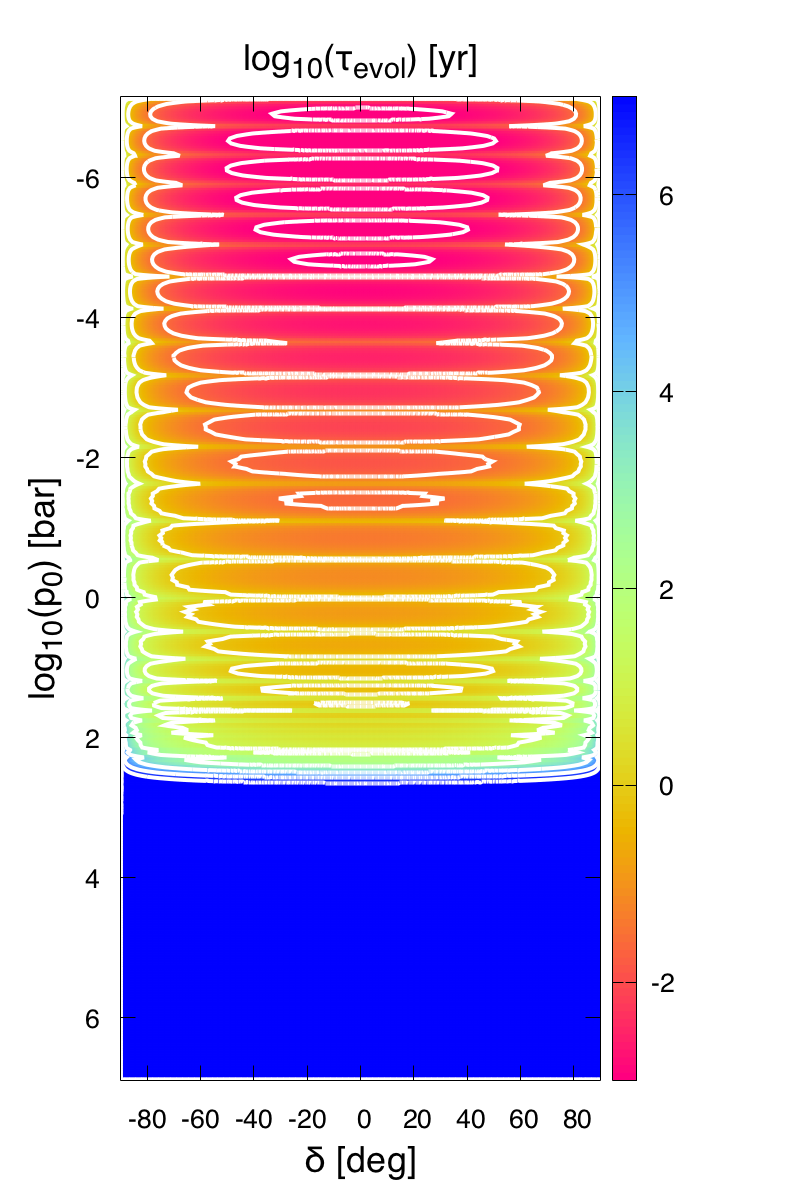} 
   \includegraphics[width=0.18\textwidth,trim = 2.5cm 2.2cm 3.cm 0.cm, clip]{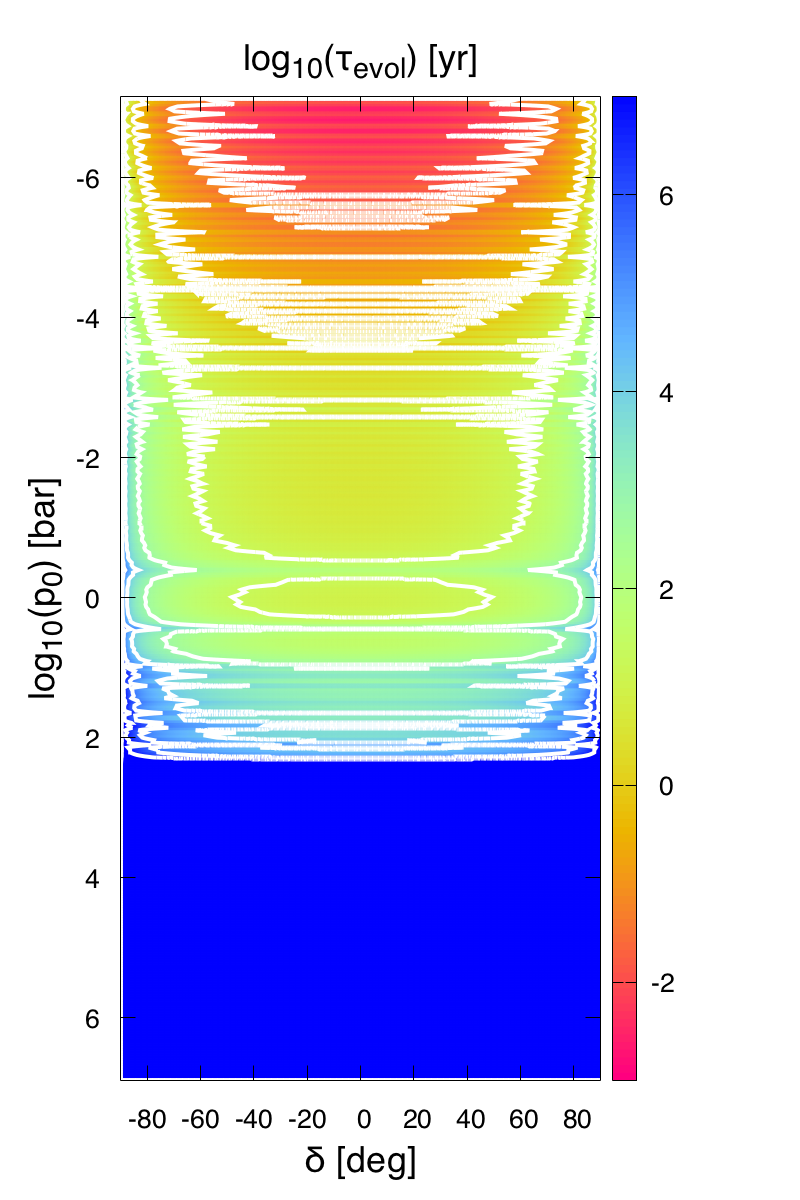}\\
   \raisebox{1.0cm}{\includegraphics[width=0.02\textwidth]{auclair-desrotour_fig6m.pdf}} \hspace{0.1cm}
   \raisebox{1.5\height}{\includegraphics[width=0.015\textwidth]{auclair-desrotour_fig6g.pdf}}
   \includegraphics[width=0.18\textwidth,trim = 2.5cm 2.2cm 3.cm 0.cm, clip]{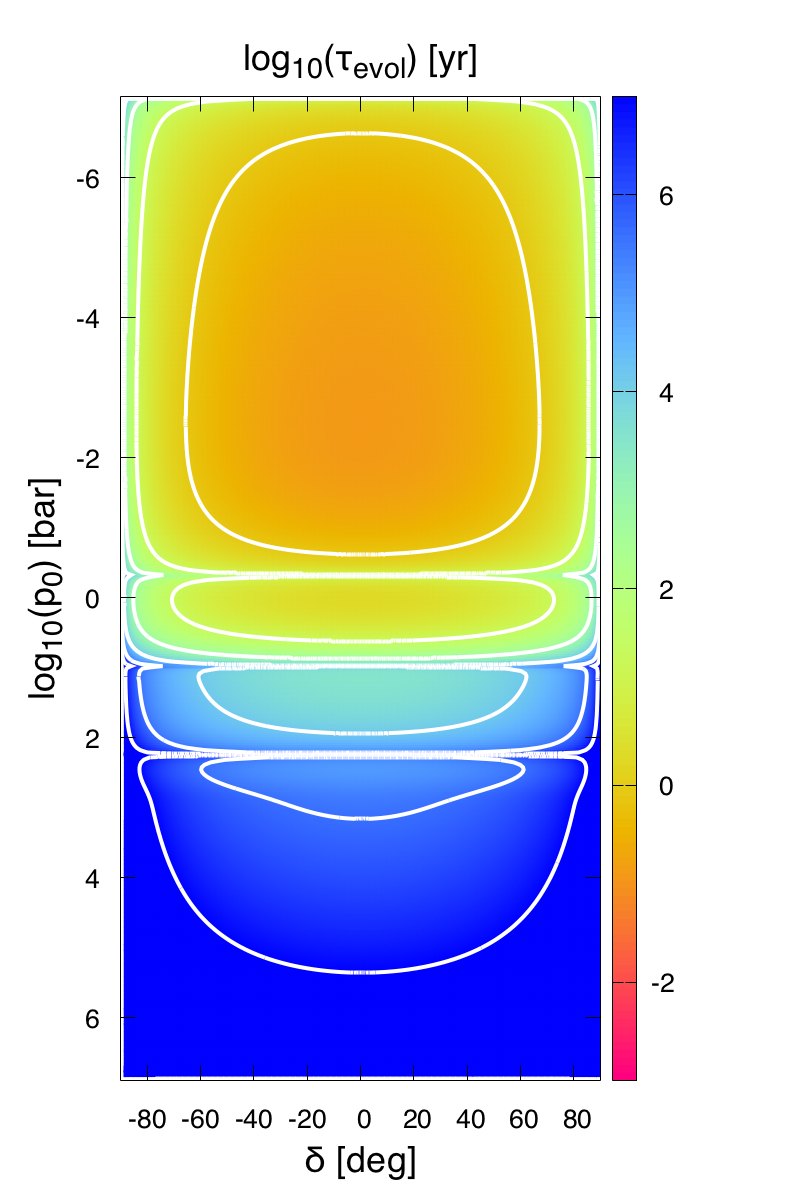}
   \includegraphics[width=0.18\textwidth,trim = 2.5cm 2.2cm 3.cm 0.cm, clip]{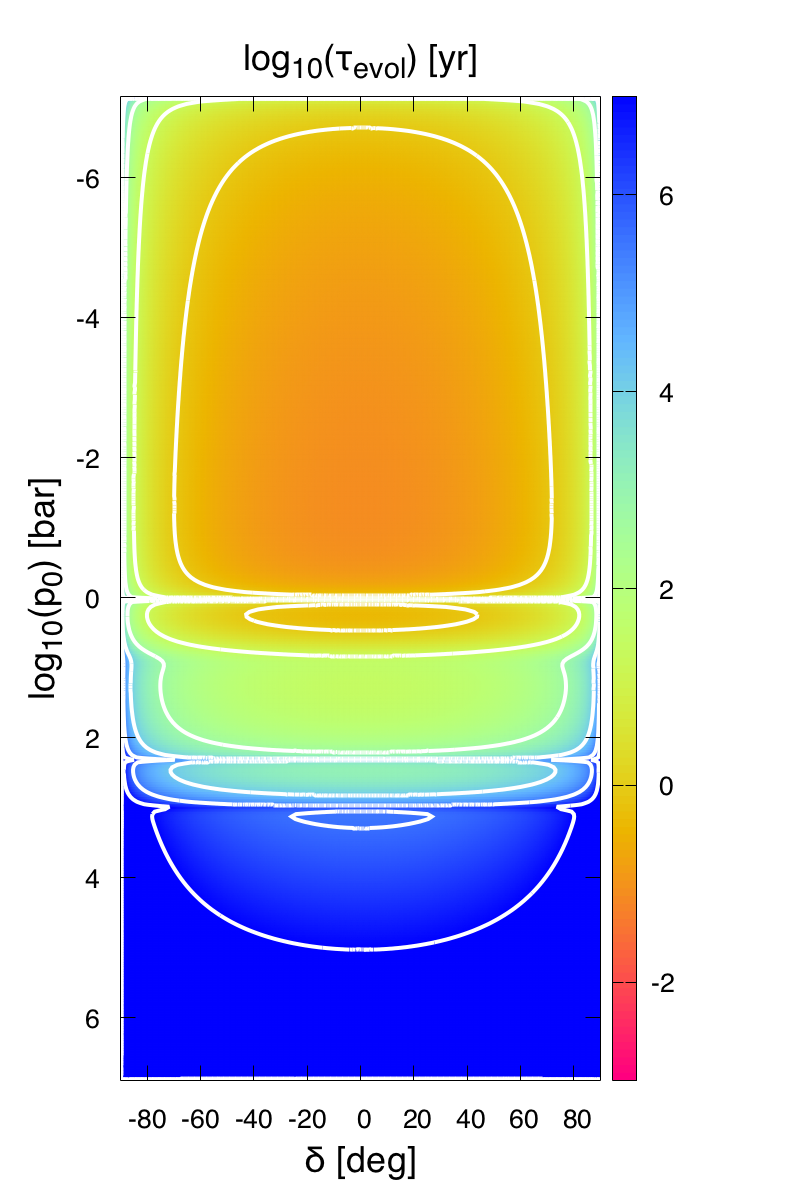}
   \includegraphics[width=0.18\textwidth,trim = 2.5cm 2.2cm 3.cm 0.cm, clip]{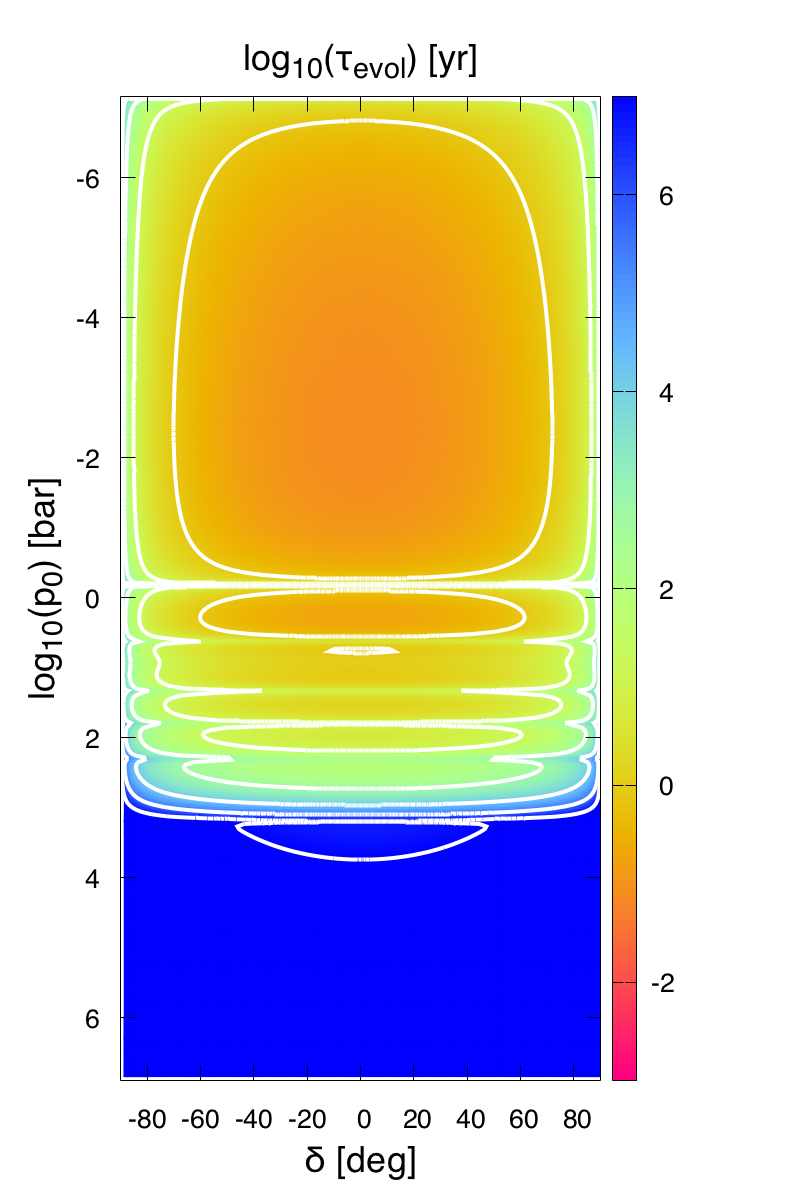} 
   \includegraphics[width=0.18\textwidth,trim = 2.5cm 2.2cm 3.cm 0.cm, clip]{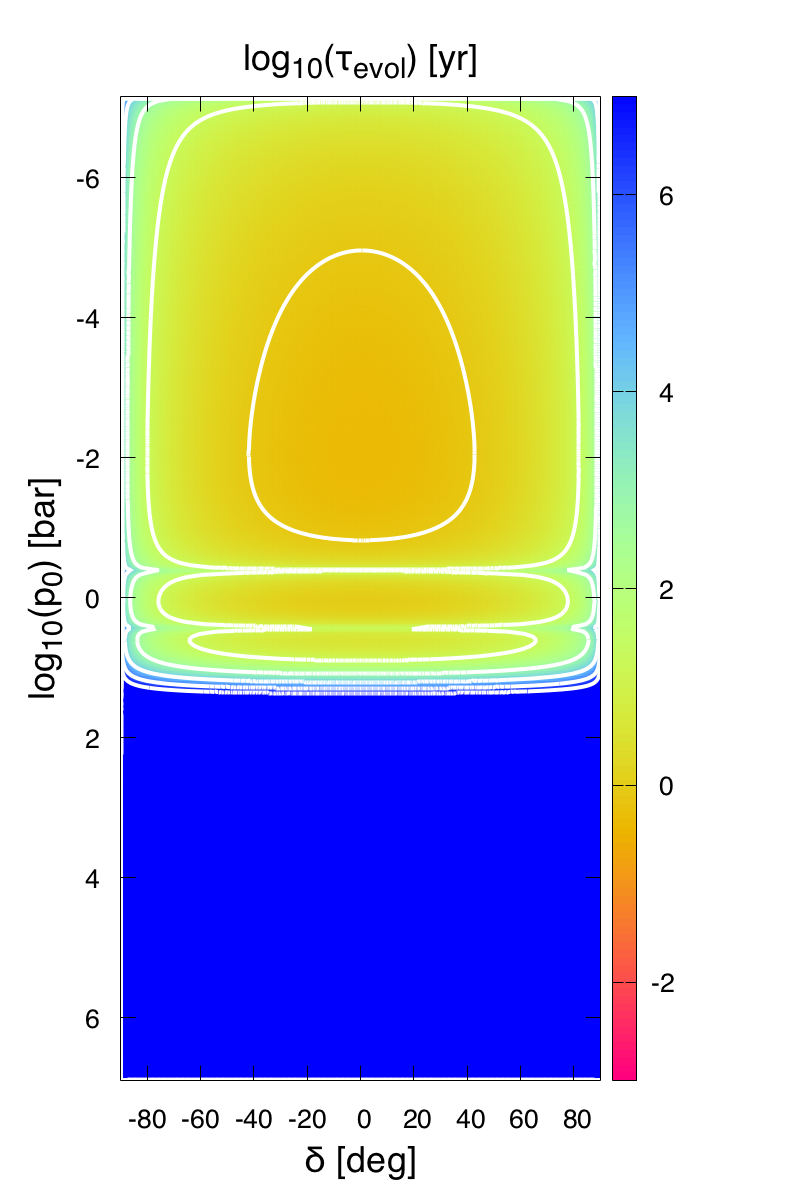} 
   \includegraphics[width=0.18\textwidth,trim = 2.5cm 2.2cm 3.cm 0.cm, clip]{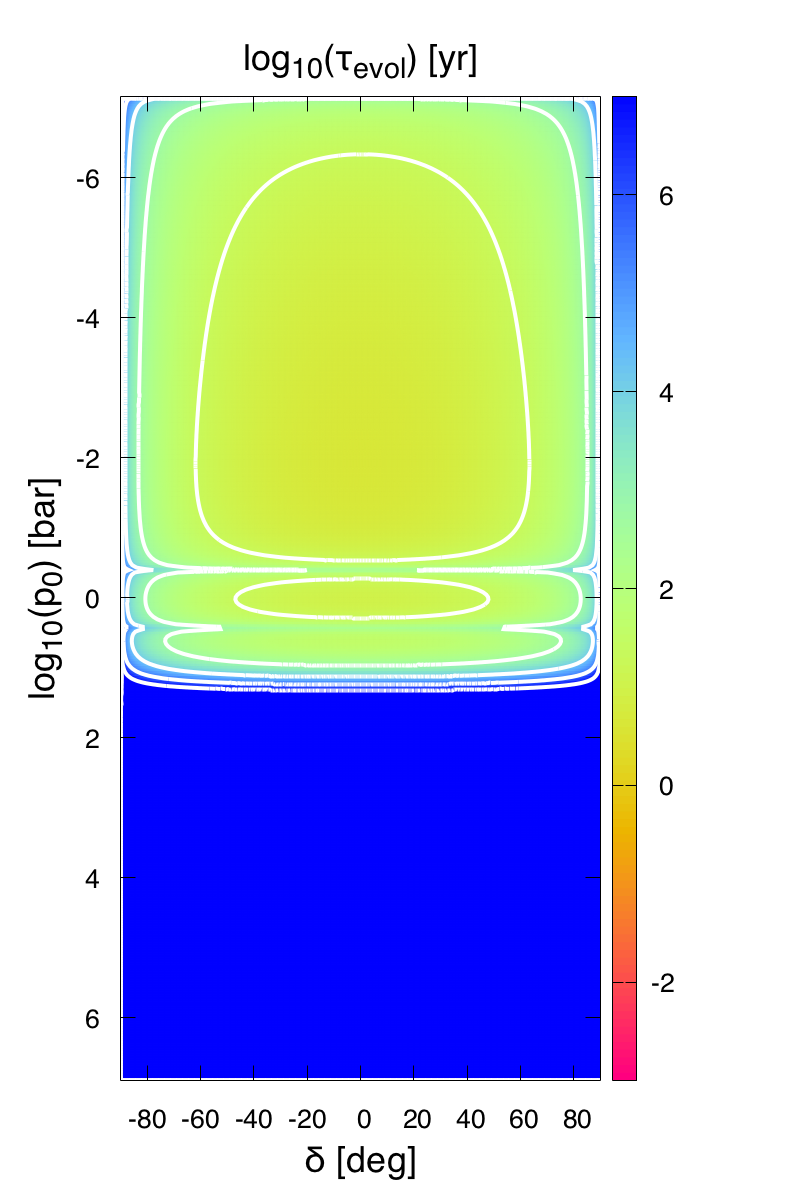} \\
   \raisebox{1.0cm}{\includegraphics[width=0.02\textwidth]{auclair-desrotour_fig6s.pdf}} \hspace{0.1cm}
   \raisebox{1.5\height}{\includegraphics[width=0.015\textwidth]{auclair-desrotour_fig6g.pdf}}
   \includegraphics[width=0.18\textwidth,trim = 2.5cm 2.2cm 3.cm 0.cm, clip]{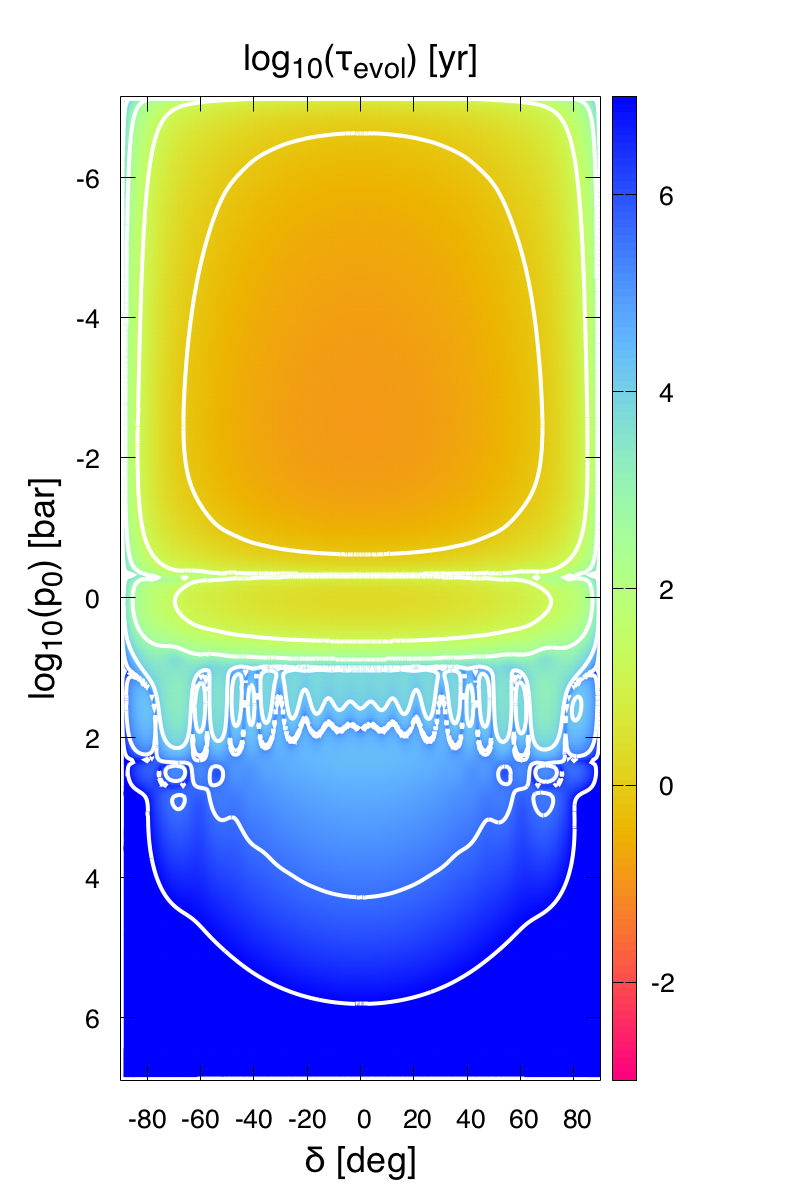}
   \includegraphics[width=0.18\textwidth,trim = 2.5cm 2.2cm 3.cm 0.cm, clip]{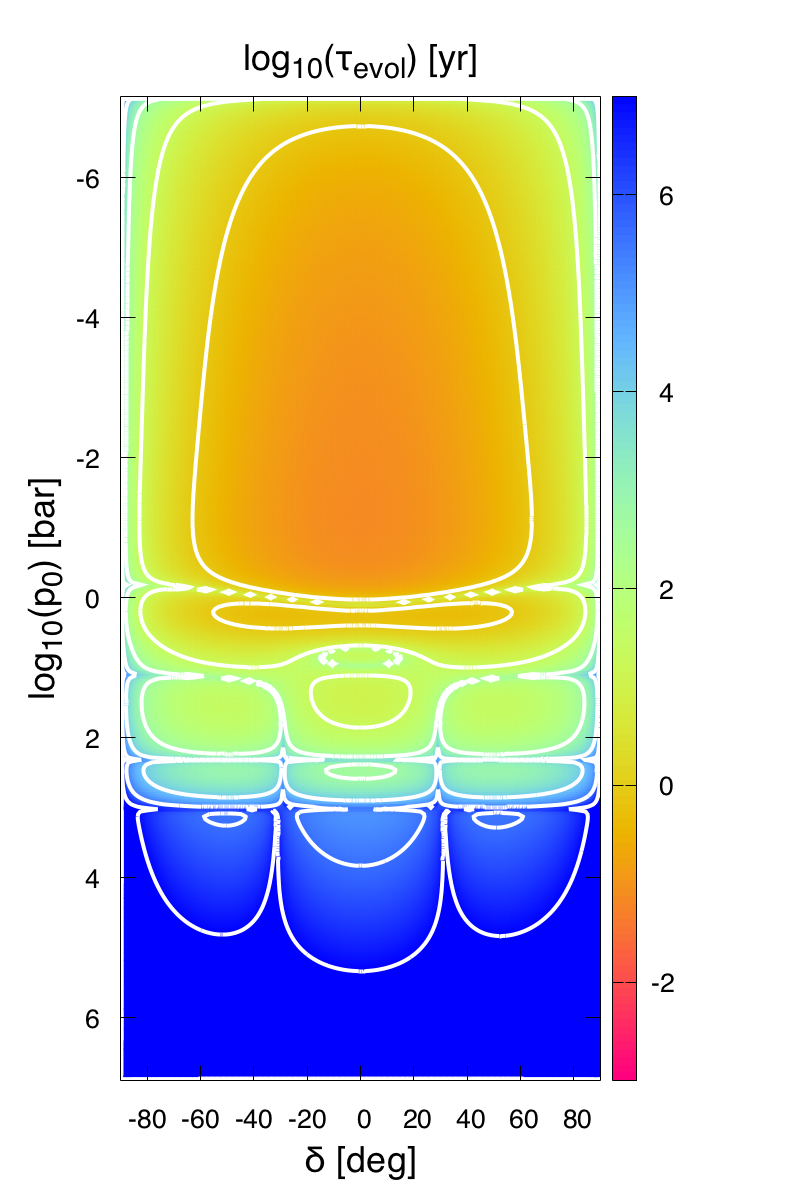}
   \includegraphics[width=0.18\textwidth,trim = 2.5cm 2.2cm 3.cm 0.cm, clip]{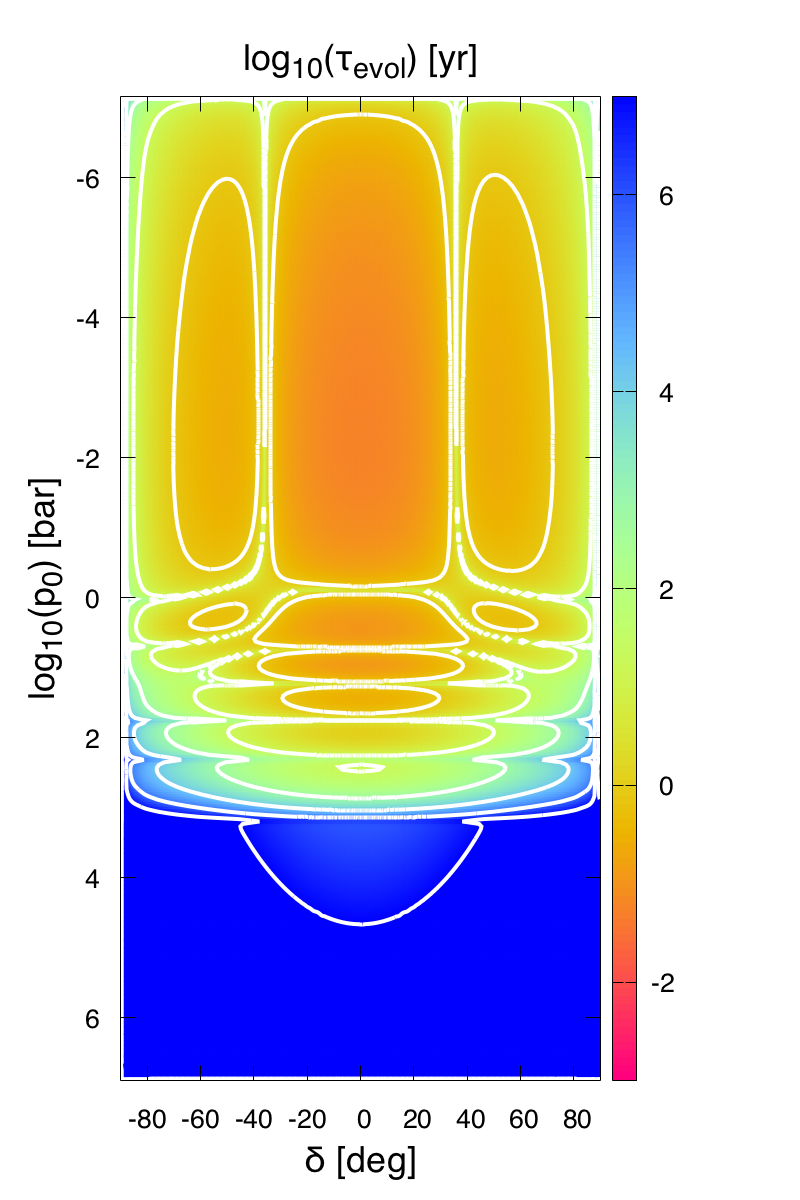} 
   \includegraphics[width=0.18\textwidth,trim = 2.5cm 2.2cm 3.cm 0.cm, clip]{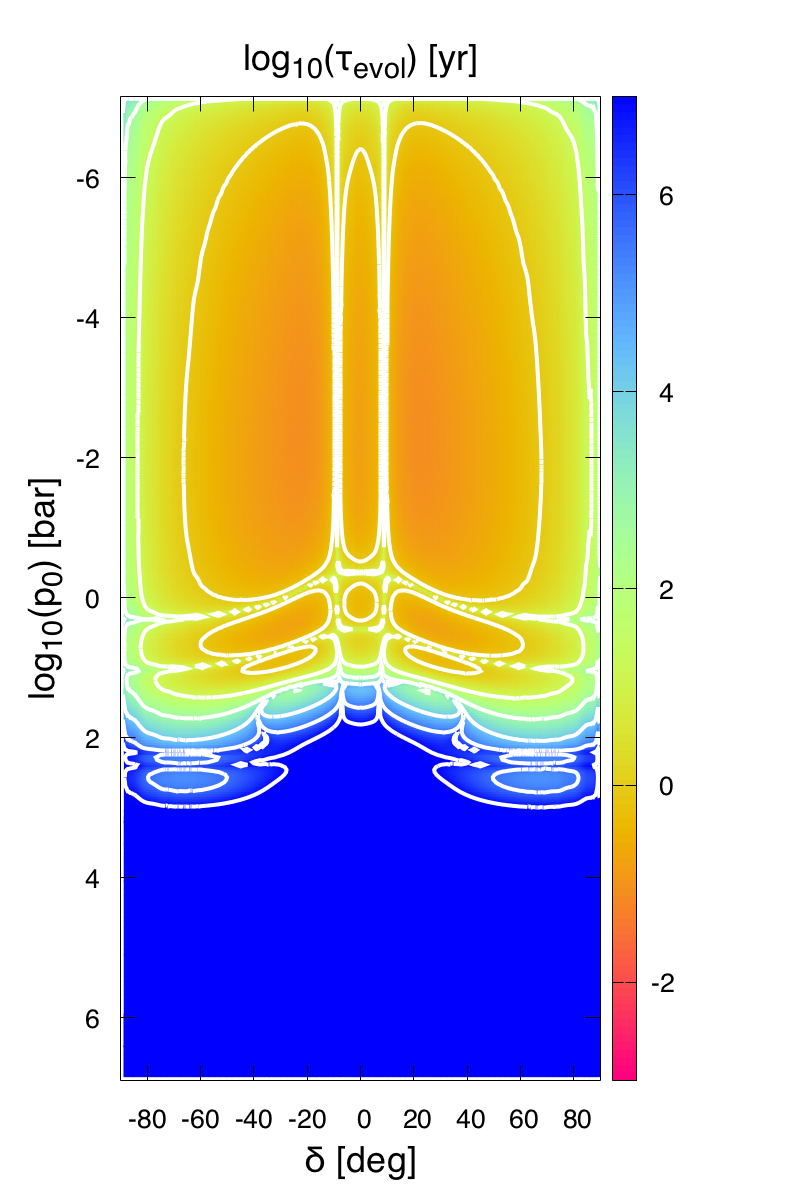} 
   \includegraphics[width=0.18\textwidth,trim = 2.5cm 2.2cm 3.cm 0.cm, clip]{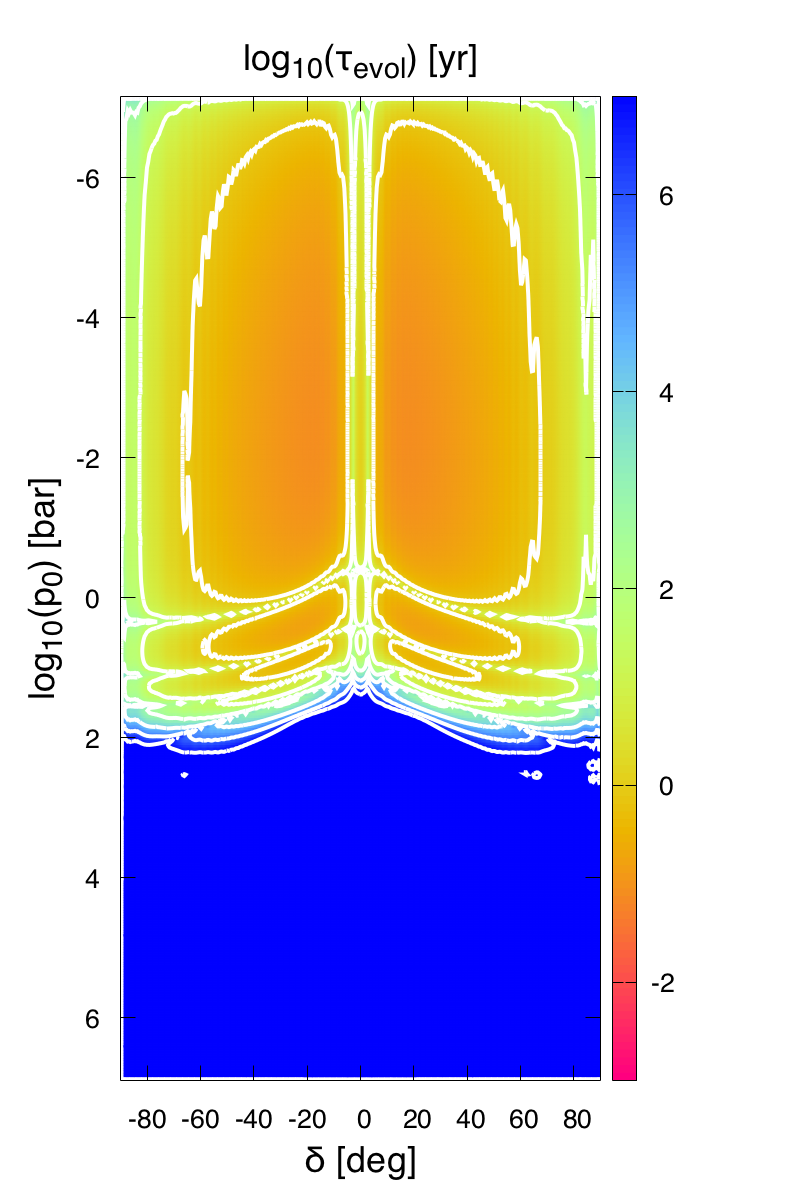} \\
   \hspace{1.9cm}
   \includegraphics[height=0.3cm]{auclair-desrotour_fig6y.pdf} \hspace{1.6cm}
   \includegraphics[height=0.3cm]{auclair-desrotour_fig6y.pdf} \hspace{1.6cm}
   \includegraphics[height=0.3cm]{auclair-desrotour_fig6y.pdf} \hspace{1.6cm}
   \includegraphics[height=0.3cm]{auclair-desrotour_fig6y.pdf} \hspace{1.6cm}
   \includegraphics[height=0.3cm]{auclair-desrotour_fig6y.pdf} \hspace{1.3cm}
   \caption{Characteristic time scale necessary for the quadrupolar semidiurnal thermal tide to generate an azimuthal jet of velocity $ V_{\rm jet}  = 1 \ {\rm km.s^{-1}} $. The logarithm of $ \tau_{\rm evol} $ (yr) is plotted using Eq.~(\ref{tau_jet}) for several decades of the forcing period ($ \tau_{\rm tide} = 2 \pi / \sigma $), $ \log \left( \tau_{\rm tide} \right) = -1, \ldots , 3 $ (from left to right), and the three studied cases. {\it Top:} case treated by \cite{AS2010}, i.e. \jlc{adiabatic without rotation} (no-Coriolis approximation). {\it Middle:} \jlc{non-adiabatic without rotation} ($ \tau_\star= 1 $~day). {\it Bottom:} \jlc{non-adiabatic} (\jlc{$ \tau_\star = 1 $~day}) with rotation (traditional approximation). The horizontal structure of the tidal response (Eq.~(\ref{Laplace_eq})) is computed for 250 Hough modes using the spectral method described by \cite{Wang2016}, with projections on 375 associated Legendre polynomials. The vertical structure equation (Eq.~(\ref{vertical_structure})) is integrated on regular mesh composed of 1000 points (10 000 points for the case $\tau_\star = \infty$) using the implicit fourth order finite differences scheme detailed in Appendix~\ref{app:num_scheme}.   }
       \label{fig:structure_Tevol}%
\end{figure*}

\jlc{Using the values of parameters given by Table~\ref{parameters}, we treat three cases:}
\begin{enumerate}
  \item \jlc{\emph{adiabatic without rotation} -- This is the case treated by \cite{AS2010}, where the rotation of the planet is ignored ($ \Omega = 0 $, $ \nu = 0 $), as well as dissipative processes ($ \tau_\star = + \infty $). }
  \item \jlc{\emph{non-adiabatic without rotation} -- The thermal time at the base of the heated layer is set to $ \tau_\star = 1 $~d, which is the order of magnitude predicted by radiative transfer modelings for the hot Jupiter HD 209458b \citep[][]{SG2002,Iro2005}. }
  \item \jlc{\emph{non-adiabatic with rotation} -- The effect of rotation is taken into account by introducing the Coriolis acceleration in the framework of the traditional approximation.}
\end{enumerate}

Since the tidal torque is directly proportional to the imaginary part of the density fluctuation (see Eq.~(\ref{Q22})), we use this quantity as a proxy to identify regions that be accelerated by the thermal tide. The imaginary part of $ \delta \rho $ is thus plotted on Fig.~\ref{fig:structure_ondes} as a function of the latitude and pressure level in each case for a wide range of forcing periods ($ \tau_{\rm tide} = 10^{-1},10^{0},10^{1},10^{2},10^{3} $~days) in the regime of sub-synchronous rotation ($ \sigma < 0 $). Hence, a positive (negative) $ \Im \left\{ \delta \rho \right\} $ is associated to an eastward (westward) accelerated zonal-mean flow. The corresponding timescale necessary to generate a jet of velocity $ V_{\rm jet} = 1 \ {\rm km.s^{-1}} $ \citep[order of magnitude of velocities of atmospheric winds in HD~209458b, e.g.][]{SG2002} is plotted on Fig.~\ref{fig:structure_Tevol} by using Eq.~(\ref{tau_jet}).

As may be observed, the thermal tide essentially affects the stably-stratified region, where the Archimedean force restores internal gravity waves and allows them to propagate. \rec{We note here that the quality of the solution depends on the condition $ \sigma^2 \ll N^2 $ mentioned above. This means that solutions involving density variations around the base of the stably-stratified region ($p_{\rm b} \approx 10^2$ bar) can be deteriorated given that the Brunt-Väisälä frequency falls down at this pressure level (see Fig.~\ref{fig:bgd}).}

\rec{In addition to gravity waves,} acoustic waves can also contribute to the tidal response, but only for forcing frequencies greater than acoustic cutoff frequencies of the horizontally propagating Lamb modes, defined by Eq.~(\ref{sigmasn}). Typically, in the non-rotating case, \jlc{only one mode} is forced by the quadrupolar thermal heating, the gravity mode of meridional degree $ n = 0 $ (black line in top panels of Fig.~\ref{fig:Hough}). This mode is associated with the eigenvalue $ \Lambda_0^{2,0} = 6 $, which sets its characteristic Lamb period ($ \tau_{\rm s ; n} = 2 \pi / \sigma_{\rm s ; n} $) to $ \tau_{\rm s ; 0} \approx 1.4~{\rm days}$ in the radiative zone. Hence, the tidal fluctuations that can be observed in the left panels of Figs.~\ref{fig:structure_ondes} and \ref{fig:structure_Tevol} are partly due to compressibility. \rec{However, the tidal frequency is not important enough to enable the propagation of internal waves in the convective region. Note that the propagation of internal waves in the convective region would not be realistic since the traditional approximation assumed in the present work is strongly violated in this region. Therefore, we will not compute solutions for tidal periods lower than $ 0.1 $ days.}

In the first case (top panels of Figs.~\ref{fig:structure_ondes} and \ref{fig:structure_Tevol}), dissipation is ignored. Therefore, in the stably stratified region and for tidal periods $ \tau_{\rm tide} \gg \tau_{\rm s;0} $, the dispersion relation of internal waves associated with the $ n $-mode given by Eq.~(\ref{kz2}) approximately reduces to 

\begin{equation}
\hat{k}_{\perp ; n}^2 +  \hat{k}_{r ; n}^2 = \frac{N^2}{\sigma^2} \hat{k}_{\perp ; n}^2, 
\label{dispersion_relation}
\end{equation}

\noindent with $ n = 0 $. This highlights the two possible asymptotic regimes. On the one hand, if the forcing period is short, the vertical wavelength is of the same order of magnitude as the typical thickness of the atmosphere. The tidal response thus exhibits larges scales patterns characterized by a small number of oscillations (top left panels in Figs.~\ref{fig:structure_ondes} and \ref{fig:structure_Tevol}). On the other hand, when tidal periods are long \jlc{($ N^2 / \sigma^2 \gg 1 $),} $ \hat{k}_{r ;n} \gg  \hat{k}_{\perp ; n} $, which explains the wave-like oscillatory response that can be observed for $ \tau_{\rm tide} \gtrsim 10 $~\rec{days}. 

\begin{figure*}
   \centering
   \includegraphics[width=0.30\textwidth,trim = 1.4cm 2.6cm 6.cm 1.cm, clip]{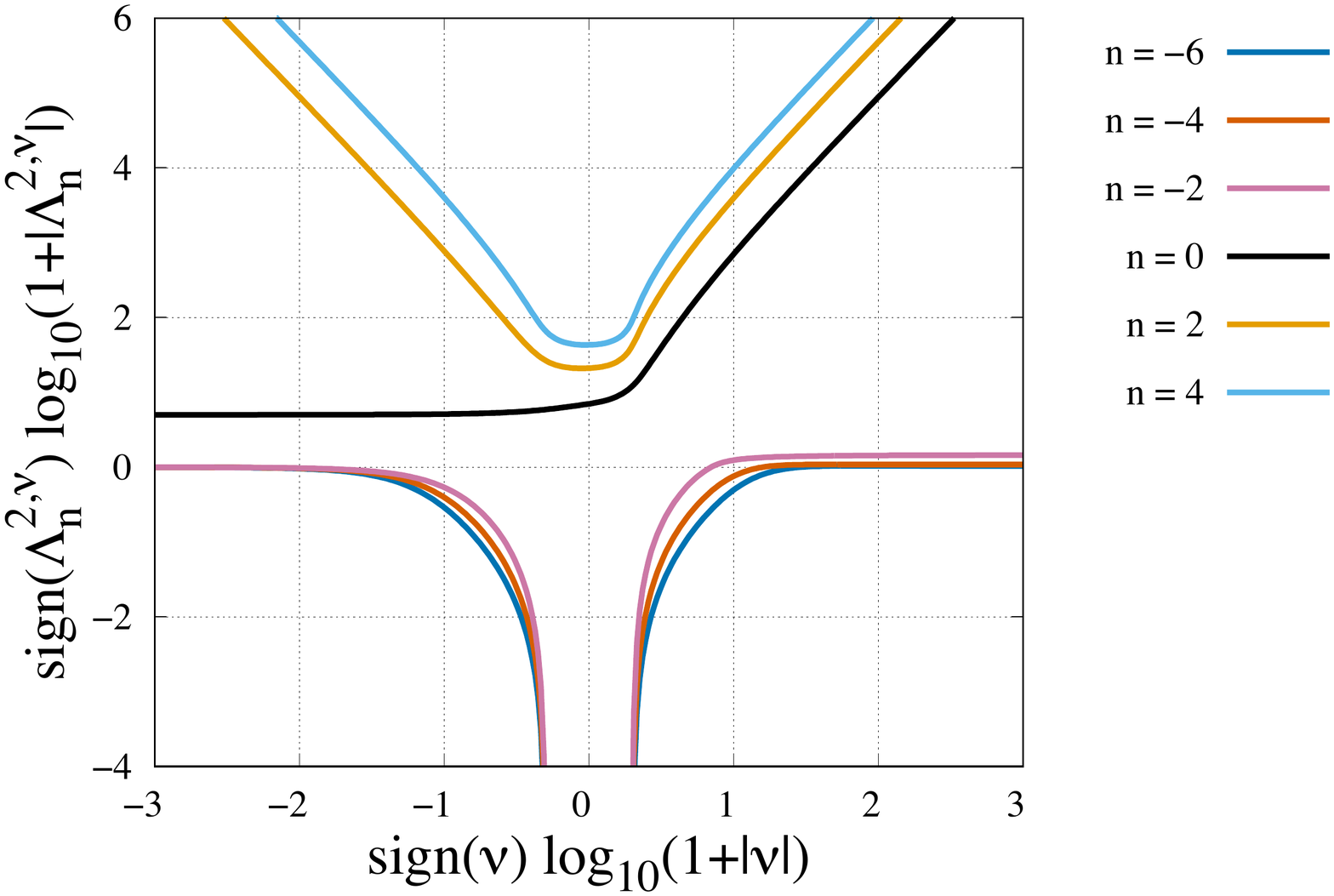}
   \includegraphics[width=0.30\textwidth,trim = 1.4cm 2.6cm 6.cm 1.cm, clip]{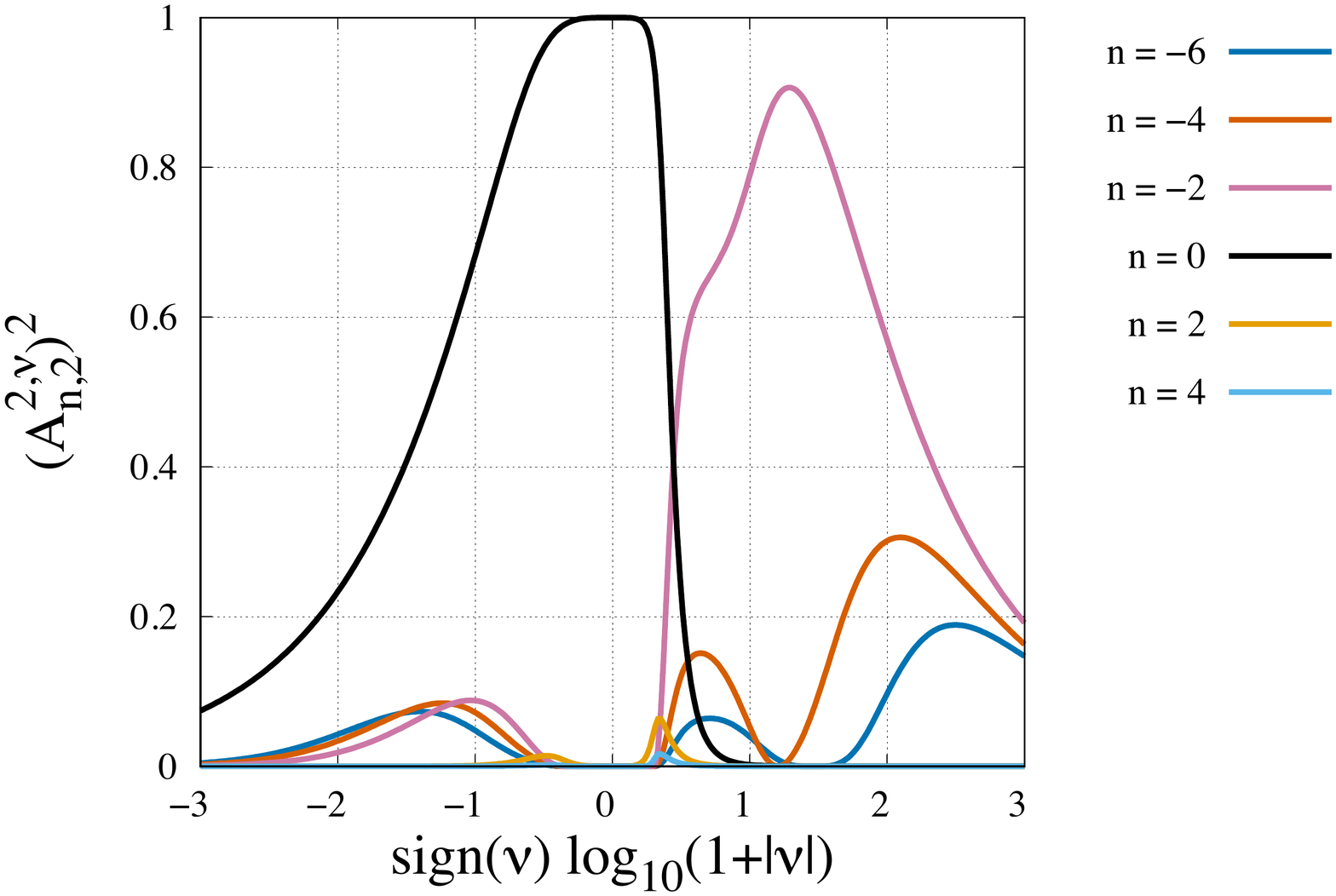}
   \includegraphics[width=0.30\textwidth,trim = 1.4cm 2.6cm 6.cm 1.cm, clip]{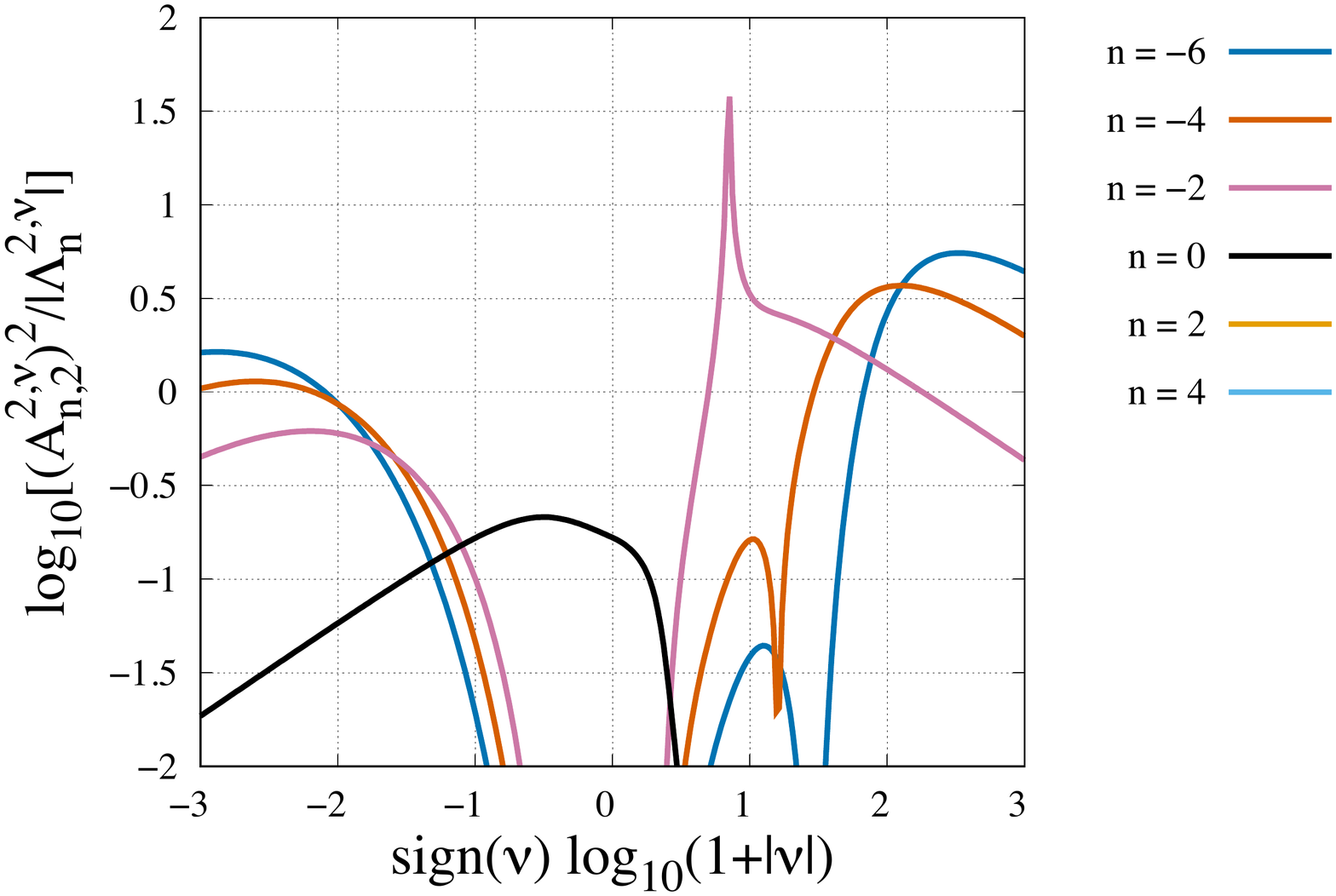}
   \includegraphics[width=0.07\textwidth,trim = 22cm 5.0cm 2.0cm 1.cm, clip]{auclair-desrotour_fig8c.pdf}
   \caption{Eigenvalues $ \Lambda_n^{2,\nu} $ associated with even Hough modes (left), coupling coefficients $ C_{2,n,2}^{2,\nu} = \left( A_{n,2}^{n,\nu} \right)^2 $ (middle) and amplification factors $ C_{2,n,2}^{2,\nu} / \left| \Lambda_n^{2,\nu} \right| $ (right) as functions of the spin parameter ($ \nu $) for $ n = \left\{ - 6 , - 4 , - 2 , 0 , 2 , 4 \right\} $ (three Rossby modes and three gravity modes). To represent quantities varying over a large range of magnitude and with sign changes, the function $ f \left( X \right) = {\rm sign} \left( X \right) \log \left( 1 + \left| X \right| \right) $ is used. It enforces $ f \left( X \right) \approx X $ for $ X \ll 1 $ and $ f \left( X \right) \approx {\rm sign} \left( X \right) \log \left( \left| X \right| \right) $ with a continuous transition. The peak observed in the right panel corresponds to the equality between the tidal frequency and the characteristic frequency of the Rossby mode of degree $ n = -2 $ \citep[at this frequency, $ \Lambda_{-2}^{2,\nu} = 0 $, see][]{LS1997}.  }
       \label{fig:Lambda_nu}%
\end{figure*}

Without dissipation, waves reach the bottom of the radiative zone and the turning point $ N^2 = \sigma^2 $, where $ \hat{k}_{r ; n} \approx 0 $. The skin thickness of their penetration in the convective region decreases while the forcing period increases. Similarly, the vertical wavelength decays folowing the scaling law $ \hat{k}_{r ; n}^2 \sim   \left( N^2 / \sigma^2 \right) \hat{k}_{\perp ; n}^2 $ derived from Eq.~(\ref{dispersion_relation}). This leads to a highly oscillating response in the vicinity of spin-orbit synchronous rotation (see top right panels of Figs.~\ref{fig:structure_ondes} and \ref{fig:structure_Tevol}). Note that such a behaviour is a source of complications for numerical calculations. Indeed, following the decay of wavelengths requires to increase the resolution by a factor 10 at each decade of $ \tau_{\rm tide} $. This strong numerical constraint prevents us to compute solutions beyond a critical value of the forcing period. Here, with the $ 10^4 $ points mesh that we use for the vertical coordinate in the first case, this value approximately corresponds to \rec{$ \tau_{\rm tide} = 10^{2}~{\rm days}$}, which means that solutions plotted in the top right panels of Figs.~\ref{fig:structure_ondes} and \ref{fig:structure_Tevol} are sub-resolved. However, this behaviour is not realistic because waves of such small scales \jlc{are in reality} damped by dissipative processes. 

This is exactly what is observed in the second case, where the radiative cooling is introduced (middle panels of Figs.~\ref{fig:structure_ondes} and \ref{fig:structure_Tevol}). Beyond the transition regime corresponding to $ \tau_{\rm tide} \sim \tau_\star $, tidal waves are first strongly damped. Second, their wavelengths remain rather large, which eliminates the numerical complications mentioned above. Third, their penetration in the atmosphere is limited by the damping. As in the previous case, the wavelike structure of the tidal response implies that the thickness of regions where the tidal force is applied scales as half of the wavelength of the dominating mode. As a consequence, the forcing tends to generate superposed zonal-mean flows of alternate directions. Figure~\ref{fig:structure_Tevol} shows the narrow separation there is between the convective envelope and the radiative atmosphere with respect to this forcing. The first layer is hardly affected by the thermal tidal torque, with time scales greater than 100 millions years to generate jets, while the second one is submitted to a strong forcing. In this region, the time scale can reach values below 1 year. In the absence of dissipation, the linear response diverges, which explains why $ \tau_{\rm evol} $ is very short in the top right panels of Fig.~\ref{fig:structure_Tevol}. The observed contrast between the two layers is due to the fact that the whole energy of the semidiurnal thermal tide is deposited in the radiative zone, which is very thin compared to the planet radius (Fig.~\ref{fig:bgd}, left panel) and far less dense than the convective region (see Fig.~\ref{fig:bgd}). 

As shown by the bottom panels of Figs.~\ref{fig:structure_ondes} and \ref{fig:structure_Tevol}, introducing rotation distorts the regular structure associated with the quadrupolar mode, which was the only forced mode in the non-rotating cases. The fluid response is now formed by a series of Hough modes that describe the propagation of gravito-acoustic-inertial waves of different wavelengths. In order to facilitate the interpretation of results in the rotating case, we plot in Fig.~(\ref{fig:Lambda_nu}) the eigenvalues associated to even Hough functions of degrees $ n = \left\{ - 6 , - 4 , - 2 , 0 , 2 , 4 \right\} $, the corresponding coupling coefficients with the quadrupolar forcing, $ C_{2,n,2}^{2,\nu} =  \left( A_{n,2}^{2,\nu} \right)^2 $, and the ratios $ C_{2,n,2}^{2,\nu} / \left| \Lambda_n^{2,\nu} \right| $ that will be identified as amplification factors resulting from rotation in the next section. The values of spin parameters, computed using Eq.~(\ref{nu_sigma}), and the \rec{associated $ \Lambda_n^{2,\nu} $ and $ C_{2,n,2}^{2,\nu} $}, are summarized in Table~\ref{tab:nu_values}. 

\begin{table}[h]
\centering
 \textsf{\caption{\label{tab:nu_values} Values of the spin parameter ($ \nu $), eigenvalues of Hough modes ($ \Lambda_n $) and coupling coefficients ($C_{2,n,2}$) associated with the values of $ \tau_{\rm tide} $ used in Section~\ref{sec:properties_waves} for $ n = \left\{ -6 , -4 , -2 , 0 , 2 , 4 \right\} $.  }}
\begin{small}
    \begin{tabular}{ l c c c c c}
      \hline
      \hline
      \textsc{Case} & (a) & (b) & (c) & (d) & (e) \\ 
      \hline 
      $ \tau_{\rm tide} \ \left[ {\rm d} \right]$ & $ 0.1 $ & $ 1 $ & $ 10 $ & $ 100 $ & $ 1000 $ \\
      $ \nu $ & $ 0.951 $ & $ 0.510 $ & $ -3.90 $ & $ -48.0 $ & $ -489 $ \\ 
      \hline
      $ \Lambda_{-6} $ & -- & -- & $ -14.0 $ & $ - 0.142$ & $-6.27 \, {\rm E}^{-3}$ \\
      $ \Lambda_{-4} $ & -- & -- & $-8.30$ & $ - 0.106 $ & $-5.58 \, {\rm E}^{-3}$ \\
      $ \Lambda_{-2} $ & -- & -- & $-4.16$ & $ -7.60 \, {\rm E}^{-2} $ & $-4.95 \, {\rm E}^{-3}$ \\
      $ \Lambda_{0} $ & $10.6$ & $7.55$ & $4.26$ & $ 4.02 $ & $4.00 $ \\
      $ \Lambda_{2} $ & $ 38.0$ & $23.7$ & $159$ &$ 2.09 \, {\rm E}^{4}$ & $ 2.15 \, {\rm E}^{6} $ \\
      $ \Lambda_{4} $ & $ 82.2 $  & $ 49.1 $ & $759$ & $1.13 \, {\rm E}^{5}$ & $1.27 \, {\rm E}^{7}$ \\
      \hline
      $ C_{2,-6,2} $ & -- & -- & $1.40 \, {\rm E}^{-2}$ & $6.35 \, {\rm E}^{-2}$ & $1.00 \, {\rm E}^{-2}$ \\
      $ C_{2,-4,2} $ & -- & -- & $2.68\, {\rm E}^{-2}$ & $5.82 \, {\rm E}^{-2}$ & $6.32 \, {\rm E}^{-3}$ \\
      $ C_{2,-2,2} $ & -- & -- & $5.01\, {\rm E}^{-2}$ & $3.79 \, {\rm E}^{-2}$ & $2.68 \, {\rm E}^{-3}$ \\
      $ C_{2,0,2} $ & $0.954$ & $0.998$ & $0.875$ & $0.330$ & $0.106$ \\
      $ C_{2,2,2} $ & $ 3.81 \, {\rm E}^{-2}$ & $1.86 \, {\rm E}^{-3}$ & $7.21 \, {\rm E}^{-3}$ & $9.65 \, {\rm E}^{-6}$ & $9.66 \, {\rm E}^{-9}$ \\
      $ C_{2,4,2} $ & $ 5.55 \, {\rm E}^{-3} $  & $9.69\, {\rm E}^{-6}$ & $7.26 \, {\rm E}^{-4}$ & $7.59 \, {\rm E}^{-7}$ & $9.83 \, {\rm E}^{-10}$ \\
      \hline
    \end{tabular}
\end{small}
 \end{table}

In the asymptotic case of rapid rotation, the \jlc{predominant} modes are the gravity modes of smallest horizontal wavenumbers. Hence, for $ \tau_{\rm tide} = 10^{-1} $~days (bottom left panel), the $ P_2^2 $-like meridional variation of $\delta \rho $ and $\tau_{\rm evol}$ corresponds to the gravity mode of degree $ n =0 $. In the next case ($ \tau_{\rm tide} = 1 $~day), we note that the amplitude of the harmonic of degree 2 is of the same order of magnitude as that of degree 0, while $ \Theta_0^{2,\nu} $ is far better coupled to $ P_2^2 $ than $ \Theta_2^{2,\nu} $ in this case ($ \nu = 0.51 $). This results from a resonant amplification of the harmonic. While $ \tau_{\rm tide} $ increases, gravity modes tend to be confined equatorially. We observe this effect in the bottom middle panel of Figs.~\ref{fig:structure_ondes} and \ref{fig:structure_Tevol}, where the meridional structure of the tidal response is essentially shaped by the main gravity mode ($ n = 0 $, see Fig.~\ref{fig:Hough}, top left panel). Because of the equatorial confinement, the coupling between gravity modes and the quadrupolar forcing decays. It follows that the family of modes composing the response switches from gravity modes to Rossby modes around $ \tau_{\rm tide} = 10~{\rm day}$, that is the transition between the inertial and the sub-inertial regimes illustrated by Fig.~\ref{fig:nu_Ttide}. 

In the vicinity of spin-orbit synchronous rotation, the tidal response converges toward the asymptotic behaviour of the zero-frequency limit ($\tau_{\rm tide} \rightarrow + \infty $), that is the so-called \emph{equilibrium tide}. The aspect of spatial distributions plotted in Figs.~\ref{fig:structure_ondes} and \ref{fig:structure_Tevol} does not evolve any more. This behaviour is studied analytically in Appendix~\ref{app:low_frequencies}, where the vertical profiles of perturbed quantities associated with the equilibrium thermal tide are given. By comparing the dissipative cases with and without rotation (right panels of cases 1 and 2), we notice that rotation strongly affects this regime. In the static approximation ($\Omega = 0$), the amplitude of the tidal response decreases while $ \tau_{\rm tide} $ increases, following the scaling law $ \delta \rho \propto \sigma $. It is not the case any more when rotation is taken into account. Instead of decreasing, the amplitude of the density fluctuations saturates. Thus, the tidal response is enhanced by rotation. This saturation typically occurs for $  \left| \sigma \right| \ll \left| 2 \Omega \right| $, i.e. when the Coriolis effects exceed the thermally forced advection in the momentum equation. We will see in the next section that it results from the amplification of Rossby modes by the \rec{factor $ C_{2,n,2}^{2,\nu} /  \left| \Lambda_n^{2,\nu} \right|  $} mentioned above. As shown by Fig.~\ref{fig:Lambda_nu}, this factor plummets for gravity modes in the limit $ \nu \rightarrow - \infty $, while it increases in the case of Rossby modes. Therefore, the coupling of the tidal response with Rossby modes is accentuated and the coupling with gravity modes annihilated. This explains why a gap can be observed at the equator, where gravity modes are confined (bottom right panels of Figs.~\ref{fig:structure_ondes} and \ref{fig:structure_Tevol}). 


\section{Frequency spectra of the tidal torque}
\label{sec:spectra_torque}

\begin{figure*}
   \centering
   \includegraphics[width=0.49\textwidth,trim = 0.5cm 0cm 1.cm 1.cm, clip]{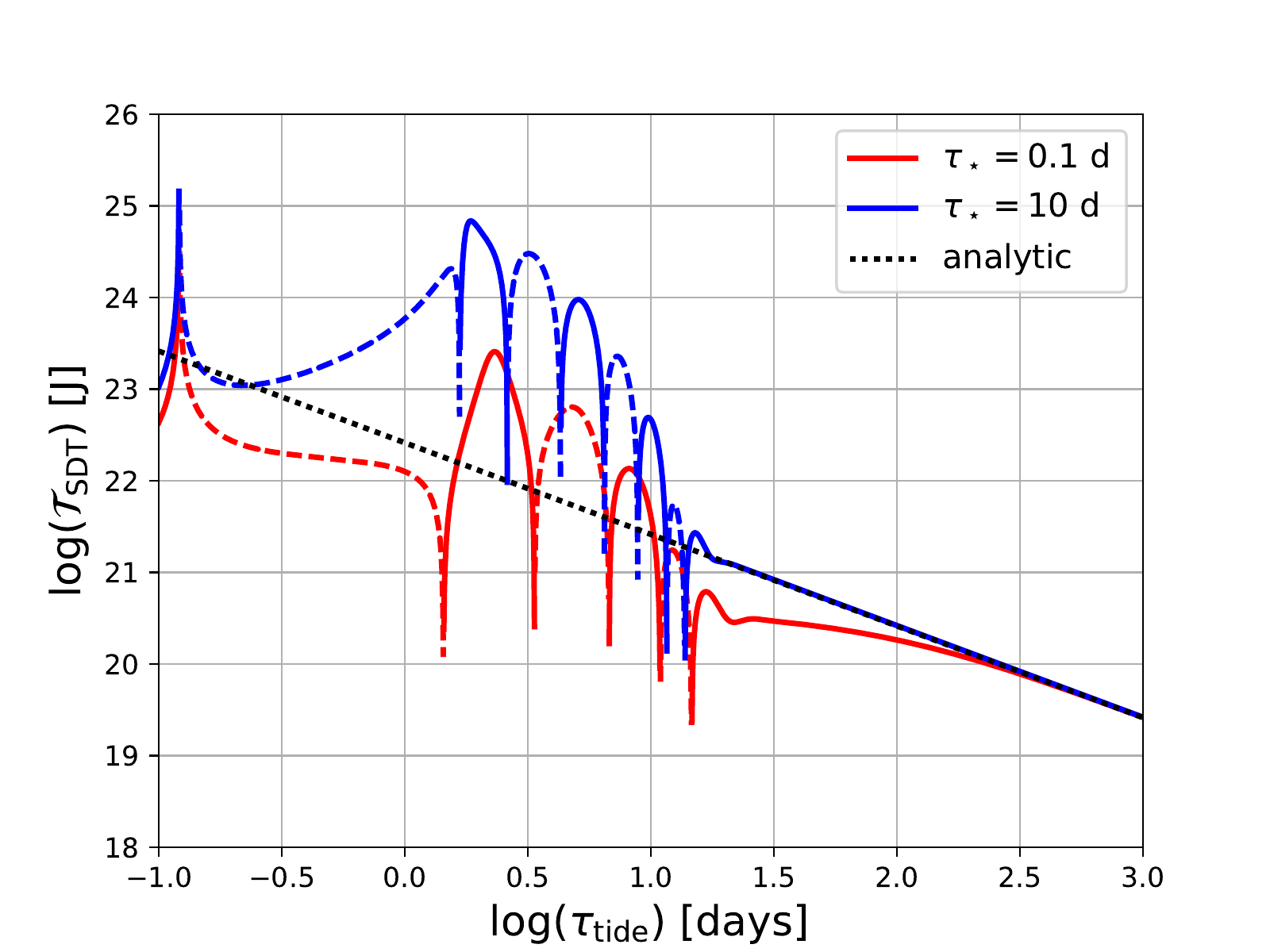}
   \includegraphics[width=0.49\textwidth,trim = 0.5cm 0cm 1.cm 1.cm, clip]{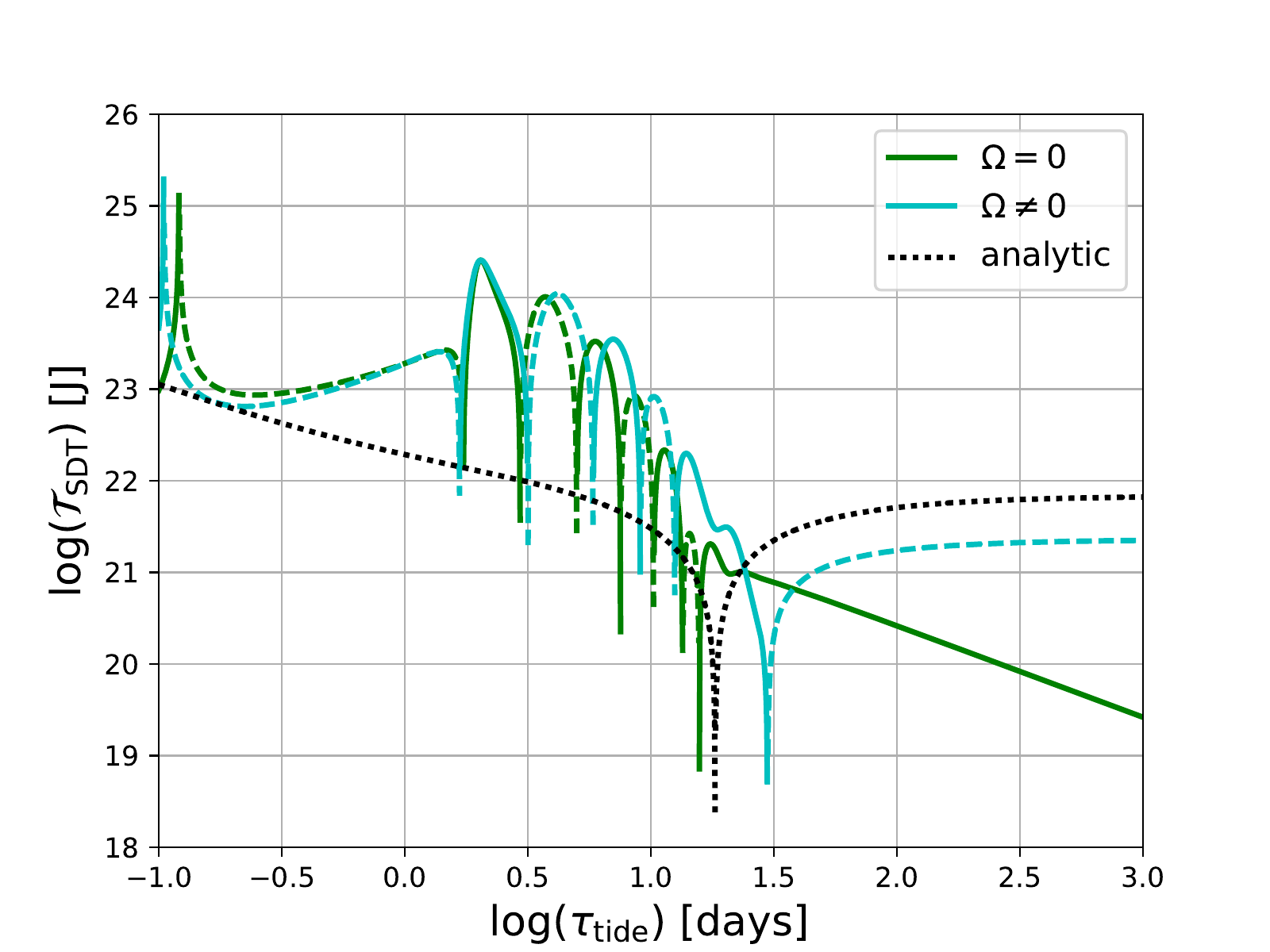}
   \caption{Total tidal torque exerted of the planet (N.m) as a function of the tidal period (days). {\it Left:} for two extremal value of the thermal time ($ \tau_\star = 0.1 , 10 $ days) in the static approximation ($ \Omega = 0 $). {\it Right:} for $ \tau_\star = 1 $ ~day in the static ($ \Omega = 0 $) and traditional ($ \Omega \neq 0 $) approximations \rec{(in this case, $\Omega$ is given by $ \Omega = n_{\rm orb} + \sigma / 2 $ with $ \left| \sigma \right| = 2 \pi / \tau_{\rm tide} $ and $n_{\rm orb} = 2 \pi / \tau_{\rm orb}$, see Table~\ref{parameters})}. In both panels, the torque resulting from the quadrupolar component of the semidiurnal tide is plotted in logarithmic scale (vertical axis) as a function of the logarithm of $\tau_{\rm tide} $ (horizontal scale). \jlc{Solid} (dashed) lines correspond to negative (positive) torques, pushing the planet away (toward) spin-orbit synchronous rotation. The black dotted line designates the equilibrium tidal torque plotted using the analytic formula of Eq.~(\ref{Q22_eq}). \jlc{The grey zone corresponds to the region of the spectrum where the rotation rate exceeds the critical Keplerian angular frequency, defined in Section~\ref{sec:tidal_waves_dyn}.} Values of parameters used for these calculations are summarized in Table~\ref{parameters}. The vertical structure of the tidal response is integrated on a regular mesh composed of $ 10^4 $ points for the left panel and $ 10^3 $ points for the right panel.  }
   \label{fig:spectre_couple_diss}%
\end{figure*}

We end this study by examining how the total tidal torque exerted on the planet depends on the tidal frequency and how it is affected by the radiative cooling and rotation. As done in the previous section, we isolate the component of the fluid response associated with the thermal tides by setting $ U = 0 $ in the equations of tidal waves, and use the parameters of Table~\ref{parameters}. First, following \cite{AS2010}, we consider the static approximation ($ \Omega = 0 $) and compute the tidal response for a strong radiative cooling ($ \tau_\star = 0.1 $ day) and a weaker one ($ \tau_\star = 10 $ day). Second, we compare the static ($ \Omega = 0 $) and rotating cases ($ \Omega = n_{\rm orb} + \sigma / 2 $) for $ \tau_\star = 1 $~day. The obtained torques are plotted on Fig.~\ref{fig:spectre_couple_diss} as functions of the tidal period. 

In the static approximation (left panel of Fig.~\ref{fig:spectre_couple_diss}), we reproduce the results previously obtained by \cite{AS2010}. Three regimes can be observed. At small tidal periods, the spectrum of the tidal torque exhibits \rec{a resonance} due to the propagation of internal gravito-acoustic waves in the radiative zone. For extremal values of the tidal period, tidal waves are mainly restored by the fluid compressibility, which allows them to cross the lower limit of the stably-stratified atmosphere and propagate in the central convective region. Note however that these frequencies correspond to rotation rates greater than the critical Keplerian frequency ($ \Omega_{\rm c} $), meaning that the planet should be destroyed by centrifugal distortion in this frequency range. In the range $ \tau_{\rm tide} \approx 1 - 30 $~days, \rec{a batch of resonances can be observed}. It results from the excitation of the pure gravity waves observed in Figs.~\ref{fig:structure_ondes} and \ref{fig:structure_Tevol} (middle columns). Contrary to those associated with gravito-acoustic waves, these resonances are damped by the radiative cooling. Their amplitude decays while $ \tau_\star $ increases. The non-dissipative asymptotic case, $ \tau_\star = + \infty $ corresponds to the spectrum of Fig.~(5) in the work by \cite{AS2010}. 

Finally, in the limit of long tidal periods (zero-frequency limit), the tidal response converges toward the regime of the equilibrium thermal tide. For more details about solutions in this regime, we refer the reader to the analytic study detailed in Appendices~\ref{app:low_frequencies} and \ref{app:multipole_moment}. Particularly, we derive in this study an analytic approximation of the tidal quadrupole moment $ Q_{2}^{2,\sigma} $ (Eq.~(\ref{Q22})), which is expressed as (see Eq.~(\ref{Qnlm_app}))

\begin{align}
 \label{Q22_eq}
Q_{2}^{2,\sigma} = & \sum_n A_{n,2}^{2,\nu} \left\{   \left[ 1 - \frac{30}{\Lambda_n^{m,\nu}} \right] \int_0^{R_{\rm e}}  \rho_0 r^{4} \left( \frac{\sigma^2}{N^2} \right) \frac{J_n}{i \sigma T_0 C_{\rm p}} \dd r  \right. \\
 & \left. + \frac{R^{5}}{\Lambda_n^{m,\nu}} \left[ \left( \frac{\sigma^2}{N^2} \right)  \rho_0 \left( 6 + \frac{R}{H} \right) \frac{J_n}{i \sigma T_0 C_{\rm p}}  \right]_{r = R_{\rm e}} \right\}, \nonumber
\end{align}

\noindent \rec{where $ R_{\rm e}$ designates the radius of the upper limit of the atmosphere introduced in Section~\ref{subsec:setup}.} This formula is the generalization of the scaling law given by \cite{AS2010} in Eq.~(45) to the rotating and dissipative case. It shows that the radiative cooling does not intervene in the asymtotic regime of the equilibrium thermal tide since the quadrupole moment does not depend on $ \sigma_0 $. Hence, whatever the efficiency of the radiative cooling, the fluid tidal response associated with the thermal tide invariably converges toward the same asymptotic regime and the same spatial distributions of perturbed quantities. This corresponds to what is observed on Fig.~\ref{fig:spectre_couple_diss} (left panel), where the equilibrium tidal torque is plotted by using Eq.~(\ref{Q22_eq}) (dotted black line). In the asymptotic zero-frequency regime, the quadrupole scales as $ Q_{n,2}^{2,\sigma} \, \propto \, \sigma $\jlc{, a frequency-dependence corresponding to that described by the \emph{constant time lag model} \citep[][]{Mignard1979,Mignard1980,Hut1981}}. Moreover, we can notice that the sign of the of the tidal torque varies with the tidal frequency depending of the internal variation of mass distribution generated by tidal waves (see Fig.~\ref{fig:structure_ondes}, middle and to panels). As the rotation is sub-synchronous in the studied case ($ \sigma < 0 $), a positive torque (dashed line) pushes the planet toward spin-orbit synchronization in a global way while a negative torque (continuous line) tends to drive it away from this state of equilibrium. Note that this diagnosis is only valid as a zero-order approximation given that fluid layers are differentially forced by the thermal tide. The tidal forcing will generate zonal flows in the radiative zone as shown by early studies \citep[e.g.][]{GO2009} and the previous section, rather than modify the mean solid rotation of the planet. These flows can nevertheless provide angular momentum to deeper layers through viscous coupling \citep[e.g.][]{TDD2014}.

Let us now examine the effect of rotation on the frequency spectrum (Fig.~\ref{fig:spectre_couple_diss}, right panel). As seen in the previous section, the rotation increases the number of excited modes by inducing a coupling in the momentum equations of tidal waves through the Coriolis acceleration. The strength of this effect is related to the absolute value of the spin parameter ($ \nu = 2 \Omega / \sigma $), the asymptotic limit $ \nu \rightarrow 0 $ corresponding to the static case. Thus, in the regime of rapid rotation $ \nu \approx -1 $, and the number of resonances due to gravito-acoustic waves is increased. The transition between retrograde and prograde rotation occurs in the period range $ \tau_{\rm tide} \approx 1 - 10 $~days (see Fig.~\ref{fig:nu_Ttide}). As a consequence, the regime of the fluid response is super-inertial and the quadrupolar forcing is essentially coupled with the gravity mode of degree $ n = 0 $ (Fig.~\ref{fig:Lambda_nu}, middle panel). This explains why the resonances associated with gravity waves in the range $ \tau_{\rm tide} \approx 1 - 30 $~days are weakly modified by Coriolis effects.

It is not the case of the low-frequency range, where the tidal torque ceases to decay beyond $ \tau \approx 30 $~days. To interpret this behaviour, let us come back to the analytic formula of the quadrupole, given by Eq.~(\ref{Q22_eq}). Unlike dissipative processes, the rotation strongly affects the equilibrium thermal tide through the coupling coefficients $A_{n,l}^{m,\nu} $ and the eigenvalues $\Lambda_n^{m,\nu} $. In the regime of sub-inertial waves, where $ \left| \nu \right| \gg 1  $, the quadrupolar perturbation of the semidiurnal tide is mainly coupled with Rossby modes, characterized by very small eigenvalues in absolute value \rec{\citep[see Fig.~\ref{fig:Lambda_nu}, left panel, and][for scaling laws of the $ \Lambda_n^{m,\nu} $]{Townsend2003}}. Therefore, the amplitude of the tidal response associated with the $n $-Hough mode is enhanced by a factor 

\begin{equation}
\frac{\left( A_{n,2} \right)^2 }{\left| \Lambda_n^{m,\nu} \right|} \gg \frac{1}{ l \left( l + 1 \right) } = \frac{1}{6},
\end{equation}

\noindent plotted in Fig.~\ref{fig:Lambda_nu} (right panel). This corresponds to the fact that Coriolis effects dominate the forced advection terms in the momentum equation (Eq.~(\ref{momentum_equation})). \jlc{As $ \tau_{\rm tide} \rightarrow + \infty $, $ \Omega \rightarrow n_{\rm orb} \neq 0 $, which lets time to the Coriolis force to affect tidal motions.} As a consequence, the tidal response reaches the observed saturation plateau (Fig.~\ref{fig:spectre_couple_diss}, right panel, cyan curve) for $ \left| \sigma \right| \ll \left| 2 \Omega \right| $ while the tidal frequency decreases. The global quadrupole does not scale as $ Q_{n,l}^{m,\sigma} \propto \sigma $ as in the static case. 

The analytic formula of Eq.~(\ref{Q22_eq}) allows us to approximate this behaviour. We can notice however a \jlc{factor 2-3} difference between the numerical results and the analytic approximation. This difference is due to limitations of the analytic study of Appendix~\ref{app:low_frequencies}, where the eigenvalues associated with Hough modes are considered as weakly sensitive to the tidal frequency. Actually, these parameters strongly depend on $ \sigma $ in the sub-inertial regimes. Particularly, those associated with the Rossby modes decrease in absolute value while $ \nu \rightarrow - \infty $ \citep[Fig.~\ref{fig:Lambda_nu}, left panel, see also][]{LS1997,Townsend2003}. They \jlc{thus tend} to compensate the decay of the tidal frequency in the zero-frequency asymptotic limit. \jlc{We however recover the correct functional form.}

\rec{The observed behaviour of the tidal torque in the low-frequency range is due to the absence of friction in the model. This absence is responsible for the widening of the spectrum of Hough modes when $ \sigma \rightarrow 0 $. In calculations, one has to take a large number of modes into account to compute the tidal response because their amplitude decays very slowly while the meridional degree ($n$) increases. As it is not possible to compute an infinite number of modes, this leads to truncature effects, which tends to degrade the obtained solution. In reality, friction modifies the tidal response in the low frequency range by introducing an additional timescale, denoted $ \tau_{\rm friction} $. If $ \tau_{\rm tide} \ll \tau_{\rm friction} $, the obtained solution is not modified. If $ \tau_{\rm tide} \gtrsim \tau_{\rm friction} $, it has a strong impact on the spectrum of excited Hough modes. In the asymptotic limit $ \tau_{\rm tide} / \tau_{\rm friction} \rightarrow + \infty $, Rossby modes merge with gravity modes and converge toward the associated Legendre polynomials \citep[see e.g.][]{VM1972a,Volland1974a}, which are the Hough functions in the absence of rotation. This means that the spectrum of Hough modes coupled with the quadrupolar thermal forcing tends to reduce to one function, as in the static case, when $ \tau_{\rm tide} \gtrsim \tau_{\rm friction} $, instead of broadening as observed in the present study. Hence, we expect that introducing friction in further works will change the aspect of the frequency-spectrum of the tidal torque in the low-frequency range. }

\section{Spin evolution}
\label{sec:spin_evolution}

The tidal torque due to the semidiurnal thermal tide is exerted on the radiative zone, where all of the tidal heating is absorbed. As showed by Section~\ref{sec:properties_waves}, there is a net separation between this zone and the convective interior as regards the direct impact of thermal tides on the dynamical evolution of the fluid. In this Section, we investigate the possibility for thermal tides to generate a dynamical decoupling between the radiative zone and the convective interiors.  
 
As a first step, let us compare the contributions of the tidal thermal and gravitational forcings in the radiative zone. For that, we introduce the ratio between the amplitude of the tidal thermal ($C_{\rm therm}$) and gravitational ($C_{\rm grav}$) components, $ \eta = C_{\rm therm} / C_{\rm grav} $.  In the general case, these two components depend on the complex tidal response of the radiative zone and cannot be simply expressed. However, $ \eta $ can be derived analytically in the zero-frequency limit using the analysis detailed in Appendices~\ref{app:low_frequencies} and \ref{app:multipole_moment}. In this case, the expression of the total multipole moment given by Eq.~(\ref{Qnlm}) shows that $ C_{\rm therm} $ and $ C_{\rm grav} $ are proportional to the two terms of the equilibrium vertical displacement, which reads for a given Hough mode (see Eq.~(\ref{xirn_eq}))

\begin{equation}
\xi_{r ; n}^{\rm (eq)} = \frac{1}{i \sigma} \left( \frac{\sigma_0 U_n}{g} + \frac{g J_n}{N^2 T_0 C_p}  \right).
\label{xirn_eq1}
\end{equation}

To simplify the problem, we consider the heated layer as isothermal, leading to $ N^2 = g^2 / C_{\rm p} T_0 $, set the gravity to $ g = \mathscr{G} M_{\rm p} / R^2 $, and assume $ \tau_\star = \left( p_\star C_{\rm p} \right) / \left( g F_\star \right) $ with $ F_\star = 4 \sigma_{\rm SB} T_0^4 $ \citep[e.g.][]{SG2002,AS2010}. Hence, by substituting in Eq.~(\ref{xirn_eq1}) the expressions of $ U_2 $ and $ J_2 $ given by Eqs.~(\ref{U22}) and (\ref{J22}) and using the third Kepler law $ r_\star^3 n_{\rm orb}^2 = \mathscr{G} \left( M_\star + M_{\rm p} \right) $ to eliminate $ n_{\rm orb} $, we finally obtain 
 
\begin{equation}
\eta \sim \frac{C_{\rm p} T_\star r_\star^3}{\mathscr{G} M_\star R^2} \left( \frac{R_\star}{r_\star} \right)^{\frac{1}{2}},
\label{eta_ratio}
\end{equation}

\noindent where the constant factor has been ignored. 

\begin{figure}
   \centering
   \includegraphics[width=0.49\textwidth,trim = 1.0cm 2.5cm 5.5cm 1.0cm, clip]{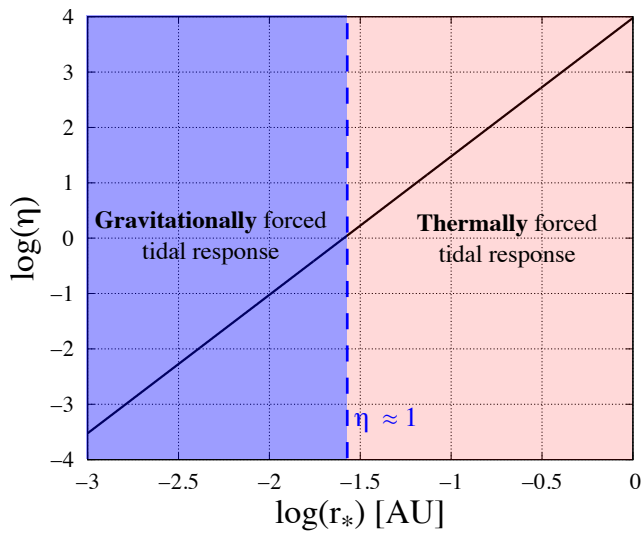}
   \caption{Regions where the tidal response of the radiative layer is dominated by the gravitational (blue area) and thermal (red area) components of the forcing in the zero-frequency limit ($ \sigma \rightarrow 0 $). The logarithm of the ratio $ \eta $ given by Eq.~(\ref{eta_ratio}) (black solid line) is plotted as a function of the star-planet distance $ r_\star $ (AU) in logarithmic scale. The blue dashed line designates the critical distance at which the two components are of the same order of magnitude ($ \eta \approx 1$).  }
   \label{fig:eta_schema}%
\end{figure}

The ratio $ \eta $ is plotted on Fig.~\ref{fig:eta_schema} as a function of the star-planet distance. This figure shows the regions where the tidal response of the radiative layer is due to gravitational ($ \eta \ll 1 $, blue area) and thermal ($ \eta \gg 1 $, red area) forcings. In the case treated here, with the values of Table~\ref{parameters} and $ C_{\rm p} = 7.49 \times 10^3 \ {\rm J.kg^{-1}.K^{-1}} $, $ \eta \sim 1 $ corresponds to $ r_\star \sim 0.03~{\rm AU}$, which means that the gravitational and thermal components are comparable in the zero-frequency limit. The thermal component can nevertheless be greater than the gravitational one in the frequency range of resonant internal gravity waves, where it is increased by two orders of magnitude (see Fig.~\ref{fig:spectre_couple_diss}). Moreover, one should bear in mind that $ \eta $ measures a mean ratio at the base of the heated layer. This ratio varies with the altitude from the base of the stably stratified layer, where gravitational tides dominate, to the high levels of the atmosphere, mainly submitted to the stellar heating. 

We now compare the global rotational evolution of the spin in the convective interior and radiative layer. The convective interior is submitted to the gravitational forcing of the star only, which generates an equilibrium elongation and a dynamical response taking the form of inertial waves \citep[][]{OL2004}. The lag due to the energy dissipated tidally by the convective zone is thus specified by the tidal quality factor $ Q $, and the corresponding tidal torque is given by \citep[e.g.][]{GS1966}

\begin{equation}
\mathcal{T}_{\rm grav} = \frac{3 \mathscr{G} M_\star^2 R^5}{2 r_\star^6 Q},
\end{equation}

\noindent from which the synchronization time scale of the interior $ \tau_{\rm syn}^{\rm CZ} $ is deduced. Ignoring the constant factor, we get \citep[e.g.][]{SG2002}

\begin{equation}
\tau_{\rm syn}^{\rm CZ} \approx  Q \left( \frac{R^3}{\mathscr{G} M_{\rm p}} \right) \left( \frac{M_{\rm p}}{M_\star} \right)^2 \left( \frac{a}{R} \right)^6 \left| \sigma \right|,
\label{tsynCZ}
\end{equation}

Similarly, by considering the radiative layer as a thin shell corresponding to the mass fraction $ f_{\rm p} $ of the planet and ignoring the constant factor, we obtain the global synchronization time scale of the spin due to semidiurnal thermal tides,

\begin{equation}
\tau_{\rm syn}^{\rm RZ} \approx   f_{\rm p} M_{\rm p} R^2 \left| \frac{\sigma}{\mathcal{T}_{\rm SDT} \left( \Omega , \sigma \right)} \right|,
\label{tsynRZ}
\end{equation}

\noindent where $ \mathcal{T}_{\rm SDT} $ is the total tidal torque given by Eq.~(\ref{torque_SDT}). The mass fraction of the radiative layer is determined by the pressure level at which the transition between neutral and stable stratification is located.   Considering the background profiles computed in Section~\ref{ssec:structure_regimes_waves}, we estimate it to $ f_{\rm p} \approx 2 \times 10^{-5} $ in the studied case.

\begin{figure}
   \centering
   \includegraphics[width=0.49\textwidth,trim = 0.5cm 0cm 1.cm 1.cm, clip]{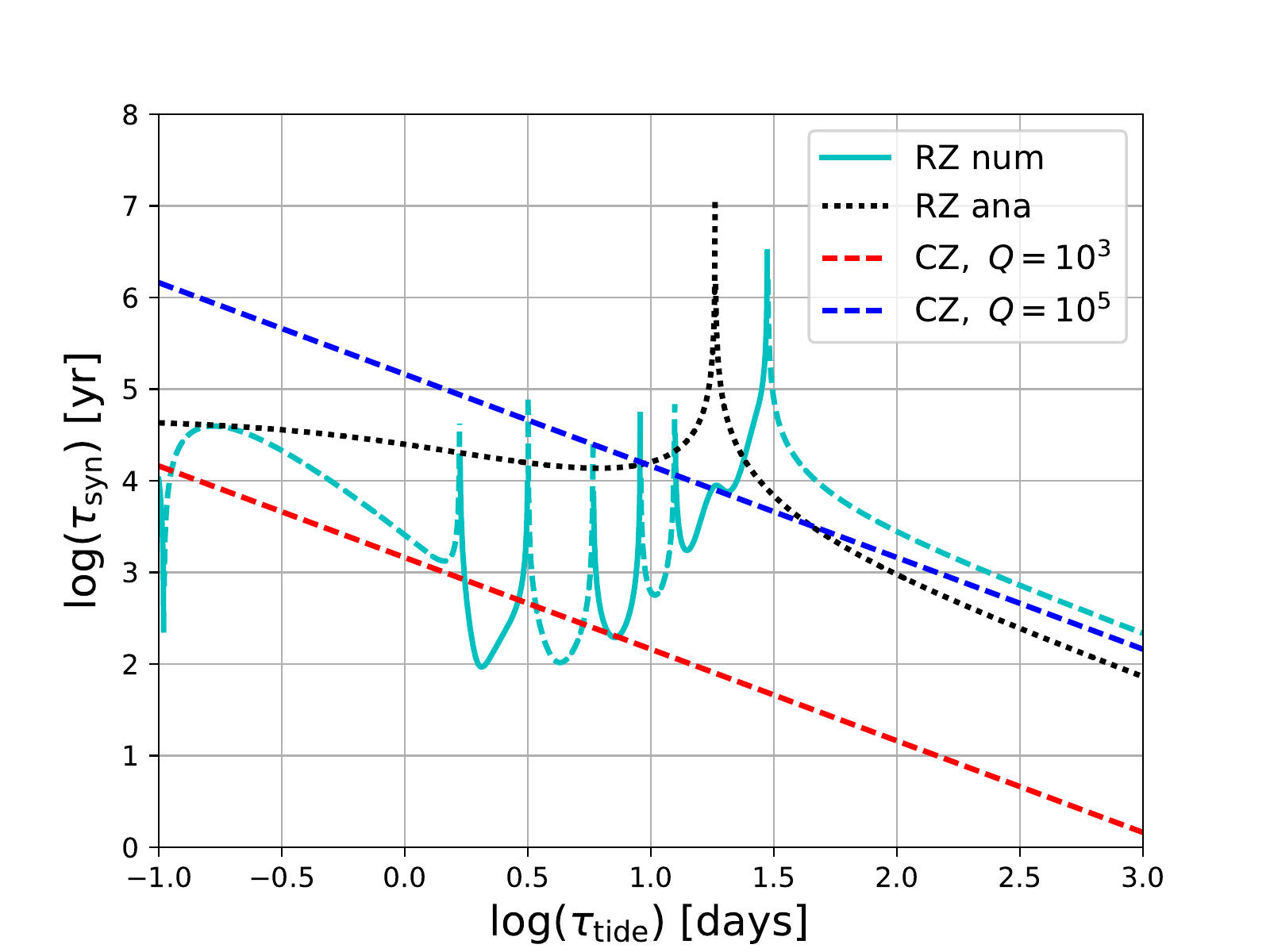}
   \caption{Synchronisation time scale (yr) of the radiative (RZ) and convective (CZ) zones as a function of the tidal period (days) in logarithmic scales. The cyan line designates the synchronization time scale computed numerically with the model of the uniformly rotating spherical shell using Eq.~(\ref{tsynRZ}). This line is solid (dashed) in regions where the radiative layer is torqued away from (toward) synchronization. The dotted black line designates the synchronization time scale resulting from the torque derived analytically in the zero-frequency limit. Red and blue dashed lines correspond to the case of the convective interior with constant tidal quality factors set to $ Q = 10^3 $ and $ Q = 10^5 $ respectively. They are plotted using Eq.~(\ref{tsynCZ}).}
   \label{fig:spectre_tsyn}%
\end{figure}

Hence, using Eqs.~(\ref{tsynCZ}) and (\ref{tsynRZ}) with the values of parameters given by Table~\ref{parameters}, we plot on Fig.~\ref{fig:spectre_tsyn} the synchronization time scale of the radiative and convective zones as a function of the tidal period. The tidal quality factor is not well constrained because it partly results from the dynamical tide within the convective region, which can vary over several orders of magnitude depending both on the properties of internal friction and on resonances associated with the propagation of inertial waves \citep[][]{OL2004}. For Jupiter, the magnitude of $ Q $ inferred from Io's dissipation rate is in the range $ 6 \times 10^{4} < Q < 2 \times 10^{6} $ \citep[][]{YP1981}. \jlc{For Saturn, \cite{Lainey2017} estimate the Love number to $ k_2 = 0.390 \pm 0.24 $ and the ratio $ k_2 / Q $ to $ k_2 / Q =  \left( 1.59 \pm 0.74 \right) \times 10^{-4} $, which gives $ 1 \times 10^{-3} < Q < 5 \times 10^{-3} $.} Therefore, to calculate the synchronization time scale of the convective interior, we employ a simplified constant-$Q$ model, with two different values of $Q $. In the first case, considered as weakly dissipative, the tidal quality factor is set to $ Q = 10^5 $ (blue dashed line). In the second case, which corresponds to a stronger tidal dissipation, $ Q = 10^3 $ (red dashed line). \jlc{Concerning the} radiative layer, we use the torque computed in the non-adiabatic case with rotation and its analytic approximation in the zero-frequency limit (see Fig.~\ref{fig:spectre_couple_diss}, right panel). 

The figure shows that the synchronization time scale of the radiative layer forced by thermal tides is similar to that of the convective interior for $ Q = 10^5 $ (weak tidal dissipation) in the zero-frequency limit. In this case, thermal tides do not induce differential rotation between the radiative and convective regions. Resonances associated with the propagation of internal gravity waves can decrease the synchronization time scale of the radiative layer by two orders of magnitude in the period range 1-30~days. However, their effects on mean flows remain local, the high dissipation rate in the vicinity of a resonance tending to drive the rotation of the layer out of it. The time scale corresponding to a strong tidal dissipation in the radiative zone is of the same order of magnitude than that associated with $ Q = 10^3 $ in the convective interior. 

\section{Discussion}
\label{sec:discussion}


The linear analysis shows that semidiurnal thermal tides are able to generate strong asynchronous zonal flows in the radiative zone, where all of the tidal heating is absorbed. Because of the propagation of internal waves, the fluid tidal response can be enhanced by several orders of magnitude. This reinforces the conclusions of \cite{AS2010} for the semidiurnal tide, and of \cite{GO2009} who demonstrated that the diurnal thermal tide could drive the upper layers of the atmosphere into asynchronous rotation by transporting angular momentum upward. Moreover, it emphasizes the necessity to consider the possibility of tidally-driven asynchronous zonal flows in the modeling of the general circulation of hot Jupiters \jlc{beyond a critical orbital radius, which is approximately $ r_\star \approx 0.03 $~AU in the treated case. Thermal tides can be difficult to take into account with GCMs because \padc{of two reasons. First,} they are negligible at time scales characterizing the global flows. \padc{Second, the atmospheric tidal response depends at the inner boundary condition on the distortion of the convective enveloppe. Thus, to compute the tidal atmospheric tidal response properly, one should know how the system of coordinates is modified by the gravitationally excited distortion of the convective interior.} However, owing to their ability to drive the radiative region into asynchronous rotation, \padc{the effects of thermal tides} on the general circulation should be included in models when the thermally excited component of the tidal response is greater than the gravitationally excited component.} 

In the present work, it emerges that dissipative processes play a key role in the fluid tidal response. First, they affect its magnitude, particularly for tidal periods $ \tau_{\rm tide} \gtrsim \tau_\star $. Second, they regularize the resonant behaviour obtained by \cite{AS2010} in the vicinity of synchronous rotation, which is due to the reflection of internal gravity waves on the boundaries of the stably stratified layer \citep[compare Fig.~\ref{fig:spectre_couple_diss} with Fig~4 of][]{AS2010}. The radiative cooling drains away the energy injected by the tidal heating. In that sense, it acts similarly \jlc{to} the radiation condition used by \cite{AS2010}, which allows the energy to propagate upwards at the upper boundary. We thus retrieve qualitatively the spectrum plotted on Fig.~5 of their study in the static approximation. The effect of rotation on the tidal response is strong when $ \left| 2 \Omega \right| \gg \left| \sigma \right| $. It induces the excitation of a large number of modes behaving in a frequency-resonant way. For this reason, the tidal response would be much more altered by rotation in the range $ 0.1 \ {\rm days} < \tau_{\rm tide} < 1 \ {\rm day}  $ in the case of super-synchronous rotation than in the case of sub-synchronous rotation treated in this study. 

We shall discuss here the two approximations that we \jlc{make} concerning rotation to simplify calculations. As regards background flows, we consider that the planet rotates uniformly at the angular frequency $ \Omega $. Thus we ignore the coupling between tidal waves and mean flows. Nevertheless, this coupling could significantly modify the tidal response of the radiative region. Indeed, taking into account winds introduces an advection term in the momentum equation. This term is weighted by the Rossby number, i.e. the parameter measuring the departure of the flow to solid rotation. Hence, in the regime of high Rossby numbers, the Coriolis acceleration is dominated by wind-driven advective accelerations, which implies a new tidal regime in Fig.~\ref{fig:spectre_regimes}. The effect of a vertical shear on internal waves has been studied in the framework of the Earth atmospheric tides \citep[e.g.][]{Chiu1952,BB1967} and oceanic turbulence \citep[e.g.][]{Worthem1983}. Beyond a critical value of the local Richardson number $ R_{\rm i} = N^2 \left| d \textbf{V}_0 / dr \right|^{-2} $ ($ \textbf{V}_0  $ standing for the velocity vector of the wind), internal gravity are attenuated by the vertical shear at the level at which their horizontal phase speed is equal to the zonal velocity $ V_{\varphi ; 0} $ \citep[][]{BB1967}. As the horizontal phase speed scales as $ v_{\varphi} \sim  \sigma R $ in the radiative zone, we expect that the vertical shear plays an important role in the low-frequency range. Typically, for $ R = 1.27 \, R_{\rm J} $ and $ V_{\varphi ; 0} = 1 \ {\rm km.s^{-1}}  $, the frequency at which $ v_{\varphi} \sim V_{\varphi , 0} $ is $ \sigma = 1.1 \times 10^{-5} \ {\rm s^{-1}}  $, which corresponds to the tidal period $ \tau_{\rm tide} = 6.5 $ days. 

The second most important simplification as regards rotation is the traditional approximation \citep[e.g.][]{Unno1989}. In this approximation, the horizontal component of the rotation vector, $ - \Omega \sin \theta \, \textbf{e}_\theta $, is neglected (see Eqs.~(\ref{eq1}) to (\ref{eq3})). Physically, this means that the we ignore the radial component of the Coriolis force and the horizontal component associated with radial motions. Therefore, the traditional approximation is appropriate in the case where the fluid displacement is dominated by horizontal motions. As the Archimedean force acts as a restoring force on fluid particles in the vertical direction, a sufficiently strong stable stratification limits vertical motions. The \jlc{buoyancy} thus dominates the radial component of the Coriolis force, and makes the horizontal component associated with radial motions negligible compared to that associated with horizontal motions. Given that the strengths of the Coriolis and Archimedean forces are measured by the inertia and Brunt-Väisälä frequencies respectively, this condition is mathematically expressed by the hierarchy of frequencies $ \left| 2 \Omega \right| \ll N $ \citep[e.g.][]{ADLM2017a}. The validity of the traditional approximation as regards the modeling of internal waves has been examined for various media, such as oceanic layers \citep[e.g.][]{GS2005,GZ2008}, planetary atmospheres and envelopes \citep[e.g.][]{OL2004,Tort2014}, and stellar interiors \citep[e.g.][]{Savonije1995,Mathis2008,Mathis2009,Prat2017}. The traditional approximation prevents us to extend the analytic approach used in this work to weakly of neutrally stratified layers, where radial and horizontal motions are of the same order of magnitude. In such cases, it is necessary to keep all of the components of the Coriolis force. \jlc{This} requires to use numerical or semi-analytical methods \jlc{like the Chebyshev pseudospectral method used by \cite{OL2004}.}

\section{Conclusions}
\label{sec:conclusions}

Motivated by the understanding of the effect of stellar semidiurnal thermal tides on the general circulation of hot Jupiters, we revisited the early study by \cite{AS2010} by including the effects of radiative cooling and rotation. We derived the equations describing the tidal response of a radially stratified fluid planet in the framework of the traditional approximation, where the horizontal and vertical structure of waves can be solved separately. For the background structure of the planet, we used the equation of state proposed by \cite{AS2010}, which mimics a bi-layer planet composed of a central neutrally stratified region and a thin superficial stably-stratified radiative zone. As regards the radiative cooling, we applied the prescription given by \cite{Iro2005} to set the scaling law of the thermal time in the heated layer. We thus derived the equations of tidal waves forced both by the stellar incoming flux and the tidal gravitational potential of the star, as well as the associated tidal torque and time scale of evolution of tidally forced mean zonal flows. We then solved these equations numerically for the case treated by \cite{AS2010} in three configurations: (a) in the static and adiabatic approximations \citep[][]{AS2010}, (b) in the static approximation with radiative cooling, (c) with rotation and radiative cooling. In each case, we computed both the internal structure of the tidal response due to the thermal forcing and the total tidal torque exerted on the planet. 

\jlc{The rotation strongly modifies the structure of the tidal waves by coupling the forcing with Hough modes of different wavenumbers. The better coupled modes can change depending on the frequency. In the super-inertial regime ($ \left| \nu \right| \ll 1 $), the main contributor is the gravity mode of degree $ n = 0 $. In the sub-inertial regime ($ \left| \nu \right| \gg 1 $), the structure of the tidal perturbation is shaped by a large number of Rossby modes.} In the framework of the static and adiabatic approximations we recover the results obtained by \cite{AS2010}. The absence of dissipation allows gravito-acoustic waves to propagate over the  whole thickness of the radiative zone and to be reflected by its boundaries, which amplify the tidal response and allows a strongly oscillating behaviour to arise in the vicinity of spin-orbit synchronous rotation. Introducing the radiative cooling leads to a regularized response where such oscillations do not exist any more. In this case, the response is characterized by large scale patterns tending to generate zonal flows of alternate directions. However, the tidal response in the zero frequency limit remains the same as in the adiabatic case. It does not depend on the thermal time of the radiative zone. 

\jlc{In the zero-frequency limit, the obtained frequency spectrum of the tidal torque reaches a saturation plateau when rotation in included. It thus qualitatively differs from the static case treated by \cite{AS2010}. In their case, the torque follows the scaling law $ \mathcal{T}_{\rm SDT} \propto \sigma $, \jlc{which corresponds to the frequency behaviour described by the constant time lag model}. Both behaviours are well approximated by the analytic formula derived for the equilibrium tide in this work, which generalizes that given by \cite{AS2010} to non-adiabatic and rotating atmospheres. Particularly, this formula allows us to identify the amplification factors at the origin of the saturation in the rotating case.} \rec{These amplification factors are due to the absence of friction in the model, which allows rotation to couple the perturbation with a large number of Rossby modes. In reality, friction would prevent this coupling in the low-frequency range and drive the perturbation toward a quadrupolar pattern, similarly as the forcing.} \jlc{As observed by \cite{AS2010},} the obtained spectra of the total tidal torque exhibit resonances corresponding to gravito-acoustic waves in the high-frequency range ($ \tau_{\rm tide} \lesssim 1 $~day) and to gravity waves in the medium-frequency range ($ 1 \lesssim  \tau_{\rm tide} \lesssim 30 $~days).

As the sign of the tidal torque varies with the tidal frequency, the semidiurnal tide is likely to drive strong asynchronous zonal flows over time scales comparable to the year in order of magnitude. This impact results from both the weak density of the radiative zone and the fact that it absorbs all of the incoming stellar flux. \jlc{It} highlights the fact that the dynamics of the radiative zone are far more sensitive to stellar thermal tides than the central convective region. \jlc{However, the time scales associated with the growth of tidally forced jets remains larger than those characterizing the general circulation in the heated zone. In the zero-frequency limit, the thermal tidal response of the radiative layer dominate the gravitational component beyond a critical orbital radius, estimated to $ r_\star \sim 0.03 $~AU in this work. The associated synchronization time scale is of the same order as that of the convective interior if this later has a Jupiter-like tidal dissipation rate, $ Q = 10^5 $. It can be decreased by several orders of magnitude by resonances due to internal gravito-inertial waves in the period range 1-30 days.}

A subject of interest for forthcoming works would be to investigate how winds affect the tidal response. The intense stellar heating to which hot Jupiters are submitted by their host star force strong zonal jets inducing an important departure to the idealized solid rotation approximation. Therefore, the interplay between the general circulation and tidal waves appears as a key question in the understanding of the effect of tides on the dynamical evolution of hot Jupiters.

\begin{acknowledgements}
      The authors acknowledge funding by the European Research Council through the ERC grant WHIPLASH 679030. \rec{They wish to thank the anonymous referee for helpful suggestions and remarks.} P. Auclair-Desrotour acknowledges Anthony Caldas for his help with the use of Python plotting tools. 
   \end{acknowledgements}

\bibliographystyle{aa} 
\bibliography{ADL2017a} 


\appendix

\section{Polarization relations of the tidal response}
\label{app:polarization_relations}

The polarization relations of the tidal response (Section~\ref{ssec:structure_regimes_waves}) are computed by substituting $ \Psi_n $ in primitive equations. \rec{Omitting the superscripts $m$ and $ \sigma $,} we obtain, for latitudinal motions,

\begin{align}
& \xi_{\theta ; n} = \frac{1}{\sigma^2 r} \left( \Phi_n \Psi_n - U_n \right), \\
& V_{\theta ; n} = \frac{i}{\sigma r} \left( \Phi_n \Psi_n - U_n \right),
\end{align}

\noindent for longitudinal motions,

\begin{align}
& \xi_{\varphi ; n} = \frac{i}{\sigma^2 r} \left( \Phi_n \Psi_n - U_n \right), \\
& V_{\varphi ; n} = - \frac{1}{\sigma r} \left( \Phi_n \Psi_n - U_n \right),
\end{align}

\noindent for vertical motions,

\begin{align}
 \xi_{r ; n} = & - \frac{1}{H} \left( \frac{\sigma}{\sigma - i \sigma_0} N^2 - \sigma^2 \right)^{-1} \left[ \Phi_n \left( \frac{d \Psi_n}{dx} + \mathcal{A}_n \Psi_n \right) - \frac{d U_n}{dx} \right. \nonumber \\
  & \left. + i \frac{\kappa}{\sigma - i \sigma_0} J_n  \right],
\end{align}

\begin{align}
 V_{r ; n} = & - \frac{i \sigma}{H} \left( \frac{\sigma}{\sigma - i \sigma_0} N^2 - \sigma^2 \right)^{-1} \left[ \Phi_n \left( \frac{d \Psi_n}{dx} + \mathcal{A}_n \Psi_n \right) - \frac{d U_n}{dx} \right. \nonumber \\
  & \left. + i \frac{\kappa}{\sigma - i \sigma_0} J_n  \right],
\end{align}

\noindent and for scalar quantities,

\begin{equation}
\delta p_n  = \rho_0 \Phi_n \Psi_n,
\end{equation}

\begin{align}
\delta \rho_n = & - \frac{\rho_0}{gH} \frac{N^2}{N^2 - \sigma \left( \sigma - i \sigma_0 \right)} \left[ \Phi_n \left( \frac{d \Psi_n}{dx} + \mathcal{B}_n \Psi_n \right)   \right. \\
 & \left. - \frac{d U_n}{dx} + i \frac{\kappa \sigma}{N^2} J_n \right], \nonumber
\end{align}

\begin{equation}
\delta T_n =  \frac{N^2}{N^2 - \sigma \left( \sigma - i \sigma_0 \right)} \left[ \Phi_n  \left( \frac{d \Psi_n}{dx} + \mathcal{C}_n \Psi_n \right)  - \frac{d U_n}{dx} + i \frac{\kappa \sigma}{N^2} J_n \right] ,
\end{equation}

\noindent where we have introduced the coefficients

\begin{equation}
 \mathcal{A}_n \left( x \right) =   \frac{1}{2} \left( \frac{1}{\Gamma_1} \frac{\sigma - i \, \Gamma_1 \sigma_0}{\sigma - i \sigma_0}  - \frac{\sigma }{\sigma - i \sigma_0} \frac{H N^2 }{g} + K_\circ \right),
\end{equation}

\begin{align}
 \mathcal{B}_n \left( x \right) = & \frac{1}{2} \left[ \frac{1}{\Gamma_1} \frac{\sigma - i \, \Gamma_1 \sigma_0}{\sigma - i \sigma_0} \left( 2 \frac{\sigma \left( \sigma - i \sigma_0 \right)}{N^2} - 1 \right) \right. \\ 
  & \left. - \frac{\sigma}{\sigma - i \sigma_0} \frac{H N^2}{g} + K_\circ \right], \nonumber
\end{align}

\begin{equation}
\mathcal{C}_n \left( x \right) = \mathcal{B}_n + 1 - \frac{\sigma \left( \sigma - i \sigma_0 \right)}{N^2}.
\end{equation}

\section{Impact of the lower boundary condition on the solution}
\label{app:lower_boundary}

\rec{As discussed in Section~\ref{ssec:BC}, the lower limit of the fluid region where the tidal perturbation is computed can affect the obtained solution if it is set to close to the base of the stably-stratified zone. We show here an example of the kind of artefacts that can occur in these cases. In plots displayed by Fig.~\ref{fig:structure_ondes}, the lower limit is set as far as possible from the perturbed region, that is at the center of the planet. This allows us to avoid artificial interactions of the perturbation with the lower boundary. In the present case, we set the lower limit of the fluid region at the pressure level $ p_0 = 10^4 $~bar, much more closer to the basis of the stably-stratified zone (the region now stands for 6\% of the planet radius), and do the calculations again for two cases treated in Fig.~\ref{fig:structure_ondes}: (i) $ \tau_{\rm tide} = 0.1 $~days and (ii) $ \tau_{\rm tide} = 10 $~days, with rotation and radiative cooling. The parameters used for this study are those given by Table~\ref{parameters}. The lower boundary condition is a rigid-wall condition, as defined in Section~\ref{ssec:BC}.}

\rec{In case (ii), we exactly obtain the same solution as in Fig.~\ref{fig:structure_ondes}. Tidal density variations are not affected by the lower boundary. We cannot observe any artefact. The other case is different. In case (i), we clearly note that the solution exhibit a new pattern with respect the the plot of Fig.~\ref{fig:structure_ondes}. This pattern corresponds to an artefact resulting from the interactions of the tidal perturbation with the lower boundary of the fluid region. It highlight the way this boundary can affects the solution and justifies that we set the lower limit of the fluid region ``at the infinite''. }

\begin{figure}[htb]
   \centering
   \includegraphics[width=0.24\textwidth,trim = 0.5cm 0cm 1.cm 1.cm, clip]{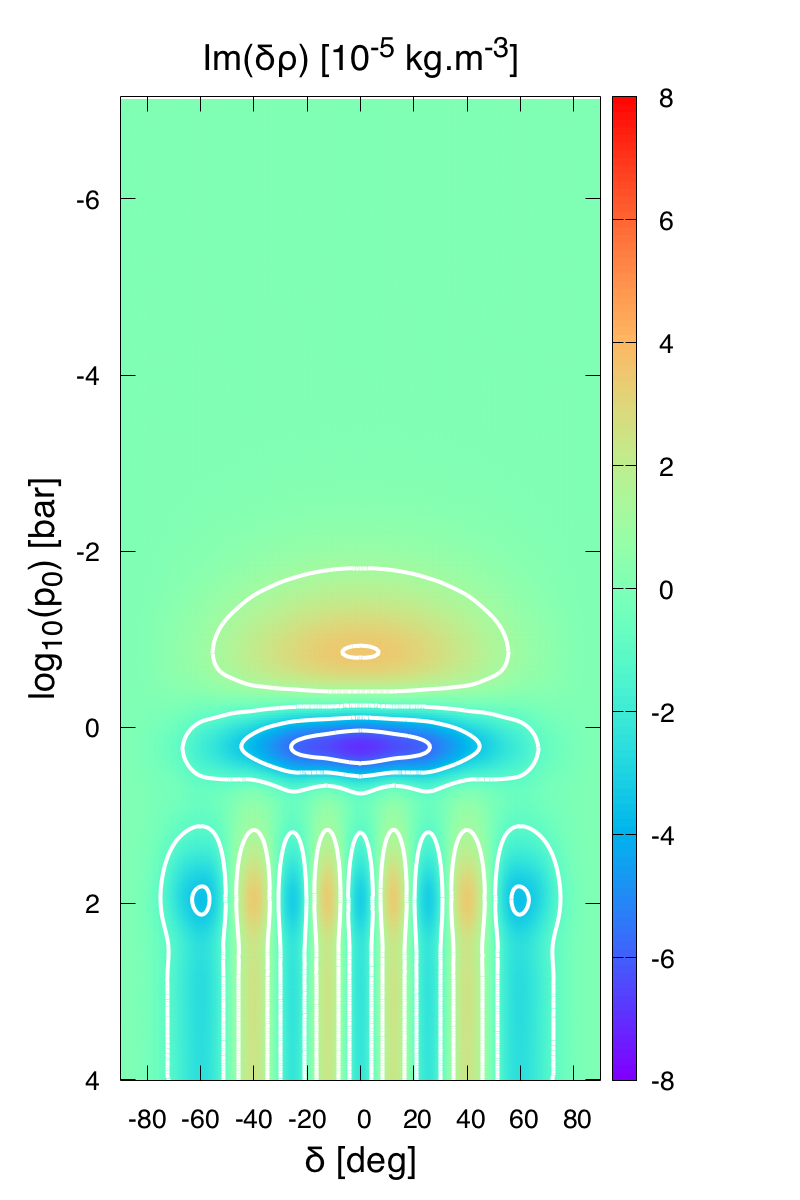}
   \includegraphics[width=0.24\textwidth,trim = 0.5cm 0cm 1.cm 1.cm, clip]{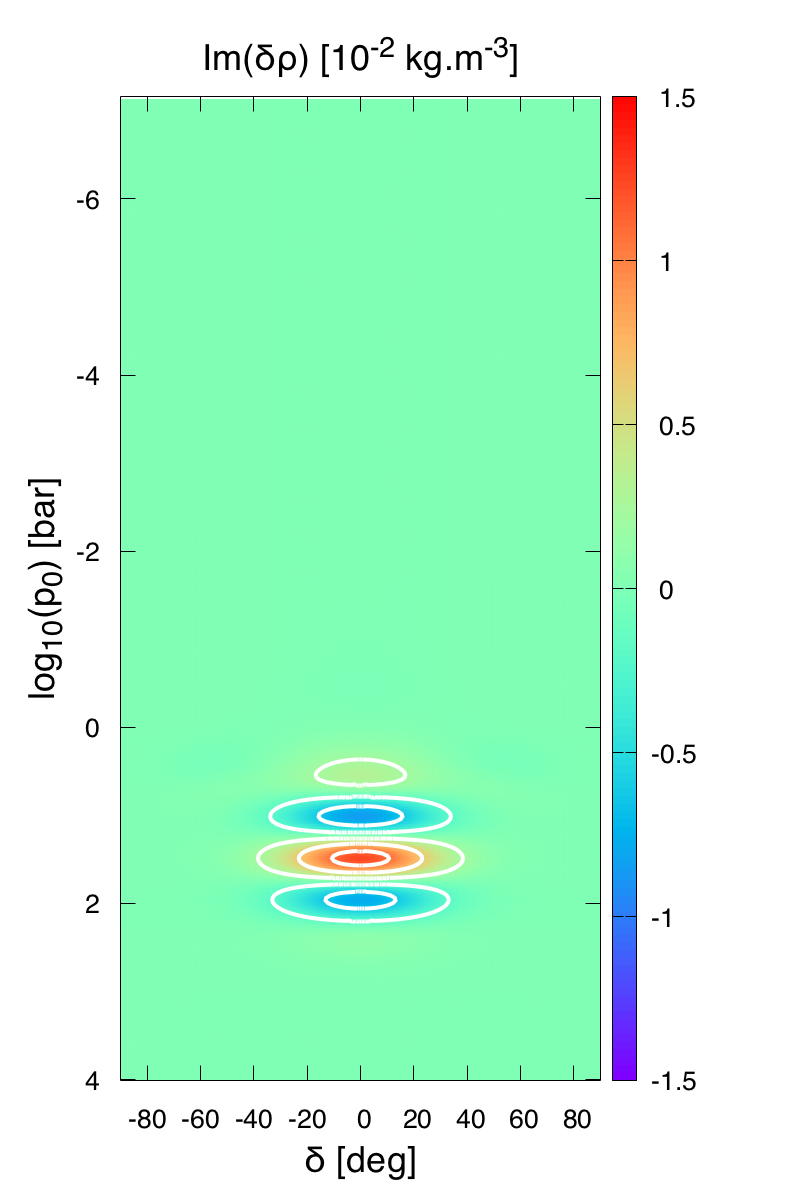}
   \caption{\rec{Imaginary part of density fluctuations in the case treated in Fig.~\ref{fig:structure_ondes} for a lower boundary located at the pressure level $ p_0 = 10^4 $~bar instead of the center of the planet as in Fig.~\ref{fig:structure_ondes}. The density fluctuations are plotted as a function of latitude (degrees, horizontal axis) and pressure in logarithmic scale (bars, vertical axis) in the case with rotation and radiative cooling and for two different tidal periods, (i) $ \tau_{\rm tide} = 0.1 $~day (left) and (ii) $ \tau_{\rm tide} = 10 $~day (right). The corresponding rotation rates are given by $ \Omega = \sigma/2 + n_{\rm orb} $ with $ \left| \sigma \right| = 2 \pi / \tau_{\rm tide} $ and $ n_{\rm orb} = 2 \pi / \tau_{\rm orb}$ (see Table~\ref{parameters}). }}
   \label{fig:side_effects}%
\end{figure}

\section{Low frequencies asymptotic regime}
\label{app:low_frequencies}

We derive here the asymptotic tidal response of the radiative zone in the low frequency regime, which corresponds to $ \sigma \ll \sigma_0 , 2 \Omega , N, \sigma_{\rm s} $ (see Fig.~\ref{fig:spectre_regimes}). Following \cite{AS2010}, we first establish the zero-frequency limit, the so-called \emph{equilibrium tide}, and then a finite frequency correction to obtain the trend in the vicinity of spin-orbit synchronous rotation. Our goal is study the role played by rotation and radiative cooling in this regime and to explain the results obtained numerically in Sections~\ref{sec:properties_waves} and \ref{sec:spectra_torque}. 

\subsection{The equilibrium tide ($ \sigma \rightarrow 0 $)}

The equilibrium tide is defined as the zero-frequency tidal response. We thus consider the primitive equations given by Eqs.~(\ref{eq1}) to (\ref{eq5}) and let $ \sigma $ tend to zero. We begin with the equilibrium displacement. In the static case, the displacement associated with the $ \left( l , m, \sigma \right)$-mode can be expanded in the poloidal harmonics \citep[e.g.][]{AS2010}

\begin{equation}
\boldsymbol{\xi}_{l}^{m,\sigma} = \left[ \xi_{r ; l}^{m,\sigma} \left( r \right) \textbf{e}_r + \xi_{\perp ; l}^{m,\sigma} \left( r \right) r \boldsymbol{\nabla}_\perp \right] P_l^m \left( \cos \theta \right) \ed^{i \left( \sigma t + m \varphi \right)}.
\label{xi_poloidal}
\end{equation}

\noindent where $ \boldsymbol{\xi}_{\perp ; l}^{m,\sigma} $ designates the vertical profile of the horizontal displacement and $ \boldsymbol{\nabla}_\perp =  \left( 1 / r \right) \partial_\theta \textbf{e}_\theta + 1 / \left( r \sin \theta \right) \partial_\varphi \textbf{e}_\varphi $ the horizontal gradient operator in spherical coordinates. When the rotation is introduced (using the traditional approximation), the solution is a series of Hough modes identified by the subscript $ n $. The previous expression thus generalizes into

\begin{equation}
\boldsymbol{\xi}_{n}^{m,\sigma} = \left[ \xi_{r ; n}^{m,\sigma} \left( r \right) + \xi_{\perp ; n}^{m,\sigma} \left( r \right) \boldsymbol{\mathcal{L}}_\perp^{m,\nu} \right] \Theta_n^{m,\nu} \left( \theta \right) \ed^{i \left( \sigma t + m \varphi \right)},
\end{equation}

\noindent the notation $ \boldsymbol{\mathcal{L}}_\perp^{m,\nu} $ referring to the horizontal operator

\begin{align}
\boldsymbol{\mathcal{L}}_\perp^{m,\nu} = & \frac{1}{1 - \nu^2 \cos^2 \theta} \left[  \left( \frac{d}{d \theta} + m \nu \cot \theta \right) \textbf{e}_\theta \right.  \\
   & \left. + \left( \nu \cos \theta \frac{d}{d \theta} + \frac{m}{\sin \theta} \right) \textbf{e}_\varphi   \right]. \nonumber
\end{align}

\noindent In the static case, $ \boldsymbol{\mathcal{L}}_\perp^{m,0} = r \boldsymbol{\nabla}_\perp $ and we recover the expansion in poloidal harmonics given by Eq.~(\ref{xi_poloidal}). In the following, we will omit the superscript $ \left( m , \sigma \right) $ in order to lighten equations.

With the notations introduced above, the equation of conservation of the horizontal momentum is expressed as 

\begin{equation}
\sigma^2 \xi_{\perp ; n} = \frac{1}{r} \left( \frac{\delta p_n}{\rho_0} - U_n \right). 
\label{horizontal_momentum}
\end{equation}

\noindent As a consequence, the equilibrium pressure fluctuation writes

\begin{equation}
\delta p_n^{\rm (eq)} = \rho_0 U_n. 
\label{deltapn_eq}
\end{equation}

\noindent Substituting Eq.~(\ref{deltapn_eq}) in the equation of conservation of the vertical momentum, 

\begin{equation}
- \sigma^2 \xi_{r ; n} = - \frac{1}{\rho_0} \frac{d \delta p_n}{dr} - \frac{g}{\rho_0} \delta \rho_n + \frac{d U_n}{dr}, 
\end{equation}

\noindent we get the density fluctuation associated with the equilibrium tide,

\begin{equation}
\delta \rho_n^{\left( eq \right) } = - \frac{d \rho_0}{dr} \frac{U_n}{g}. 
\label{deltaqn_eq}
\end{equation}

\noindent We then consider the equation of heat transport, 

\begin{equation}
\frac{\delta \rho_n}{\rho_0} \left( 1 + \frac{\sigma_0}{i \sigma} \right) = \frac{\delta p_n}{\rho_0 c_{\rm s}^2} \left( 1 + \frac{\sigma_0}{i \sigma} \right) + \frac{N^2}{g} \xi_{r ; n} - \frac{J_n}{i \sigma T_0 C_{\rm p}},
\end{equation}

\noindent where we substitute the pressure and density fluctuations associated with the equilibrium tide, Eqs.~(\ref{deltapn_eq}) and (\ref{deltaqn_eq}). It follows that 

\begin{equation}
N^2 \xi_{r ; n}^{\rm (eq)} = \frac{N^2}{g} \frac{\sigma_0}{i \sigma} U_n + \frac{ gJ_n}{i \sigma T_0 C_{\rm p}}. 
\end{equation}

As discussed by \cite{AS2010}, we note that no solution exist in the central neutrally stratified region, since $ N^2 \rightarrow 0 $ while the thermal forcing is nonzero. This results from the vanishing of the Archimedean force, which is the only restoring force of the system in the vertical direction. In the neutrally stratified region, fluid particles can move freely along the vertical direction. As a consequence, no equilibrium can be reached when a forcing is applied. We shall also highlight here the fact that the traditional approximation is not valid in neutrally stratified region in the zero-frequency limit since $ 2 \Omega \gtrsim \sigma, N $.

In the stably-stratified radiative zone however, the vertical displacement associated with the equilibrium tide is expressed as 

\begin{equation}
\xi_{r ; n}^{\rm (eq)} =  \frac{\sigma_0 }{i \sigma} \frac{U_n}{g} + \frac{g}{N^2} \frac{J_n}{i \sigma T_0 C_{\rm p}}. 
\label{xirn_eq}
\end{equation}

\noindent By substituting Eqs.~(\ref{deltapn_eq}), (\ref{deltaqn_eq}) and (\ref{xirn_eq}) in the equation of mass conservation,

\begin{equation}
\frac{\delta \rho_n}{\rho_0} = - \frac{1}{r^2 \rho_0} \frac{d}{dr} \left( r^2 \rho_0 \xi_{r ; n} \right) + \frac{\Lambda_n}{r} \xi_{\perp ; n},
\end{equation}

\noindent we finally get the horizontal displacement,

\begin{equation}
r \rho_0 \xi_{\perp ; n}^{\rm (eq)} = \frac{1}{\Lambda_n} \left[ \frac{d}{dr} \left( r^2 \rho_0 \xi_{r ; n}^{\rm (eq )} \right) + r^2 \delta \rho_n^{\rm (eq)} \right]. 
\label{xiperp}
\end{equation}

Hence, the pressure and density fluctuations associated with the equilibrium tide are determined by the tidal gravitational potential only while the displacement results from both gravitational and thermal forcings. Looking at Eq.~(\ref{xirn_eq}), we note that the radiative cooling only affects the gravitationally forced tidal component. This explains why the tidal torques computed for different thermal times all converge toward the same asymptotic law at $ \tau_{\rm tide} \rightarrow + \infty $. The role played by rotation is of greater complexity. It is first expressed through the eigenvalues associated with Hough modes ($\Lambda_n$), which take very different values from those of gravity modes in the static case. Typically, for $ \left| \nu \right| \gg 1 $, \jlc{predominant} symmetric ($ n $ even) Rossby modes coupled with the quadrupolar forcing ($ m = 2 $) are characterized by $  \left| \Lambda_n^{2,\nu} \right| \ll \Lambda_0^{2,0} = 6 $. Second, the rotation modifies the coupling coefficients weighting Hough modes (the $A_{n,l}^{m,\nu} $ introduced in Eq.~(\ref{Thetan_Pln})) in a non trivial way. Thus, predominant Hough modes can change depending on $\nu$. In the super-inertial regime ($ \left| \nu \right| < 1 $) the predominant mode is the gravity mode of index $ n = 0 $. For  $ \nu = 10 $, it is the Rossby mode of index $ n = -2 $. 


\subsection{Finite frequency correction}

We now consider the first order correction in $\sigma^2 $, in order to derive scaling laws describing the behaviour of the density fluctuation as a function of $ \sigma$ in the asymptotic regime of low tidal frequencies. First, by identifying $\delta p_n^{\rm (eq)} $ (Eq.~(\ref{deltapn_eq})) in Eq.~(\ref{horizontal_momentum}), we get

\begin{equation}
\delta p_n = \delta p_n^{\rm (eq)} + \sigma^2 r \rho_0 \xi_{\perp ; n}^{\rm (eq)}. 
\end{equation}

\noindent We then substitute this equation in the vertical momentum equation to obtain the density fluctuation,

\begin{equation}
\delta \rho_n = \delta \rho_n^{\rm (eq)} + \frac{\sigma^2}{g} \left[ \rho_0 \xi_{r ; n}^{\rm (eq)} - \frac{d}{dr} \left( r \rho_0 \xi_{\perp ; n}^{\rm (eq) } \right) \right]. 
\end{equation}

\noindent Finally, by using the expression of the horizontal displacement associated with the equilibrium tide (Eq.~(\ref{xiperp})) in this equation and ignoring the gravitational forcing, we end up with the expression of $\delta \rho_n $ as a function of $\xi_{r ; n}^{\rm (eq)} $,

\begin{equation}
\delta \rho_n = \frac{\sigma^2}{g} \left[ \rho_0 \xi_{r ; n}^{\rm (eq)} - \frac{1}{\Lambda_n} \frac{d^2}{dr^2} \left( r^2 \rho_0 \xi_{r ; n}^{\rm (eq)} \right) \right].  
\end{equation}

\section{Tidal multipole moment}
\label{app:multipole_moment}

The tidal multipole moment $ Q_l^{m,\sigma} $ introduced in Eq.~(\ref{Qlmsigma}) stands for the global variation of mass distribution associated with the $ \left( l , m , \sigma \right) $-mode. It quantifies the energy tidally dissipated by the mode and is defined by \citep[e.g.][]{AS2010}

\begin{equation}
Q_l^{m,\sigma} = \int_0^{R_{\rm e}} r^2 \delta \rho_l^{m,\sigma} \dd r
\label{Qlm_def}
\end{equation}

\noindent where $ \delta \rho_l^{m,\sigma} $ designates the projection of the distribution of density fluctuations on the $ \left( l , m \right) $-harmonic, \rec{and $R_{\rm e} $ the radius of the upper boundary introduced in Section~\ref{subsec:setup}}. The tidal multipole moment is thus the sum of the contributions of Hough modes, denoted $ Q_{n,l}^{m,\sigma} $, weighted by the projections coefficients of Hough functions on the associated Legendre polynomials, i.e.

\begin{equation}
Q_l^{m,\sigma} = \sum_n A_{n,l}^{m,\nu} Q_{n,l}^{m,\sigma}
\end{equation}

To improve the convergence of the numerical integration, \cite{AS2010} separate the equilibrium and the dynamical tide \jlc{\citep[see][Appendix~A]{AS2010}}. Following their method, we substitute $ \delta \rho_l^{m,\sigma} $ in Eq.~(\ref{Qlm_def}) using Eq.~(\ref{eq3}). After two integrations by parts, and omitting the superscripts $ \left( m , \sigma \right) $, we end up with

\begin{align}
Q_{n,l}^{m,\sigma} = & \int_0^{R_{\rm e}} r^{2+l} \delta \rho_n^{\rm (eq)} dr +  \sigma^2 \left[ \frac{r^2}{\Lambda_n} \frac{d}{dr} \left( \frac{r^{2+l}}{g} \right)  \rho_0 \xi_{r ; n}  - \frac{r^{3+l}}{g} \rho_0 \xi_{\perp ; n} \right]_0^{R_{\rm e}} \nonumber \\
&  + \sigma^2 \int_0^{R_{\rm e}}  \left\{ \rho_0 \xi_{r ; n} \left[ \frac{r^{2+l}}{g} - \frac{r^2}{\Lambda_n}  \frac{d^2}{dr^2} \left( \frac{r^{2+l}}{g} \right)  \right] \right. \nonumber \\
  & \left. + \frac{r^2}{\Lambda_n} \frac{d}{dr} \left( \frac{r^{2+l}}{g} \right) \delta \rho_n \right\} dr.
  \label{Qnlm}
\end{align}

\noindent where $\xi_{\perp ; n} $ and $\delta \rho_n^{\rm (eq)} $ are defined by Eqs.~(\ref{horizontal_momentum}) and (\ref{deltaqn_eq}) respectively. We recover here the multipole moment given by \cite{AS2010} modified by rotation (the $ l \left( l + 1 \right) $ of Legendre polynomials have been replaced by the associated eigenvalues of Hough modes, $ \Lambda_n $). In Eq.~(\ref{Qnlm}), we have conserved the boundary term resulting from the integrations by part (second term of the right member). \cite{AS2010} ignore this term by arguing that it is negligible with respect to the others. They thus compute the quadrupole moment associated with \rec{the $ l = m = 2 $} tidal component by using a simplified expression and find that it leads to a better convergence than integrating the density variations directly. \jlc{However, we notice in our calculations that the boundary term is negligible in the zero-frequency limit and in the non-adiabatic case. In the adiabatic case, tidal waves can be reflected backward at the upper boundary, leading to a non-negligible boundary term. We verify that we obtain numerically the same results by using either Eq.~(\ref{Qlm_def}) or Eq.~(\ref{Qnlm}), which suggests that writing the tidal quadrupole moment in the way of Eq.~(\ref{Qnlm}) does not improve the precision of the computation.}

In the zero-frequency limit ($\sigma \rightarrow 0$), the above expression can be simplified using the equations of the equilibrium thermal tide ($ U_n = 0 $) established in Appendix~\ref{app:low_frequencies}. First, let us note that $ \delta \rho_n^{\rm (eq)} = 0 $, $ \xi_{r ; n} \sim \xi_{r ; n}^{\rm (eq)} $ and $ \delta \rho_n = O \left( \sigma^2 \right) $ in this regime, which means that the term in $\delta \rho_n $ in Eq.~(\ref{Qnlm}) tends to become negligible compared to that in $ \xi_{r ; n} $ while $ \sigma \rightarrow 0$. Moreover, as tidal waves cannot propagate in the central convective region, the perturbation at the lower boundary can be neglected. It follows that 

\begin{align}
 \label{Qnlm_simp1}
Q_{n,l}^{m,\sigma} = & \ \sigma^2 \int_0^{R_{\rm e}}  \rho_0 \xi_{r ; n} \left[ \frac{r^{2+l}}{g} - \frac{r^2}{\Lambda_n} \frac{d^2}{dr^2} \left( \frac{r^{2 + l}}{g} \right)  \right] \dd r \\
 & + \sigma^2 \left[ \frac{1}{\Lambda_n} \frac{d}{dr} \left( \frac{r^{2+l}}{g} \right) r^2 \rho_0 \xi_{r ; n}  - \frac{r^{3+l}}{g} \rho_0 \xi_{\perp ; n} \right]_{r = R_{\rm e}} . \nonumber
\end{align}

\noindent The thermal tide only affects a thin superficial layer of the planet. We thus approximate the internal mass by $ M \approx M_{\rm p} $, which makes the gravity scale as $g \propto r^{-2} $ in this region. This allows us to simplify the derivatives of $ r^{2 + l} / g $ in Eq.~(\ref{Qnlm_simp1}) and to obtain for the integrand

\begin{equation}
\frac{r^{2+l}}{g} - \frac{r^2}{\Lambda_n} \frac{d^2}{dr^2} \left( \frac{r^{2+l}}{g} \right) = \frac{r^{2 + l}}{g} \left[ 1 - \frac{\left( l + 4 \right) \left( l + 3 \right)  }{\Lambda_n} \right]. 
\end{equation}

We now consider the boundary term of Eq.~(\ref{Qnlm_simp1}). Without the contribution of the gravitational tidal potential the horizontal displacement of the $ n$-Hough mode given by Eq.~(\ref{xiperp}) simplifies into

\begin{equation}
\xi_{\perp ; n}^{\rm (eq)} = \frac{1}{\Lambda_n r \rho_0} \frac{d}{dr} \left( r^2 \rho_0 \xi_{r ; n}^{\rm (eq )} \right).  
\end{equation}

\noindent In this equation, the derivative can be expanded as

\begin{equation}
\frac{d}{dr} \left( r^2 \rho_0 \xi_{r ; n}^{\rm (eq )} \right)  = \frac{d}{dr} \left( \frac{r^2 g \rho_0}{i \sigma N^2 T_0 C_{\rm p}} \right) J + \left( \frac{r^2 g \rho_0}{i \sigma N^2 T_0 C_{\rm p}}  \right) \frac{d J}{dr}
\label{derivJ}
\end{equation}

\noindent First, with the approximation done above for the internal mass, $ r^2 g $ is a constant. Second, the background structure that we use of the radiative zone corresponds to an isothermal atmosphere (see Fig.~\ref{fig:bgd}). As a consequence, $ T_0 $ and $ C_{\rm p} $ are constant and $ N^2 = \kappa g /H $ varies slowly with the vertical coordinate compared to the density, which is approximated by the exponential law $ \rho_0 \propto \exp \left( - z / H \right) $, the notation $ z $ standing for the altitude with respect to the base of the isothermal region. Besides, $ J_n  \propto J_{\star} \exp \left[ - \exp \left(- z / H  \right) \right] $, where $ J_{\star} $ designates the tidal heating power per unit mass at $ z = + \infty $. It follows that 

\begin{equation}
\begin{array}{rlc}
  \displaystyle J_n \propto J_{\star} & \mbox{and} & \displaystyle \frac{d J_n}{dr} \propto \frac{J_{\star}}{H} \ed^{-z / H}.
\end{array}
\end{equation}

\noindent Therefore, the ratio between the second and the first term of Eq.~(\ref{derivJ}), denoted $ \varsigma $, scales as $ \varsigma \sim \ed^{-z/H} $, which allows us to approximate $ \xi_{\perp ; n}^{\rm (eq)} $ at $ r = R_{\rm e} $ by

\begin{equation}
\left. \xi_{\perp ; n}^{\rm (eq)} \right|_{r = R_{\rm e}} \approx - \frac{1}{\Lambda_n} \frac{R}{H} \left(  \frac{g J_n}{i \sigma N^2 T_0 C_{\rm p}} \right)_{r = R_{\rm e}}, 
\end{equation}

\noindent and to obtain finally

\begin{align}
\label{Qnlm_app}
Q_{n,l}^{m,\sigma} = & \left[ 1 - \frac{\left( l + 3 \right) \left( l + 4 \right)}{\Lambda_n} \right] \int_0^{R_{\rm e}}  \rho_0 r^{2 + l} \left( \frac{\sigma^2}{N^2} \right) \frac{J_n}{i \sigma T_0 C_{\rm p}} \dd r  \\
 & + \frac{R^{3+l}}{\Lambda_n} \left[ \left( \frac{\sigma^2}{N^2} \right)  \rho_0 \left( 4 + l + \frac{R}{H} \right) \frac{J_n}{i \sigma T_0 C_{\rm p}}  \right]_{r = R_{\rm e}}. \nonumber
\end{align}

\section{Numerical scheme used to integrate the vertical structure equation of tidal waves}
\label{app:num_scheme}

Integrating the vertical structure equation of tidal waves, Eq.~(\ref{vertical_structure}), consists in seeking solutions of the boundary conditions problem defined by the equations

\begin{equation}
\left\{ 
\begin{array}{ll}
\displaystyle \frac{d^2 y}{dx^2} + P \frac{d y}{dx} + Q y = R, & {\rm for} \ x \in \left[ x_{\rm inf} , x_{\rm sup} \right] , \\[0.3cm]
\displaystyle a_{\rm inf} \frac{dy}{dx} + b_{\rm inf} y= c_{\rm inf}, & {\rm at} \ x = x_{\rm inf}, \\[0.3cm]
\displaystyle a_{\rm sup} \frac{dy}{dx} + b_{\rm sup} y = c_{\rm sup}, & {\rm at} \ x = x_{\rm sup}. 
\end{array}
\right.
\label{EDOf}
\end{equation}

\noindent where $ P $, $ Q $ and $ R $ are $ r $-dependent coefficients and $ \left( a_{\rm inf} , b_{\rm inf}, c_{\rm inf} \right) $ and $ \left( a_{\rm sup} , b_{\rm sup}, c_{\rm sup} \right) $ triplets of coefficients defining the lower and upper boundary conditions respectively. To solve this system, we use \padc{a procedure of forward and backward substitution \citep[cf.][Section~2.4]{NumericalRecipes}, also called \textit{Thomas' algorithm}.} This method is interesting because \padc{it takes only $ O \left( N \right) $ operations}, the parameter $ N $ being the size of the mesh. \padc{It is used by \cite{CL70} with a finite difference scheme of the second order.} Here, we adapt it to a finite difference numerical scheme of the $4^{\rm th}$ order. This allows us to improve the stability of the scheme, and thus the treatment of highly oscillating tidal waves\footnote{This regime typically corresponds to gravity waves propagating in a stably-stratified radiative zone characterized by weak dissipative mechanisms (see e.g. Fig.~\ref{fig:structure_ondes}, top right panels).}, for which the curvature terms can be very important.

The domain $ \left[x_{\rm inf} , x_{\rm sup} \right] $ is divided into $ N $ intervals of size $ h $. The points are indexed with the subscript $ n $, such that $ 0 \leq n \leq N $, $ n =0 $ corresponding to the left boundary $ x = x_{\rm inf} $ and $ n = N $ to the upper boundary $ x = x_{\rm sup} $. At the $ 4^{\rm th} $ order, the centered finite differences approximations of the first and second derivatives of $ f $ at the point of index $ n $ are expressed as 

\begin{equation}
\frac{d f}{dx} = \frac{y_{n-2} - 8 y_{n-1} + 8 y_{n+1} - y_{n+2}}{12 h},
\label{dfn}
\end{equation}

\noindent and

\begin{equation}
\frac{d^2 f}{dx^2} = \frac{- y_{n-2} + 16 y_{n-1} - 30 y_{n} + 16 y_{n+1} - y_{n+2}}{12 h^2}.
\label{d2fn}
\end{equation}

Therefore, by substituting Eqs.~(\ref{dfn}) and (\ref{d2fn}) into the differential equation given by Eq.~(\ref{EDOf}), we obtain, for $ 2 \leq n \leq N-2 $, the recurrence relation 

\begin{equation}
A_n y_{n+2} + B_n y_{n+1 } + C_n y_n + D_n y_{n-1} + E_n y_{n-2} = F_n,
\label{rec4}
\end{equation}

\noindent with the coefficients

\begin{equation}
\begin{array}{ll}
  A_n = - 1 - h P_n, & 
  B_n = 8 \left( 2 + h P_n \right), \\[0.3cm]
  C_n = 6 \left( 2 h^2 Q_n - 5 \right), &
  D_n = 8 \left( 2 - h P_n \right), \\[0.3cm]
  E_n = h P_n - 1 , & 
  F_n = 12 h^2 R_n 
\end{array}
\end{equation}

\noindent At the points $ n = 1 $ and $ n = N - 1 $, we use the centered finite difference scheme of the second order, where derivatives are written

\begin{equation}
\frac{df }{dx} = \frac{y_{n+1} - y_{n-1}}{2 h} ,
\end{equation}

\noindent and 

\begin{equation}
\frac{d^2 f}{dx^2} = \frac{y_{n-1} - 2 y_n + y_{n+1}  }{h^2}. 
\end{equation}

\noindent The recurrence relation is thus given by

\begin{equation}
B_n y_{n+1} + C_n y_n + D_n y_{n-1} = F_n,
\label{rec2}
\end{equation}

\noindent with the coefficients 

\begin{equation}
\begin{array}{ll}
   \displaystyle B_n = 1 + \frac{h}{2} P_n , & 
   \displaystyle C_n = h^2 Q_n - 2, \\[0.3cm]
   \displaystyle D_n = 1 - \frac{h}{2} P_n, &
   \displaystyle F_n = h^2 R_n.
\end{array}
\end{equation}

\noindent We introduce then the triplet $ \left( \alpha_n , \beta_n , \gamma_n \right) $ such that 

\begin{equation}
y_n = \alpha_n y_{n+1} + \beta_n y_{n+2} + \gamma_n.
\label{fn_rec}
\end{equation}

\noindent Substituting Eq.~(\ref{fn_rec}) into Eq.~(\ref{rec4}), we express the triplet $ \left( \alpha_n , \beta_n , \gamma_n \right) $ as a function of triplets of smaller indices (i.e. $ n -1 $ and $ n-2 $), 

\begin{align}
& \alpha_n = - K_n^{-1} \left( B_n + D_n \beta_{n-1} + E_n \alpha_{n-2} \beta_{n-1} \right), \\
& \beta_n = - K_n^{-1} A_n, \\
& \gamma_n = K_n^{-1} \left[ F_n - D_n \gamma_{n-1} - E_n \left( \alpha_{n-2} \gamma_{n-1} + \gamma_{n-2} \right)  \right],
\end{align} 

\noindent where $ K_n $ is expressed as

\begin{equation}
K_n = C_n + D_n \alpha_{n-1} + E_n \left( \alpha_{n-2} \alpha_{n-1} + \beta_{n-2} \right). 
\end{equation}

To initialize the series $ \left( \alpha_n , \beta_n , \gamma_n \right) $, we consider the left boundary condition. This condition implies that 

\begin{equation}
\begin{array}{ll}
\displaystyle \alpha_0 = - \frac{2 a_{\rm inf}}{h b_{\rm inf}  - \frac{3}{2} a_{\rm inf}} , & 
\displaystyle \beta_0 = \frac{1}{2} \frac{a_{\rm inf}}{h b_{\rm inf}  - \frac{3}{2} a_{\rm inf}}, \\[0.3cm]
\displaystyle \gamma_0 = \frac{h c_{\rm inf}}{h b_{\rm inf}  - \frac{3}{2} a_{\rm inf}}. 
\end{array}
\end{equation}

\noindent Besides, Eq.~(\ref{rec2}) expressed at $ n = 1 $ provides the coefficients

\begin{equation}
\begin{array}{lll}
 \displaystyle  \alpha_1 = - \frac{B_1 + D_1 \beta_0}{C_1 + D_1 \alpha_0}, & \beta_1 = 0, & \displaystyle \gamma_1 = \frac{F_1 - D_1 \gamma_0}{C_1 + D_1 \alpha_0}. 
\end{array}
\end{equation}

\noindent Hence, the coefficients $ \left( \alpha_n , \beta_n , \gamma_n \right) $ are computed forward from $ n = 2 $ to $ n = N-2 $. 

The terms $ y_{\rm N-1} $ and $ y_{\rm N} $ of the series $ y_n $ shall now be determined. We use the upper boundary condition and Eq.~(\ref{rec2}) at $ n = N-1 $ to obtain the algebraic system

\begin{equation}
\left\{
\begin{array}{l}
 \displaystyle \mathcal{A}_1 y_{N-1} + \mathcal{B}_1 y_N = \mathcal{C}_1  , \\[0.3cm]
  \displaystyle \mathcal{A}_2 y_{N-1} + \mathcal{B}_2 y_N = \mathcal{C}_2,
\end{array}
\right.
\end{equation} 

\noindent where the coefficients $ \mathcal{A}_1 $, $ \mathcal{B}_1 $, $ \mathcal{C}_1 $, $ \mathcal{A}_2 $, $ \mathcal{B}_2 $, and $ \mathcal{C}_2 $ are given by

\begin{equation}
\begin{array}{ll}
 \displaystyle \mathcal{A}_1 = a_{\rm sup} \left( \frac{1}{2} \alpha_{N-2} - 2 \right) , & 
 \displaystyle \mathcal{A}_2 = C_{N-1} + D_{N-1} \alpha_{N-2}, \\[0.3cm]
 \displaystyle \mathcal{B}_1 = \frac{1}{2} a_{\rm sup} \left( 3 + \beta_{N-2} \right) + h b_{\rm sup}, &
 \displaystyle \mathcal{B}_2 = B_{N-1} +D_{N-1} \beta_{N-2}, \\[0.3cm]
 \displaystyle \mathcal{C}_1 =  h c_{\rm sup} - \frac{1}{2} a_{\rm sup} \gamma_{N-2} , &
 \displaystyle \mathcal{C}_2 = F_{N-1} - D_{N-1} \gamma_{N-2}.
\end{array}
\end{equation}

\noindent It follows that 

\begin{equation}
\begin{array}{lcl}
\displaystyle  y_{N-1} = \frac{\mathcal{B}_2 \mathcal{C}_1 - \mathcal{B}_1 \mathcal{C}_2}{\mathcal{A}_1 \mathcal{B}_2 - \mathcal{A}_2 \mathcal{B}_1}, & \mbox{and} & \displaystyle y_N = \frac{\mathcal{A}_1 \mathcal{C}_2 - \mathcal{A}_2 \mathcal{C}_1}{\mathcal{A}_1 \mathcal{B}_2 - \mathcal{A}_2 \mathcal{B}_1}. 
\end{array}
\end{equation}

\noindent The $ y_n $ are finally integrated backward from $ n = N-2 $ to $ n = 0 $ with the recurrence relation given by Eq.~(\ref{fn_rec}).


\end{document}